\documentclass[12pt,cite,epsfig]{article}
\usepackage{times,epsfig}
\usepackage{caption}
\usepackage{wasysym}
\newcommand{\nc}{\newcommand}
\nc{\bb}{\bibitem}
\nc{\be}{\begin{equation}}
\nc{\ee}{\end{equation}}
\nc{\pa}{\partial}
\nc{\parsym} {\stackrel{\leftrightarrow}{\pa}}
\nc{\ra}{\rightarrow}
\nc{\la}{\leftarrow}
\nc{\etp}{{\eta^\prime}}
\nc{\omg}{\omega}
\nc{\ggam}{\gamma \gamma}
\nc{\gam}{\gamma }
\nc{\I}{{\rm i} }
\nc{\beas}{\begin{eqnarray*}}
\nc{\eeas}{\end{eqnarray*}}
\nc{\ba}{\begin{eqnarray}}
\nc{\ea}{\end{eqnarray}}
\nc{\non}{\nonumber}
\nc{\mv}{{\rm MeV }}
\nc{\second}{{\prime\prime}}

\def\hhhd{\rule[-3.mm]{0.mm}{2.mm}}
\def\hhhe{\rule[-3.mm]{0.mm}{4.mm}}

\def\hhhv{\rule[-3.mm]{0.mm}{9.mm}}
\def\hhhw{\rule[-1.mm]{0.mm}{5.mm}}
\def\hhhq{\rule[-5.mm]{0.mm}{12.mm}}

\newcommand{\epo}{\;.}

\nc{\E}{{\rm e} }
\nc{\cF}{{\cal F} }
\nc{\cM}{{\cal M} }

\newcommand{\crn}{\nonumber \\}
\newcommand{\noi}{\noindent }
\begin{document}
\begin{titlepage}
\vbox{~~~ \\
%
\title{ BHLS$_2$, a New Breaking of the HLS Model and its Phenomenology}
 \author{
M.~Benayoun$^a$, L.~DelBuono$^a$, F.~Jegerlehner$^{b,c}$ \\
\small{$^a$ LPNHE des Universit\'es Paris VI et Paris VII, IN2P3/CNRS, F--75252 Paris, France }\\
\small{$^b$ Humboldt--Universit\"at zu Berlin, Institut f\"ur Physik, Newtonstrasse 15, D--12489 Berlin,
Germany }\\
\small{$^c$ Deutsches  Elektronen--Synchrotron (DESY), Platanenallee 6, D--15738 Zeuthen, Germany}
}
\date{\today}
\maketitle
\begin{abstract}
Previous studies have shown that the  Hidden Local
Symmetry (HLS) Model, supplied with appropriate
symmetry breaking mechanisms, provides an Effective Lagrangian
(BHLS) able to encompass a large number of processes within a unified
framework. This allowed to design  a global fit procedure
which provides a fair simultaneous description of the $e^+ e^-$
annihilation  into 6 final states ($\pi^+\pi^-$, $\pi^0\gamma$,
$\eta \gamma$, $\pi^+\pi^-\pi^0$, $K^+K^-$, $K_L K_S$),
the dipion spectrum in the $\tau$ decay and some more
light meson decay partial widths. In this paper, additional breaking schemes are
defined which improve the BHLS working and extend its scope
so as to absorb spacelike processes within a new framework (BHLS$_2$).
The phenomenology previously explored with BHLS is fully
revisited in the BHLS$_2$ context with special emphasis on the $\phi$ mass region  using
all  available data samples.  It is  shown that BHLS$_2$ addresses
perfectly the close spacelike region covered by NA7 and Fermilab data; it
is also shown that the recent Lattice QCD (LQCD) information on the pion form factor
are accurately {\it predicted} by the BHLS$_2$ fit functions derived from fits to
{\it only} annihilation data. The  contribution
to the muon anomalous magnetic moment  $a_\mu^{\rm th}$ of these
annihilation channels over the range of validity  of BHLS$_2$
(up to $\simeq$ 1.05 GeV) is updated within the new BHLS$_2$ framework
and shown to strongly reduce the former BHLS systematics. The
uncertainty on  $a_\mu^{\rm th}(\sqrt{s}< 1.05$ GeV) is much
improved compared to  standard approaches relying on direct
integration  methods of measured spectra. Using the BHLS$_2$
results, the leading order HVP contribution to the muon anomalous moment
is $a_\mu^{\rm HVP-LO}= 686.65 \pm 3.01 +(+1.16,-0.75)_{\rm syst}$ in units
of $10^{-10}$. Using a conservative
estimate for the light--by--light contribution, our evaluation for the muon anomalous
magnetic moment is $a_\mu^{\rm th}=\left [11\,659\,175.96 \pm 4.17
+(+1.16,-0.75)_{\rm syst}\right] \times
10^{-10}$. The relationship between the dispersive and LQCD approaches to the
$\rho^0-\gamma$ mixing is also discussed which may amount to a shift of
$\delta a_\mu[\pi\pi]_{\rho\gamma}=+(3.10\pm 0.31) \times10^{-10}$ at LO+NLO,  
presently treated as additional systematics. Taking also this shift into account,
the difference $a_\mu^{\rm th}-a_\mu^{\rm BNL}$ exhibits a significance not smaller
than $3.8 \sigma$.
\end{abstract}
}
\end{titlepage}

\section{Introduction}
\label{introduction}
\indent \indent The Standard Model is widely recognized as the (gauge) theory which unifies
 the whole  realm  of weak, electromagnetic and strong interactions among
quarks, leptons and the various gauge bosons (gluons, photons, $W^\pm$, $Z^0$).
For the physics processes -- and quantities -- involving strong interactions, QCD is at work
under two different regimes  tightly connected with the energy involved and  the onset of the perturbative
regime is {\it a priori} expected  to occur at high energies. However, as clearly shown by
the data recently collected by  KEDR \cite{KEDR2016,KEDR2015} on the ratio
$ R(s)=\sigma(e^+e^- \ra {\rm hadrons})/\sigma(e^+e^- \ra \mu^+\mu^-)$,
 the $\{u,d,s\}$ sector  of this ratio reaches the
perturbative regime at energies as low as  $\simeq 2.0$ GeV (see   \cite{RPP2016,Jegerlehner_2017},
for instance); above this energy,  the observed departures from perturbative QCD predictions -- including, of course,
 the $c \overline{c}$ and $b \overline{b}$  threshold effects therein --
are only spikes and narrow bumps associated with the charmonium and bottomonium states
which require an additional  specific treatment in the mass range where they are located.

However, in  the low energy region  where the non-perturbative
regime of QCD is involved, getting   theoretical predictions able to compete with the accuracy of
some important experimental measurements  may be challenging. This is, in particular,  the issue met
with the photon hadronic vacuum polarization (HVP), which contributes  importantly  to the muon
anomalous magnetic moment $a_\mu$, one of the best measured particle properties. The
hadronic part of  $a_\mu$ is related to the so-called $R(s)$ ratio and is given (at leading order) by~:
\be
a_\mu^{\rm had} = \displaystyle
\left [ \frac{\alpha m_\mu}{3 \pi} \right ]^2
\int_{s_0}^\infty
\frac{ds}{s^2 } \hat{K}(s) R(s)\,,
\label{eq1}
\ee
where $s_0=m_{\pi^0}^2$ is the lowest hadronic threshold and $\hat{K}(s)$ is a known smoothly
varying positive function \cite{FredBook}. The $1/s^2$ factor strongly enhances the low energy contribution
of the spectrum, {\it i.e.} just in the validity domain of non-perturbative
QCD.  Fortunately, as obvious from  the KEDR data \cite{KEDR2016,KEDR2015}, one can undoubtedly consider
that the non-perturbative regime does not extend to energies larger than $\simeq  2.0 $ GeV.

So,  the real issue with predictions for objects such as $a_\mu$ is to get precise estimates
of the effects covered by the non-perturbative regime of QCD. Chiral Perturbation Theory
(ChPT)\cite{GL1,GL2}, the low energy limit of QCD,  is of limited help for the present purpose
as its realm does not extend much beyond the 400 $\div$ 500 MeV region and misses the quite
important  meson resonance mass region.

The most promising approach to the non-perturbative regime of QCD is certainly  Lattice QCD (LQCD)
which already provides valuable information at low energies \cite{Colangelo_final}.
The  recently derived LQCD evaluations of $a_\mu^{\rm had}$
\cite{Lattgm2-1,Lattgm2-2,Chakraborty2016, DellaMorte2017}  have been found in accord with the
experimental measurement  performed at BNL  \cite{BNL,BNL2}. However, the magnitude
of the reported uncertainties is by far too large to fruitfully compare with
the already existing BNL datum and, {\it a fortiori}, with the measurements expected from
the  Fermilab experiment \cite{LeeRoberts,Fermilab_gm2}, already running, or from
the experiment  planned to start later on at J-PARC \cite{Iinuma},
as both are expected to improve the uncertainties by a factor of 4. So much progress remains
 to be done before getting satisfactory uncertainties from LQCD.

 \vspace{1.cm}

This leaves room  for low energy Effective Resonance Lagrangian Approaches,
which can contribute to improve the knowledge of $a_\mu^{\rm had}$ by providing
a good description of the rich amount of experimental data collected in the
timelike region.
Among the richest possible Lagrangians, the elegant Hidden Local Symmetry (HLS)
Model \cite{HLSRef} is worth to be considered;
it has been proven to be equivalent to  R$\chi$PT \cite{Ecker1,Ecker2},
provided consistency with the QCD asymptotic behavior is incorporated. It thus follows that the
HLS model is  a motivated and constraining QCD rooted framework, moreover, easy enough to handle
in phenomenological applications.

The original HLS Model deals with the lowest vector meson nonet and provides a framework
for hadron production in $e^+e^-$ annihilation, naturally bounded by the $\phi$ mass region
-- {\it i.e.}  up to $\simeq 1.05$ GeV.
It thus represents  a  tool giving a handle on
a mass region contributing for $\simeq 83\%$ of the total muon hadronic VP.
The region extending from just above the $\phi$ meson mass to 2 GeV only contributes
for 7\%, slightly less than the $[2 {\rm ~GeV}, \infty]$  region (10\%).

As such, the non-anomalous HLS Lagrangian  \cite{HLSOrigin} sets up a unified framework
which encompasses the $e^+e^- \ra \pi^+  \pi^-$, $e^+e^- \ra K^+  K^-$ and
$e^+e^- \ra K^0 \overline{K}^0$ annihilation  channels;
the $\tau^\pm \ra \pi^\pm  \pi^0 \nu$ spectrum also belongs to the same framework, allowing naturally
 a simultaneous treatment of all the mentioned annihilation and decay processes. On the other hand,
the HLS Model possesses an anomalous sector \cite{FKTUY,HLSRef} which also brings
 the $e^+e^- \ra \pi^0  \gamma$, $e^+e^- \ra \eta  \gamma$ and
$e^+e^- \ra \pi^+  \pi^- \pi^0$ annihilation channels inside
the same framework, together with some radiative
decay modes and the dipion spectra in the $\eta/\etp \ra \pi^+  \pi^- \gamma$ decays
as shown in former studies \cite{ExtMod1,ExtMod2}.
The annihilation channels just listed exhaust almost completely the processes
contributing to $a_\mu^{\rm had}$ below  1.05 GeV; indeed the missing
channels\footnote{The $e^+e^- \ra \etp  \gamma$ channel has been considered
in \cite{ExtMod3,ExtMod4} and  found to contributes
to $a_\mu^{\rm had}$ at the $10^{-12}$ level only and can thus be safely discarded.}
($4 \pi$, $2 \pi \eta$, \ldots) contributes only  $\simeq 2  \permil $ of the full HVP.
Such a broken HLS (BHLS) Model has already been built up \cite{ExtMod3,ExtMod4}
and shown to provide a pretty good simultaneous  description of almost all data samples
covering the six channels listed above up to $\simeq 1.05$ GeV. This work has also proved that
there was no contradiction between the  $e^+e^- \ra \pi^+  \pi^-$ and
$\tau^\pm \ra \pi^\pm  \pi^0 \nu$ spectra.

The non-anomalous \cite{HLSOrigin} and anomalous \cite{FKTUY}
sectors of the HLS Model thus open a unified framework able to encompass
a large corpus of data and  physics processes. However, as such,
the HLS framework -- with only the universal vector coupling $g$ as
 free parameter -- cannot pretend to provide a satisfactory simultaneous
 description of the wide ensemble of high statistics  data samples collected by
several sophisticated experiments in several annihilation channels.

In order to achieve such a program,  the HLS model  must be supplied
with appropriate symmetry breaking mechanisms not present in its original
formulation \cite{HLSRef}. A first successful attempt has been done in
\cite{ExtMod3,ExtMod4,ExtMod5} which has set up a model (BHLS)
based on a breaking mechanism originally proposed in
\cite{BKY}, hereafter named BKY.

The present work reports on
a new breaking scheme for the HLS model which aims at improving the behavior
of the form factors in the dipion threshold region and also in the $\phi$ region where
the original BHLS \cite{ExtMod3} meets some difficulty leading to introduce additional
systematic uncertainties in the evaluation of the muon $g-2$. The model (BHLS$_2$)
 presented here will be shown to widen the scope of BHLS, in particular to the spacelike
 domain and improve the BHLS prediction for the muon $g-2$. Indeed, using
 variants of BHLS$_2$, together with the original BHLS should allow evaluating the model
 dependence effects in the estimates of derived physics quantities, noticeably the
 photon HVP.

The layout of the paper is as follows. Sections \ref{HLSMod} gives  a brief reminder of
the original HLS model while Section \ref{BKYbrk} briefly reminds the BKY mechanism, still involved
in BHLS$_2$. Sections \ref{Vbrk}   and \ref{HLSLag}  introduce a new breaking mechanism at the
level of the covariant derivative which is the basic object leading to the original HLS model.
Section \ref{zterms}  describes the ${\bf O}(p^4)$ terms \cite{HLSRef} of the HLS framework
which are included in the  BHLS$_2$ Lagrangian.

At this step, one has at hand the first variant of the BHLS$_2$ model, named Basic Solution (BS),
and, relying on its vector meson mass term,  Section \ref{BasicSol} defines its parameter
properties.  Remarking that there is no fundamental reason why the neutral vector fields involved in
physics processes should be their ideal combinations ($\rho_0^I$, $\omg^I$, $\phi^I$), one allows
for combinations of these via a mechanism called Primordial Mixing described in Section \ref{RefSol}.
This gives rise to a second variant of the BHLS$_2$ model, named Reference Solution (RS). In Section
\ref{propagators},  one first reminds the parametrization of the propagators for the $\omg$ and $\phi$
fields \cite{ExtMod3} in connection with form factor behaviors at $s=0$;  this part of the work also shows
that the breaking of nonet symmetry in the vector meson sector must be accompanied by
breaking of SU(3) of the same intensity.

The dynamical breaking of vector meson  \cite{taupaper,ExtMod3} is
revisited in Section \ref{DynMix} and the role of anomalous loop corrections
is emphasized, especially in order to connect smoothly the timelike and spacelike
branches of the pion form factor.
The derivation of the pion and kaon form factors is the subject of Section \ref{form_factors}.
The anomalous sector is fully analyzed in Section \ref{anomalous} and the cross sections for
the $e^+e^- \ra (\pi^0/\eta) \gam$ and $e^+e^- \ra  \pi^+  \pi^- \pi^0 $ annihilations
 are derived.

 Altogether, the above-listed Sections and the Appendix fully report on  the BHLS$_2$ model properties
 and tools. Section \ref{DataSamples} provides a comprehensive discussion of the available data
 samples falling into the BHLS$_2$ scope and their peculiarities. Relying on several preliminary studies,
 one comments here on the 3 discarded data samples; one should note that the number of data samples
 found to accommodate with each other within the BHLS$_2$ framework exceeds now 50 and the total
 number of data points reaches 1237.

Sections \ref{global_fits} and \ref{kkb_data} give a full description of all aspects of the fits
to the 6 annihilation channels performed within the global BHLS$_2$ framework. Moreover,
Section \ref{kkb_data} also provides  a comprehensive analysis
within our global framework of the existing $K \overline{K}$ data samples.
The aim is to substantiate the issue between the BaBar, CMD-3 and CMD-2
spectra revealed by fits. A summary of the various fit properties is presented in Section
\ref{genfitall}.

The description of the spacelike region for the pion and kaon form factors as coming
from BHLS$_2$ is the purpose of Section \ref{spacelike}.
The existing model-independent data
for the $\pi^\pm$ and $K^\pm$  form factors are found to naturally  accommodate our
global framework, giving support to the low energy behavior predicted by BHLS$_2$
for all meson form factors.
This fair agreement extends to the $\pi^\pm$ form factor data provided by several Lattice
QCD (LQCD) groups.
One may infer from these data and from other model-dependent data
that  BHLS$_2$ should almost certainly apply
down to $\simeq -1$ GeV$^2$. The prediction for the $K^0$  form factor is shown to cope
with expectations at $s\simeq 0$. Information on the charge radii of the $\pi^\pm$ and $K^\pm$
mesons
is provided. Section \ref{phaseshift} examines the $P$-wave $\pi\pi$ phase shift prediction and
shows its agreement with data and other predictions, including our former BHLS, over
an energy range extending up to the $\phi$ mass region.

The values found for some parameters, noticeably the $a$ HLS parameter and some BKY
breaking parameters, deserve a specific account given in Section \ref{LagVal}.

The main motivation of the present work is the study of the BHLS$_2$ predictions
and their consequences on the muon $g-2$.  This topic is addressed in Section \ref{LO-HVP}.
The part ($\simeq 83$\%) of the muon HVP contribution to $a_\mu$ covered by BHLS$_2$ is 
estimated
in several contexts to derive $a_\mu^{\rm HVP-LO}(\sqrt{s} < 1.05 ~{\rm GeV})$ and estimate
additional systematics possibly due to modeling effects and to observed tension among some data samples.
Complementing this piece by the non-HLS part of the HVP derived by more usual means, one
gives our best evaluation of the full HVP and our estimate of the muon anomalous
 magnetic moment.
We also discuss an issue in the relationship between the LO-HVP
results derived by dispersive approaches -- in particular BHLS/BHLS$_2$ -- and the
LQCD results, where the $\rho^0-\gamma$ mixing appears as a NLO effect.
Comparison is done with other published evaluations for $a_\mu^{\rm HVP-LO}$.
Finally, Section \ref{summary} summarizes our results and conclusions.

\section{Outline of the Hidden Local Symmetry (HLS) Model}
\label{HLSMod}
\indent \indent Let us briefly outline  the derivation of the HLS
model\footnote{See, for instance, \cite{HLSRef} for a full derivation.}.
 The usual ChPT Lagrangian \cite{GL1,GL2} can be written
in two different manners \cite{HLSRef,Heath1998}~:
\be
\displaystyle
{\cal L}_{\rm chiral}=\frac{f_\pi^2}{4} {\rm Tr} \left [ \pa_\mu U~  \pa^\mu U^\dagger \right ]
=-\frac{f_\pi^2}{4} {\rm Tr} \left [ \pa_\mu \xi_L ~ \xi_L^\dagger - \pa_\mu \xi_R ~ \xi_R^\dagger
\right ]^2\,,
\label{eq1-1}
\ee
where $f_\pi$ (=92.42 MeV) is the pion decay constant and~:
\be
\begin{array}{lll}
\displaystyle U(x)= \exp{\left [ 2iP(x)/f_\pi \right]}~~,
& \xi_{R/L}(x)=\exp{\left [ \pm iP(x)/f_\pi \right]}~~\Longrightarrow
U(x)=\xi_L^\dagger(x) \xi_R(x)\,,
\end{array}
\label{eq1-2}
\ee
when working in the so-called unitary gauge  which removes a scalar field term  in the definition of
$\xi_{R/L}(x)$; $P(x)$ is the  pseudoscalar (PS) field matrix.
The hidden local symmetry which gives its name
to the HLS Lagrangian has transformation properties which affect   $\xi_{R/L}(x)$   letting $U$
unchanged. Ignoring for the moment the weak sector to ease the discussion,
the HLS Lagrangian is derived by replacing  in  Equation (\ref{eq1-1}) the usual derivative  by the
covariant derivative~:
\be
\displaystyle
 D_\mu \xi_{R/L} =\pa_\mu \xi_{R/L} -ig V_\mu \xi_{R/L} +ie \xi_{R/L} A_\mu Q\,,
\label{eq1-3}
\ee
where $A_\mu$ is the photon field, $Q={\rm Diag}[2/3,-1/3,-1/3]$  the quark charge matrix and
$V_\mu$ is the  vector field matrix; the expressions for $P$ and $V$ are the usual
ones -- fulfilling the $U(3)$ flavor symmetry --
and can be found in \cite{HLSRef,Heath1998,ExtMod3}. In the expressions for $D_\mu \xi_{R/L}$,
the universal vector  coupling $g$
occurs beside the unit electric charge $e$. The $\omg$ and $\phi$ fields occurring in the
diagonal of the $V$ matrix, also denoted $V^I$ below, are the so-called ideal combinations
generally denoted $\omg_I$ and $\phi_I$.

Substituting the covariant derivative Equation (\ref{eq1-3})  to the usual derivative in Equation (\ref{eq1-1}),
 ${\cal L}_{\rm chiral}$ becomes the first HLS Lagrangian piece denoted
${\cal L}_A$ while  another piece ${\cal L}_V$ shows up which vanishes
in the reversed substitution  $D_\mu \Rightarrow  \pa_\mu$~:
\be
\displaystyle
\begin{array}{lll}
\displaystyle
{\cal L}_A=-\frac{f_\pi^2}{4} {\rm Tr} \left [ D_\mu \xi_L ~ \xi_L^\dagger - D_\mu \xi_R ~ \xi_R^\dagger
\right ]^2~~,&
\displaystyle
\displaystyle
{\cal L}_V=-\frac{f_\pi^2}{4} {\rm Tr} \left [ D_\mu \xi_L ~ \xi_L^\dagger + D_\mu \xi_R ~ \xi_R^\dagger
\right ]^2\epo
\end{array}
\label{eq1-4}
\ee
The full HLS Lagrangian is then defined by~:
\be
\displaystyle {\cal L}_{\rm HLS}={\cal L}_A +a {\cal L}_V\,,
\label{eq1-5}
\ee
where $a$ is a parameter specific of the HLS approach \cite{HLSRef}. The usual VMD
framework is obtained by setting $a=2$ \cite{Heath1998}  while the presently reported phenomenology
prefers larger values \cite{ExtMod2,taupaper}. The explicit expression for this unbroken
$ {\cal L}_{\rm HLS}$  can be found fully developed in \cite{Heath1998}. Let us note
that $ {\cal L}_{\rm HLS}$ contains a photon mass term; however this Reference also showed that
the loop dressing (in the HLS context) cancels out the physical photon mass and thus, this mass term
can be safely ignored in phenomenological studies.

So, the unperturbed ({\it i.e.} unbroken) HLS Lagrangian depends on only two parameters
($g$ and $a$)  which can be adjusted using data;
it is obviously unrealistic to expect this basic  $ {\cal L}_{\rm HLS}$ to describe
precisely the large amount of data covering its scope with so few parameters. To have any
chance to
account for real data, additional  input should enter the HLS Lagrangian  to make it more
 flexible;  this is what motivates the introduction  of breaking procedures within the HLS framework.

As already noted, both HLS Lagrangian pieces fulfill a $U(N_f) \times U(N_f)$ symmetry and
not $SU(N_f) \times SU(N_f)$. To reduce this symmetry which introduces a ninth  PS
meson, one  includes \cite{WZWChPT,ExtMod3}  the 't~Hooft determinant terms \cite{tHooft} which
break
 the axial $U(1)$ symmetry; this turns out to complement the ${\cal L}_A$ Lagrangian piece
 with~:
\be
\displaystyle
{\cal L}_{\rm 'tHooft}=\frac{\mu^2}{2} \eta_0^2+ \frac{\lambda}{2} \pa_\mu \eta_0\pa^\mu \eta_0\epo
\label{eq1-4-b}
\ee
The singlet mass term  manifestly breaks nonet symmetry in the PS sector. Nothing analogous
affects the vector sector.

\section{Breaking the HLS Lagrangian~I~: The BKY Mechanism}
\label{BKYbrk}
\indent \indent Up to now, a single breaking mechanism  for the HLS Lagrangian has been
proposed \cite{BKY}; it is named  here BKY after its proponent names. However,
undesirable properties of the original  BKY proposal have led to  modifications
of this breaking scheme  \cite{BGP,Heath} originally proposed
 to break  (only) the flavor symmetry of the Lagrangian.
 The "new scheme" variant from  \cite{Heath} has been analyzed with data
and its results shown to compare fairly well to ChPT expectations \cite{WZWChPT}.
It has been extended
to include isospin breaking effects in \cite{ExtMod3} following the lines of
\cite{Hashimoto}. This (modified) BKY breaking scheme, intended to cover isospin and
 $SU(3)$ breaking effects within the same framework, has been proved fairly
 successful, allowing for a high quality
 global fit of (almost) all available data covering the validity range  of the HLS model
 \cite{ExtMod3,ExtMod4,ExtMod5}~:  $e^+e^-$ annihilations, $\tau$ decay spectra and
 radiative decays of light flavor mesons; in particular, the dipion spectra in
 the $\eta/\etp \ra \pi^+
 \pi^- \gamma $ decays are fairly well predicted \cite{ExtMod1,ExtMod2}, especially the
 detailed  form of the drop-off in the $\rho-\omg$ interference region as recently
 measured at BESIII \cite{Ablikim_etp,Dai_etp}. Nevertheless, as already stated in
 \cite{ExtMod3,ExtMod4,ExtMod5}, the description of the threshold region in
 the dipion spectrum and of the $\phi$ mass region  in the three pion annihilation
 channel deserves improvements; this issue is addressed by the present paper.

 As BKY is one of the breaking mechanisms  used in the present work,
 let us briefly remind  how it is implemented. To lighten writing, we use the notations
 $L=D_\mu \xi_L ~ \xi_L^\dagger$ and $R=D_\mu \xi_R ~ \xi_R^\dagger$.
The (modified and extended) BKY breaking which also underlies the broken HLS (BHLS) model
  developed in \cite{ExtMod3} turns out to modify Equations (\ref{eq1-4}) as follows~:
 \be
\displaystyle
\begin{array}{lll}
 \displaystyle
{\cal L}_A=-\frac{f_\pi^2}{4} {\rm Tr} \left [ (L - R) X_A\right ]^2
~~,&
 \displaystyle
{\cal L}_V=-\frac{f_\pi^2}{4} {\rm Tr} \left [ (L + R) X_V\right ]^2
~~~,
 \end{array}
\label{eq1-6}
\ee
where $X_{A/V}=$Diag$(q_{A/V},y_{A/V},z_{A/V})$ are constant real matrices. In practice, one prefers setting
$q_{A/V}=1+(\Sigma_{A/V}+ \Delta_{A/V})/2$ and $y_{A/V}=1+(\Sigma_{A/V}- \Delta_{A/V})/2$.
As $z_{A/V}$ are affecting the $s\overline{s}$ entries, their departure from 1 can be large
compared to $q_{A/V}$ and $y_{A/V}$ -- which refer to resp. the $u\overline{u}$ and
$d\overline{d}$ entries; so,  the $\Sigma$'s and $\Delta$'s are expected small. Within
the  previous broken HLS framework -- hereafter named BHLS -- \cite{ExtMod3,ExtMod4}, one got
$z_A\simeq [f_K/f_\pi] \simeq 1.5$,  $z_V\simeq 1.2$  while  the $\Sigma$'s
and $\Delta$'s were found at the few percent level.

Once these breakings are applied, the  PS kinetic energy term contained in ${\cal L}_A$
is no longer diagonal and a PS field
redefinition has to be performed in order to restore the kinetic energy term to canonical
form. The procedure is fully described in Section 4 of \cite{ExtMod3} and is valid unchanged
in the present work; it is not  repeated here.

In the BHLS model \cite{ExtMod3}, after the $X_V$ breaking, the vector meson mass term is no
longer diagonal and writes ($m^2=a g^2 f_\pi^2$)~:

 \be
\displaystyle
{\cal L}_V^{\rm mass}= \frac{m^2}{2} \left [
(1+\Sigma_V) (\rho^2_I +\omg^2_I)+ 2 \Delta_V \rho_I \cdot \omg_I +z_V \phi_I^2+
2(1+\Sigma_V) \rho^+ \rho^-
\right]\,,
\label{eq1-7}
\ee
discarding the unessential $K^*$ sector and neglecting terms of degree greater than 1
in the breaking parameters.
This mass term can be diagonalized by a 45$^\circ$ rotation, a quite
unacceptable solution as it does not give a smooth limit in the symmetry limit $\Delta_V \ra 0$.
A smooth solution
is obtained by the following transform   to renormalized (R) fields~:
  \be
  \left (
\begin{array}{l}
 \displaystyle \rho_I \\[0.5cm]
 \displaystyle \omg_I
\end{array}
\right )
=
 \left (
 \begin{array}{l}
 \displaystyle \rho_R \\[0.5cm]
 \displaystyle \omg_R
\end{array}
\right )
- \Delta_V  \left (
 \begin{array}{c}
 \displaystyle h_V\omg_R \\[0.5cm]
 \displaystyle (1-h_V)\rho_R
\end{array}
\right )\,,
\label{eq1-8}
\ee
where $h_V$ is a parameter submitted to fit, together with $z_V$, $\Sigma_V$ and $\Delta_V$.
This solution, adopted in  \cite{ExtMod3,ExtMod4,ExtMod5}, provides a quite satisfactory
description of the data.  This transformation introduces a $\Delta_V \pa  \rho_R \pa  \omg_R$ term in
the kinetic energy term of the vector mesons; this issue is known to imply
the occurrence of wave-function renormalization factors absorbed in the effective
couplings \cite{Sakurai}; in the case of BHLS, they can be considered absorbed in the
breaking parameters.

The full account of the BHLS model also requires the dynamical breaking \cite{taupaper,ExtMod3}
generated by PS loop effects
which also calls for an additional  $s$-dependent  renormalization step. It is rediscussed below
within the context of the present work.

\section{Breaking the HLS Lagrangian~II~: The Covariant Derivative (CD) Handle}
\label{Vbrk}
\indent \indent As clear in Equations (\ref{eq1-6}),   BKY  breaks the symmetry
of the HLS Model  in the very definition of its Lagrangian. This is
quite legitimate  and strongly validated by its remarkable
description of a considerable amount of data \cite{ExtMod3,ExtMod4,ExtMod5}.
Nevertheless, it is of interest to explore other possibilities  to improve
the description of the data in some specific mass regions and, then, reduce
or cancel out  the additional systematic uncertainties reported in \cite{ExtMod5}.

Let us focus on  Equation (\ref{eq1-3}) which defines the original HLS covariant derivative~:
 $$D_\mu \xi_{R/L} =\pa_\mu \xi_{R/L} -ig V_\mu^I \xi_{R/L} +ie \xi_{R/L} A_\mu Q~~~.$$
The superscript  $I$ assigned to the vector field matrix $V$ aims at reminding that the Isospin zero mesons
are the so-called ideal combinations $\omg^I$ and $\phi^I$. These naturally occur in the $U(3)$ symmetric
expression of $V$.
\vspace{0.5cm}

The vector meson term in $D_\mu \xi_{R/L}$
can be written $g V^I= g \sum_{a=0,8} V_a T_a$ where the $V_a$'s are the vector meson
fields and  $T_a,~ (a=1,\cdots 8)$ are the Gell-Mann
matrices normalized such that ${\rm Tr}[T_a T_b]=\delta_{ab}/2 $;   $T_0$ is  the
unit matrix appropriately normalized~: $T_0=I/\sqrt{6}$. Thus, the (nonet)  $U(3)_V$  symmetry
of HLS  is obtained by~:
\begin{itemize}
\item
Plugging the appropriately constructed $V^I$ into the covariant derivative
$D_\mu \xi_{R/L}$,
\item
Assuming the universality of the vector coupling $g$ to the external world.
\end{itemize}
Therefore, we propose a direct breaking of the covariant derivative, a new tool
independent of the BKY mechanism.
Such a breaking mechanism is expressed by the modified covariant derivative~:
 \be
\displaystyle
D_\mu \xi_{R/L} =\pa_\mu \xi_{R/L} -ig \left[V_\mu^I +\delta V_\mu \right ] \xi_{R/L} +ie \xi_{R/L} A_\mu Q\,,
\label{eq1-9}
\ee
where $\delta V_\mu$ can be chosen to break the $U(3)_V$ symmetry in a controlled way.
For instance,  breaking solely the nonet symmetry  of  $V^I$ turns out to allow
the coupling to the  singlet $V^0_\mu$ to differ from those of the octet fields, preserving
in this way the $SU(3)_V$ symmetry. It will become clearer below why  a more systematic approach
must be preferred.

Identifying the field combinations associated with each of the canonical $T_a$ matrices,
one is led to define the following components which can participate
to $\delta V_\mu$ separately or together~:
\be
\left \{
\displaystyle
 \begin{array}{l}
\displaystyle \delta V_\mu^0 =\frac{\xi_0}{\sqrt{2}}
\left[ \frac{\sqrt{2} \omg_\mu^{I} + \phi_\mu^{I}}{3}\right]{\rm Diag} [1,1,1]
\,,\\[0.5cm]
 \displaystyle
\delta V_\mu^8 =\frac{\xi_8}{\sqrt{2}}
 \left[ \frac{\omg_\mu^{I} - \sqrt{2} \phi_\mu^{I}}{3\sqrt{2}}\right]{\rm Diag} [1,1,-2] \,,\\[0.5cm]
\displaystyle
 \delta V_\mu^3 =\frac{\xi_3}{\sqrt{2}} \left[\frac{\rho_I^0}{\sqrt{2}}
 \right]{\rm Diag} [1,-1,0]\,,
\end{array}
\right .
\label{eq1-10}
\ee
in terms of the usual ideal field combinations.

The (free) breaking parameters $\xi_0$, $\xi_8$ and  $\xi_3$ are  only requested to be real
in order that $\delta V_\mu$ is hermitian as $V_\mu^I$ itself. Clearly, $\delta V_\mu^0$ defines
a breaking of the nonet symmetry down to $SU(3)_V \times U(1)_V$,  $\delta V_\mu^8$ rather
expresses the breaking of the  $SU(3)_V$ symmetry, while $\delta V_\mu^3$ is related to
a direct  breaking of Isospin symmetry in the vector sector.

One could also introduce some breaking affecting the  $\rho^\pm$ entries of the $V^I$
matrix\footnote{For the sake of simplicity,
the case for  the $K^*$ sectors is left aside; substantially,  this would introduce two additional
real parameters $\xi_{4,5}$ and $\xi_{6,7}$ which may be fixed by the two radiative $K^*$ decays;
additional constraints may come from the $\tau$ decay to $K\pi\nu$.
}.
The  $\rho^\pm$ entries are given
by the sum of the Gell-Mann matrix terms  $T_1$ and $T_2$; forcing a breaking
for these entries requires two real parameters which should be equal
 ($\xi_{1,2}$) in order to preserve hermiticity.

So,  $\xi_3$ and this $\xi_{1,2}$ could
summarize the whole isospin breaking effects in the vector meson side;
however, one can choose $\xi_{1,2}=0$ as
 no theoretical value for the (unbroken) universal coupling $g$ is presently available.
Indeed, all vector couplings in the HLS Lagrangian
could then be reexpressed in terms of  $g^\prime=g(1+\xi_{1,2})$; in this case,  all physics
quantities will depend on $g^\prime$ and $\xi_i^\prime =\xi_i/(1+\xi_{1,2})$ ($i=0,8,3$) without any
other occurrence of $\xi_{1,2}$ dependency.

Therefore,
phenomenologically, the $\xi_i$'s and $g$ itself are defined up to  a normalization factor
 presently out of reach. This scaling
property has obviously no consequence on physics observables like cross-sections or form factors.

So, from now on, one assumes the maximal breaking of $U(3)_V$ experimentally accessible~:
$\delta V_\mu=\delta V_\mu^0+\delta V_\mu^8+ \delta V_\mu^3$. Compared with the original $V^I$ entries,
this turns out to modify
only the diagonal entries of $V^I$ by the following substitutions\footnote{
Remind that a  $1/\sqrt{2}$ term is factored out in the definition of $V^I$.}~:
\be
\left \{
\begin{array}{ll}
\displaystyle
\frac{\rho_I}{\sqrt{2}} \Rightarrow & \displaystyle
 \frac{\rho_I}{\sqrt{2}} [1+\xi_3] \,,\\[0.5cm]
\displaystyle
\frac{\omg_I}{\sqrt{2}} \Rightarrow & \displaystyle
 \frac{\omg_I}{\sqrt{2}} \left [
 1+\frac{2 \xi_0 +\xi_8}{3}
\right ]  + \frac{ [\xi_0 -\xi_8]}{3} ~\phi_I\,,\\[0.5cm]
\displaystyle \phi_I \Rightarrow &
\displaystyle \phi_I \left [
 1+\frac{\xi_0 +2 \xi_8}{3}
\right ]  + \frac{ \sqrt{2}[\xi_0 -\xi_8]}{3} ~\omg_I \epo
\end{array}
\right .
\label{eq1-11}
\ee
Then, the $U(3)_V$ breaking of the covariant derivative generates a breaking of
the vector coupling universality. For this purpose, one should note that a vector
mixing is generated,  except  if $\xi_0=\xi_8$.
For the sake of conciseness, this mechanism is referred to below as CD breaking.
\section{BHLS$_2$~: A New Broken HLS Lagrangian}
\label{HLSLag}
\indent \indent  Here, we define a new version of the broken HLS Lagrangian, merging
the two breaking schemes just presented -- BKY and CD -- as
these two ways of breaking are conceptually  independent.
Their contributions are thus complementing each other.
In order to fully take into account the electroweak sector,
one should modify correspondingly the covariant derivative to~:
 \be
\left \{
\begin{array}{ll}
\displaystyle
D_\mu \xi_{L} =\pa_\mu \xi_{L} -ig \left[V_\mu^I +\delta V_\mu \right
] \xi_{L} +i  \xi_{L} {\cal L}_\mu \,,\\[0.5cm]
D_\mu \xi_{R} =\pa_\mu \xi_{R} -ig \left[V_\mu^I +\delta V_\mu \right ] \xi_{R} +i  \xi_{R}  {\cal R}_\mu\,,
\end{array}
\right .
\label{eq1-12}
\ee
where, as usual\cite{HLSRef}~:
 \be
 \hspace{-1.5cm}
\begin{array}{ll}
\displaystyle
{\cal L}_\mu =   \displaystyle  e Q A_\mu
+\frac{g_2}{\sqrt{2}} (W^+_\mu T_+ + W^-_\mu T_-) \hspace{1.cm} {\rm and} & \hspace{1.cm}
{\cal R}_\mu =   \displaystyle e Q A_\mu  ~~~,
\end{array}
\label{eq1-13}
\ee
discarding the $Z^0$ boson terms of no concern for the phenomenology we address. The $T_\pm$ matrices are constructed
out of the matrix elements  $V_{ud}$ and $V_{us}$  of the Cabibbo-Kobayashi-Maskawa matrix and can be found
in \cite{HLSRef}. Finally, the weak coupling $g_2$ is related to the Fermi constant by
$g_2=2 m_W \sqrt{G_F\sqrt{2}}$.

The expression for the non-anomalous HLS Lagrangian pieces given in Section \ref{HLSMod} remains formally valid,
being understood that the covariant derivatives are modified according
to Equations (\ref{eq1-12}--\ref{eq1-13}).
As a result, the ${\cal L}_V$  piece substantially differs from its partner in \cite{ExtMod3} while
the ${\cal L}_A$ piece in the present scheme is strictly identical to those given in
\cite{ExtMod3}. The non-anomalous BHLS$_2$ Lagrangian is given expanded in Appendix \ref{AA}.

\section{The Order ${\cal O}(p^4)$  Terms of the HLS Lagrangian}
\label{zterms}
\indent \indent Beside the ${\cal L}_A$ and ${\cal L}_V$ pieces which are  ${\cal O} (p^2)$,
the HLS approach also possesses terms of order ${\cal O} (p^4) $(see Section 4.3 in \cite{HLSRef}),
which modify the $V-\gamma/W$ couplings in a specific way. As the role of such terms has never been
really examined in phenomenology\footnote{See, however,  the discussion in \cite{ExtMod1} which
is revisited here.},
it looks worthwhile examining their relevance when dealing with data of high
accuracy.

 Using ${\cal L}_\mu$ and ${\cal R}_\mu$  just given,
one first defines~:
\be
\left \{
\begin{array}{ll}
\displaystyle  {\cal L}_{\mu,\nu}= &
\displaystyle \pa_\mu {\cal L}_\nu -\pa_\nu {\cal L}_\mu
-i\left [{\cal L}_\mu,{\cal L}_\nu \right ]\,,\\[0.5cm]
\displaystyle  {\cal R}_{\mu,\nu}= &
\displaystyle \pa_\mu {\cal R}_\nu -\pa_\nu {\cal R}_\mu
-i\left [{\cal R}_\mu,{\cal R}_\nu \right ]\,,
\end{array}
\right .
\label{eq1-14}
\ee

\noi
and also~:
\be
\begin{array}{ll}
\displaystyle
\widehat{{\cal L}}_{\mu,\nu}=\xi_L {\cal L}_{\mu,\nu} \xi_L^\dagger~~,~~~&
\widehat{{\cal R}}_{\mu,\nu}=\xi_R {\cal R}_{\mu,\nu} \xi_R^\dagger~~,
\end{array}
\label{eq1-15}
\ee
where $\xi_R=\xi_L^\dagger=\exp{[iP/f_\pi]}$. Furthermore defining~:
\be
\begin{array}{ll}
\displaystyle
\widehat{{\cal V}}_{\mu,\nu}=\frac{1}{2}
\left [ \widehat{{\cal R}}_{\mu,\nu} +  \widehat{{\cal L}}_{\mu,\nu}
\right]
~~~,&
\displaystyle
\widehat{{\cal A}}_{\mu,\nu}=\frac{1}{2}
\left [ \widehat{{\cal R}}_{\mu,\nu} - \widehat{{\cal L}}_{\mu,\nu}
\right]~~~,
\end{array}
\label{eq1-16}
\ee
the ${\cal O} (p^4)$ Lagrangian writes \cite{HLSRef}~:
\be
\displaystyle
{\cal L}_z=
z_1 {\rm Tr}\left [ \widehat{{\cal V}}_{\mu,\nu}  \widehat{{\cal V}}^{\mu,\nu}\right ]+
z_2 {\rm Tr}\left [ \widehat{{\cal A}}_{\mu,\nu}  \widehat{{\cal A}}^{\mu,\nu} \right ]+
z_3 {\rm Tr}\left [  \widehat{{\cal V}}_{\mu,\nu} V^{\mu,\nu} \right ]\,,
\label{eq1-17}
\ee
where  one has generically defined~:
\be
\displaystyle
X_{\mu , \nu}=\pa_\mu X_\nu - \pa_\nu X_\mu -i [X_\mu,X_\nu]\,,
\label{eq1-18}
\ee
and where the $z_i$ are constants  not theoretically constrained. 
 The $z_3$ term, the most involved in the phenomenology one addresses, writes~:
\be
\displaystyle
{\cal L}_{z_3}= \displaystyle \frac{ge}{2}~z_3~ A_{\mu,\nu}
\left [ \rho^0_{\mu,\nu}+\frac{1}{3} \omg_{\mu,\nu}-\frac{\sqrt{2}}{3} \phi_{\mu,\nu}
\right ] + \frac{gg_2}{4}~z_3~
\left [
\overline{V}_{ud} W^-_{\mu,\nu}  \rho^+_{\mu,\nu} +V_{ud} W^+_{\mu,\nu}  \rho^-_{\mu,\nu}
\right ] 
\label{eq1-19}
\ee
at lowest order in the expansion of the $\xi_{L/R}$ fields; here one has kept the unbroken
$V_\mu$ matrix for clarity.
By integrating by part and fixing the gauge condition to
$\partial_\mu X^\mu=0$ for all vector fields, this piece becomes~:
\be
\displaystyle
{\cal L}_{z_3}=e g z_3 ~s~ A_\mu  \left [ \rho^0_\mu+\frac{1}{3} \omg_\mu
-\frac{\sqrt{2}}{3} \phi_\mu\right ]
+\frac{g_2 g z_3~s}{2} \left [\overline{V}_{ud} ~\rho^+_\mu ~W^-_\mu +
	V_{ud} ~\rho^-_\mu ~W^+ _\mu\right ]\epo
\label{eq1-20}
\ee
Thus, the ${\cal L}_{z_3}$ piece exhibits  $s$-dependent parts for the
 $V-\gamma$ and $V-W$ transition amplitudes which complement their  constant
parts provided by the usual ${\cal O} (p^2)$ HLS Lagrangian. The  'broken' version
of  Equation (\ref{eq1-20}) is given in  Appendix \ref{AA-3}.
\section{The Basic Solution (BS) and the Vector Mass Term}
\label{BasicSol}
\indent \indent In the context where  both the BKY  and CD breaking mechanisms are at work,
the vector mass term in ${\cal L}_V$ becomes~:
 \be
{\cal L}_{\rm mass}= \displaystyle a f_\pi^2 {\rm Tr}
\left [ X_V (V_\mu^I+\delta V_\mu) \right ]^2\epo
\label{eq1-21}
\ee
Ignoring the $K^*$ sector and setting  $m^2=a g^2 f_\pi^2$, it writes~:
\be
\begin{array}{ll}
{\cal L}_{\rm mass}=& \hspace*{-3mm} \displaystyle \frac{m^2}{2}
\left [(1+ \Sigma_V + 2 \xi_3) ~[\rho^0_{I}]^2 +
       (1+ \Sigma_V+ \frac{4}{3} \xi_0 + \frac{2}{3} \xi_8) ~[\omg_{I}]^2
      + z_V(1+\frac{2}{3} \xi_0 +  \frac{4 }{3} \xi_8 )~[\phi_{I}]^2 \right.
 \\[0.5cm]
\hspace{-1.5cm}
~~& \left . \displaystyle
+ 2\Delta_V~ \rho^0_{I} \cdot \omg_{I} +
\frac{2\sqrt{2}}{3}   (1+z_V) (\xi_0 - \xi_8)~\omg_{I}\cdot \phi_{I}
 \right ]
 + m^2  (1+ \Sigma_V)\,\rho^+ \cdot \rho^-\epo
\end{array}
\label{eq1-22}
\ee

Therefore the CD breaking, via its $\delta V_3$ component,  allows for a
 $\rho^0-\rho^\pm$ (HK) mass difference, provided fits favor
non-vanishing $\xi_3$ values. Moreover, all the components
of $\delta V$ defined above  contribute to generate different HK masses
for the $\rho^0$ and $\omg$ mesons as
$\Delta m^2_{\rho^0-\omg}=m^2[\xi_3-(2 \xi_0+ \xi_8)/3]$. This contrasts with the
previous BHLS model \cite{ExtMod3} where $m^2_{\rho^0}=m^2_{\rho^\pm}=m^2_\omg$
is fulfilled.
Because of the non-vanishing HK mass difference between the $\rho^0$ and $\omg$,
one can already expect an improved treatment  of the dipion threshold
and spacelike regions in the BHLS$_2$ framework compared to BHLS.

Equation (\ref{eq1-22}) is manifestly diagonalized by setting~:
\be
\begin{array}{ll}
\displaystyle \Delta_V=0~~~,& \displaystyle \xi_0 = \xi_8~~.
\end{array}
\label{eq1-25}
\ee
This solution -- which lets $z_V$ unconstrained -- defines  our
Basic Solution (BS). Within
the framework of this solution,  the breaking parameters to be derived from
fits are $\Sigma_V$,  $z_V$,  $\xi_3$, $\xi_0~(=\xi_8)$. Here one feels the issue with
assuming solely nonet symmetry breaking; indeed, as this turns out to state  $\xi_8=0$,
the CD breaking implies $\xi_0 =0$ and thus breaking HLS via the covariant
derivative intrinsically implies that nonet symmetry cannot be solely broken.

\vspace{0.5cm}

At leading order in breaking parameters, the vector meson mass term
in ${\cal L}_V$ becomes diagonal and one has~:
\be
\left \{
\begin{array}{ll}
\displaystyle m_{\rho^\pm}^2= \displaystyle m^2 ~(1+\Sigma_V)
\,,\\[0.5cm]
m_{\rho^0}^2= \displaystyle m^2 \left [
1+\Sigma_V + 2 ~\xi_3\right ]
\,,\\[0.5cm]
\displaystyle m_{\omg}^2= \displaystyle m^2 \left [
1+\Sigma_V + \frac{4}{3} ~\xi_0 +  \frac{2 }{3} ~\xi_8\right ]
=m^2 \left [
1+\Sigma_V +2 ~\xi_0\right ]
\,,\\[0.5cm]
\displaystyle m_{\phi}^2  =
\displaystyle m^2 ~z_V  \left[ 1+\frac{2}{3}~\xi_0 + \frac{4}{3}~\xi_8 \right]
=  m^2 ~z_V  \left[ 1+2~\xi_0 \right]
\epo\\[0.5cm]
\end{array}
\right .
\label{eq1-26}
\ee

So, BHLS$_2$ {\it a priori} yields different HK masses for all the vector mesons.

\section{The Primordial Mixing (PM) and the Reference Solution (RS)}
\label{RefSol}
\indent \indent
Another unused mechanism can be invoked; indeed, independently of the BKY and CD mechanisms
just defined,  one can always
consider that the neutral vector fields $\rho^0,~\omg,~\phi$ involved in physical processes
are not
directly the ideal ones but combinations of these. For this purpose,
let us define an infinitesimal rotation matrix $R(U_3)~={\bf 1}+{\cal O}(\epsilon)$, which associates to the
ideal field vector ${\cal V}_I=(\rho_{I}, \omg_{I},\phi_{I})$  a (first step) renormalized field vector
 ${\cal V}_R=(\rho_{R},\omg_{R},\phi_{R})$ via~:
 \be
 \left (
\begin{array}{ll}
\rho_{I}\\[0.5cm]
\omg_{I}\\[0.5cm]
\phi_{I}\\[0.5cm]
\end{array}
\right )
 = \left (
\begin{array}{ccc}
\displaystyle 1     &\displaystyle -\psi_\omg &  \psi_\phi \\[0.5cm]
\displaystyle \psi_\omg & \displaystyle 1 & \displaystyle \psi_0 \\[0.5cm]
\displaystyle  -\psi_\phi & \displaystyle  -\psi_0  & \displaystyle 1
\end{array}
\right )
 \left (
\begin{array}{ll}
\rho_{R}\\[0.5cm]
\omg_{R}\\[0.5cm]
\phi_{R}\\[0.5cm]
\end{array}
\right )~~~, ~~ {\rm also~ denoted}~~~ {\cal V}_I=R(U_3) \cdot {\cal V}_R~.
\label{eq1-23}
\ee
This is also a quite legitimate tool to extend the flexibility of BHLS$_2$.

The matrix $R(U_3)$  fulfills  $R(U_3)\widetilde{R}(U_3)=\widetilde{R}(U_3) R(U_3)=\bf{1}$, up to
terms of order  ${\cal O}(\epsilon^2)$ which are discarded in our ${\cal O}(\epsilon)$
approach.
As it is a rotation, this transformation preserves the canonical structure of the
vector field kinetic term provided by the Yang-Mills Lagrangian
(up to ${\cal O}(\epsilon^2)$ terms).

However, the (real) $\psi$ (Euler) angles from transformation in Equation (\ref{eq1-23})
should be chosen in such a way that ${\cal L}_{\rm mass}({\cal V}_I) $ remains canonical
in the change of fields ${\cal V}_I \rightarrow {\cal V}_R$, {\it i.e.}
 the crossed terms in ${\cal L}_{\rm mass}({\cal V}_R)$ should be canceled out.
 Restarting from the mass term given by Equation (\ref{eq1-22}),  three conditions should
 be fulfilled~:
\be
\left \{
\begin{array}{ll}
\displaystyle
\rho_R-\omg_R =0 \Rightarrow & \displaystyle  \Delta_V  + \left[
\frac{4}{3} \left\{\xi_0+\frac{1}{2} \xi_8 -\frac{3}{2} \xi_3
\right\} \psi_\omg \right]  =0 \,,\\[0.5cm]
\rho_R-\phi_R  =0 \Rightarrow &  \displaystyle   [1-z_V]~\psi_\phi
+\frac{\sqrt{2}}{3} \left \{1+z_V\right\}(\xi_0 - \xi_8)
 +\left[~\psi_\omg~\psi_0
\right]  =0\,,\\[0.5cm]
\omg_R-\phi_R  =0 \Rightarrow &  \displaystyle
3 [1-z_V]~\psi_0 + [1+z_V]  \sqrt{2}~~(\xi_0 - \xi_8) -3 \left[ \psi_\omg~\psi_\phi\right]=0\epo
\end{array}
\right .
\label{eq1-24}
\ee
The last bracketed terms in these expressions can be discarded as clearly of second order in
 the breaking parameters $\{ \xi_i,~\psi_j\}$; this already implies that $\Delta_V=0$
 and that $\psi_\omg $ becomes unconstrained.

As for the parameter $z_V$ -- generated by the BKY breaking in its $s \overline{s}$ entry  --
the situation deserves  further comments~:
\begin{itemize}
\item {\bf Either}
$z_V$ is found such
that\footnote{
\label{epsilon} One may think that a reference magnitude for any generic $\epsilon$
is ${\cal O}(\epsilon) \simeq e=\sqrt{4\pi\alpha_{\rm em }}\simeq 0.3$.}
 ${1-z_V} \simeq {\cal O}(\epsilon)$, then none of the $\psi$
angles is constrained at order ${\cal O}(\epsilon)$
and, moreover, the last two Equations (\ref{eq1-24}) then imposes $\xi_0=\xi_8$.
\item {\bf Or}
$z_V$ is returned by fits  such that $1-z_V$ is not ${\cal O}(\epsilon)$. Then,
at first order in breakings, the last two equations imply $\psi_\phi=\psi_0$ and
that they are proportional to $\xi_0 - \xi_8$.
\end{itemize}

Preliminary fits, performed using this mechanism -- named from now on 
Primordial Mixing (PM) --
indicate that $1-z_V$ is  in the range $(1 \div 2) \epsilon$ and thus,
in this case also $\xi_0 =\xi_8$ can be imposed.
Therefore, besides the Basic Solution (BS), one gets an additional one, we name it
Reference Solution (RS), which also includes  the Primordial Mixing (PM). Also considering
the former BHLS model \cite{ExtMod3,ExtMod4,ExtMod5}, one thus has
at disposal three different models.
This allows for a better insight into the systematics and model dependence effects.

\section{Vector Meson Propagators and Form Factors}
\label{propagators}
\indent  \indent
Let us anticipate  on form factor calculations using the Lagrangian given in Appendix \ref{AA}
and have a look at the form factor values at $s=0$  -- before introducing mixing effects due to
loop corrections. For this purpose, let us start with a preliminary digression on vector
meson propagators, especially those for the $\omg$ and $\phi$ mesons, as these play an important role
at the chiral point and in the physics of the (close) spacelike region which is also addressed in the present
work.

For the $\rho$ meson,  following the pioneering work \cite{Klingl},
the inverse propagator at one loop can be written (also see \cite{ExtMod1,ExtMod2,ExtMod3})~:
$$D_\rho(s)=s-m_\rho^2-\Pi(s)\,,$$
where the self-energy $\Pi(s)$ is the sum of the
loops allowed by the non-anomalous Effective Lagrangian given in Appendix \ref{AA}; in our
context,
these are essentially pion and
kaon loops\footnote{ Loops generated by the anomalous HLS Lagrangian pieces \cite{HLSRef,FKTUY}
also contribute but can be discarded \cite{ExtMod3} as reminded in the next Section;  however,
some anomalous loops, suppressed by  a factor of $e^2$, can have to
be kept as they may play some role in specific
expressions as will be
emphasized below. }.
$\Pi(s)$ is a real analytic function
($\Pi(s^*)=[\Pi(s)]^*$) which vanishes at $s=0$ because of current conservation, and is real for
negative $s$ -- in fact, this property holds already below $m_{\pi^0}^2$, the lowest energy hadronic threshold. Hence,
this implies that $D_\rho(0)=-m_\rho^2$, where $m_\rho^2$ is displayed in Equations (\ref{eq1-26}).

In principle, this also applies to the $\omg$ and $\phi$ propagators; however, taking into account
their narrow character, it looks unmotivated to enter into such complications when fitting
objects like $e^+e^-$ annihilation cross sections. In this case, phenomenology has lengthily
experienced a broad success using  Breit-Wigner (BW) lineshapes. However, as discussed
in \cite{ExtMod3}, some physics quantities like the contribution of the $ e^+e^- \ra \pi^+\pi^-$
cross-section to the estimate for the muon HVP, may require some care
about the behavior of approximations close to the chiral point,
 as this region gives an enhanced contribution  to the muon HVP. For this purpose, \cite{ExtMod3}
 proposed the following modified BW lineshape for the $\omg$ and $\phi$ mesons~:
\be
 \displaystyle D_V(s) =
 s-m_V^2 -\frac{s}{m_V^2} \left [ \widetilde{m}_V^2 -m_V^2 -i\widetilde{m}_V \widetilde{\Gamma}_V
 \right ]\,,
  \label{bw1}
\ee
where $(\widetilde{m}_V,\widetilde{\Gamma}_V)$ are parameters to be fitted and the $m_V^2$'s
are the  relevant HK square masses as displayed in Equations (\ref{eq1-26}). Numerically, this
expression for the $D_V(s)$'s is
close to usual BW lineshapes\footnote{Setting $\widetilde{m}_V=m_V$ in this approximation
gives $D_V(s)=s-m_V^2+i s \widetilde{\Gamma}_V/m_V$; more common choices for the imaginary
part here are $m_V \widetilde{\Gamma}_V$ or $\sqrt{s} \widetilde{\Gamma}_V$, quite analogous
to our own choice.}.
This BW-like parametrization imposes  the $\omg$ and $\phi$ meson inverse propagators to
fulfill  $D_V(0)=-m^2_V$  as already does  the $\rho$ meson inverse propagator;
this property of
the propagators  at $s=0$  is essential to recover the expected
values for the pion and kaon form factors at the origin  within the HLS framework.

On the other hand,  if one wishes to examine the analytic continuation of the $\omg$ and $\phi$ propagators
somewhat inside the spacelike region, it might be desirable to make it real there; this can
be achieved by simply replacing $\widetilde{\Gamma}_V$ in Equation (\ref{bw1}) by
$\theta(s) \widetilde{\Gamma}_V $ -- or rather, $\theta(s-m_{\pi^0}^2) \widetilde{\Gamma}_V $.

\vspace{0.5cm}

 Using the Basic Solution (BS) or the Reference Solution (RS) to redefine
 the physical vector meson fields, one can show, using the Lagrangian given in
 the Appendix  \ref{AA}, that~:

 \be
 \begin{array}{lll}
 \displaystyle F_\pi(0)=1~~, &
 \displaystyle ~~F_{K^\pm}(0)=1~~, &
 \displaystyle ~~F_{K^0}(0)=0\,,
\end{array}
\label{eq1-27}
\ee
up to terms of order ${\cal O}(\epsilon^2)$. One can check that the conditions
$\Delta_V=0$ and $\xi_0=\xi_8$ are essential in the derivation of these constraints.

 So,  the breaking parameters to be derived from fits are
$\Sigma_V$, $\xi_3$, $\xi_0~(=\xi_8)$ and $z_V$ when working within the
BS framework; one should also include the $\psi$ rotation angles
when extending  BHLS$_2$ to the Reference Solution.

It is worthwhile reminding that the effects of $\delta V_\mu^0$ and $ \delta V_\mu^8$
 are intimately intricated within our new Lagrangian frameworks.
 Here, indeed, a breaking of solely nonet symmetry ({\it i.e.} $\xi_0 \ne 0$) cannot
 occur if not accompanied  by  a breaking of the $SU(3)_V$ symmetry of
 the same intensity.

In conclusion, the present model (BHLS$_2$) is not a  trivial variant
of BHLS \cite{ExtMod3}.  Some important properties of BHLS$_2$ versus BHLS
mentioned below, will further substantiate this statement.

\section{Dynamical Mixing of Vector Mesons}
\label{DynMix}
\subsection{The Squared Mass Matrix at One Loop \ldots}
\indent \indent As previously noted \cite{taupaper,ExtMod3} and reminded above,
all variants of the HLS Model, including BHLS$_2$ (see Appendix \ref{AA-1}),  exhibit
$\rho^0/\omg/\phi \ra K \overline{K} $ couplings. This implies that, at one loop, the
squared mass matrix of the  $\rho^0/\omg/\phi $ system receives non-diagonal
entries, {\it i.e.} the vector fields occurring in the Lagrangian Eq (\ref{eqAA-2})
are no longer mass eigenstates.  Mass eigenstates are constructed using perturbative
 methods as performed in \cite{taupaper,ExtMod3}. As the loops are (real analytic)
functions of  $s$, the mass eigenstates become also $s$-dependent. Within the BHLS and 
BHLS$_2$ frameworks, the kaon loops produce a $s$-dependent difference between the 
$\rho^\pm - W^\pm$ and $\rho^0 - \gamma$ transition amplitudes which has provided
 the first solution \cite{taupaper,ExtMod3} to the long standing  $e^+e^-$ versus $\tau$ puzzle
\cite{DavierHoecker,DavierHoecker3,CapriMB}. Nevertheless, in another Effective Lagrangian context,
this $s$-dependent difference can be successfully generated by other means \cite{Fred11}.

Basically, the Dynamical (or Loop) Mixing of Vector Mesons has
been first defined\footnote{See also  \cite{ExtMod3}.}
and motivated in \cite{taupaper}. In order to ease the reading,  we remind it
and emphasize the new features provided by the BHLS$_2$ context.

The (squared) mass matrix of the $\rho^0/\omg/\phi$ sector can be written~:
\be
\displaystyle M^2(s)=M_0^2(s)+\delta M^2(s)\,,
\label{eq1-28}
\ee
where~:
\be
\displaystyle M_0^2(s)={\rm Diag} (m_{\rho^0}^2+ \Pi_{\pi\pi}(s),m_\omg^2,m_\phi^2)\epo
\label{eq1-29}  ~~~
\ee
The Higgs-Kibble masses $m_V$ are displayed in Equations (\ref{eq1-26}), and $\Pi_{\pi\pi}(s)$
is the pion loop weighted by the square of the $\rho^0 \pi^+\pi^-$ coupling constant.
Because all the loop functions are real analytic
function of $s$, $M_0^2(s)$, $\delta M^2(s)$, and hence $M^2(s)$, are hermitian analytic
matrices (e.g. fulfilling $[X(s)]^\dagger=X(s^*)$).

In order to construct explicitly $\delta M^2(s)$, it is worth reexpressing  some $VPP$
coupling constants in a suitable manner. One can write~:
\be
\left \{
\begin{array}{lll}
\displaystyle g^\rho_{K^+ K^-}= \widetilde{G} (1-\frac{\Delta_A}{2})~\overline{g}^\pm_\rho~~,
&
\displaystyle g^\omg_{K^+ K^-}= \widetilde{G} (1-\frac{\Delta_A}{2})~\overline{g}^\pm_\omg~~,
&
\displaystyle g^\phi_{K^+ K^-}= \widetilde{G} (1-\frac{\Delta_A}{2})~\overline{g}^\pm_\phi
\,,\\[0.5cm]
\displaystyle g^\rho_{K^0 \overline{K}^0}= \widetilde{G} (1+\frac{\Delta_A}{2})~\overline{g}^0_\rho~~,
&
\displaystyle g^\omg_{K^0 \overline{K}^0}= \widetilde{G} (1+\frac{\Delta_A}{2})~\overline{g}^0_\omg~~,
&
\displaystyle g^\phi_{K^0 \overline{K}^0}= \widetilde{G}
(1+\frac{\Delta_A}{2})~\overline{g}^0_\phi \,,
\end{array}
\right .
\label{eq1-30}
\ee
having defined $\widetilde{G}=ag/(4 z_A)$. Using Equations (\ref{eq1-25}),  common to both
the BS and RS variants, one finds~:
\be
\left \{
\begin{array}{lll}
\displaystyle
\overline{g}^\pm_\rho =
\left \{1 +\Sigma_V+\xi_3+ \psi_\omg +\sqrt{2} z_V \psi_\phi \right \}\,,\\[0.5cm]
\displaystyle
\overline{g}^\pm_\omg =
\left \{1 +\Sigma_V+\xi_8 - \psi_\omg +\sqrt{2} z_V \psi_0 \right\}\,,\\[0.5cm]
\displaystyle
\overline{g}^\pm_\phi=\left \{-\sqrt{2} z_V (1+\xi_8)
+ \psi_\phi+ \psi_0 \right\}\,,
\end{array}
\right .
\label{eq1-30b}
\ee
and~:
\be
\left \{
\begin{array}{lll}
\displaystyle
\overline{g}^0_\rho =~~
\left \{1 +\Sigma_V+\xi_3- \psi_\omg-\sqrt{2} z_V \psi_\phi\right \}\,,\\[0.5cm]
\displaystyle
\overline{g}^0_\omg =-
\left \{1 +\Sigma_V+\xi_8 + \psi_\omg+\sqrt{2} z_V \psi_0 \right\}\,,\\[0.5cm]
\displaystyle
\overline{g}^0_\phi=~~
\left \{\sqrt{2} z_V (1+\xi_8) + \psi_\phi - \psi_0 \right\}\epo
\end{array}
\right .
\label{eq1-30c}
\ee
When dealing specifically with the BS variant, these two sets of equations are used
by simply dropping out the $\psi_\alpha$ parameters.

Denoting by resp. $\Pi_\pm(s)$ and $\Pi_0(s)$
 the {\it amputated} charged and neutral kaon loops,
 the ${\cal V}_R^i \rightarrow {\cal V}_R^j$
transition amplitudes ($i,j=\rho^0,~\omg,~\phi$) are given by~:
\be
\displaystyle
\varepsilon_{i, j}(s)=g^i_{K^+ K^-}g^j_{K^+ K^-}\Pi_\pm(s)+
 g^i_{K^0 \overline{K}^0}g^j_{K^0 \overline{K}^0}\Pi_0(s)
\label{eq1-30d}
\ee
using the notations just defined.
Then, the elements of the $\delta M^2(s)$ matrix are~:

\be
\displaystyle
\delta M^2_{i, j}(s)=\varepsilon_{i, j}(s)\epo
\label{eq1-31}
\ee

As in the former BHLS \cite{ExtMod3}, $\delta M^2(s)$ is non-diagonal
for non-zero values of $s$. Therefore, at one-loop order, the field
combinations defined in both the BS and RS variants are not mass eigenstates,
as in the previous BHLS release. This calls for a mass-dependent diagonalization which
is reminded in the next Subsection.

Let us make a few more remarks about the non-diagonal entries of
 $\delta M^2$.  In the no breaking limit, where
the neutral and charged kaon loops coincide, the $\rho^0-\omg$
and $\rho^0-\phi$ entries identically vanish; however, the $\omg-\phi$
entry,  proportional to the sum of the neutral and charged kaon loops, does
not vanish, indicating that the $\omg-\phi$ mixing is always at work
at one loop order within the HLS framework.

In the former BHLS, however, if no breaking is applied in the ${\cal L}_A$
sector and if there is no mass breaking among the kaons, there was
no $\rho^0-\omg$ and $\rho^0-\phi$ transitions. This remains true  within the
BS framework defined above. However, within the RS framework, the $\psi_\alpha$'s
come in such a way that none of the entries of $\delta M^2(s)$ vanishes.

\subsection{\ldots And its Diagonalization}
\label{angle}
\indent \indent The diagonalization of ${\cal L}_{\rm mass}$
at one loop order is performed by means
of another ($s$-dependent) rotation matrix to a second step of vector field
renormalization  $R \Rightarrow R^\prime$~:
  \be
 \left (
\begin{array}{ll}
\rho_{R}\\[0.5cm]
\omg_{R}\\[0.5cm]
\phi_{R}\\[0.5cm]
\end{array}
\right )
 = \left (
\begin{array}{ccc}
\displaystyle 1     &\displaystyle -\alpha(s) &  \beta(s) \\[0.5cm]
\displaystyle \alpha(s) & \displaystyle 1 & \displaystyle \gamma(s) \\[0.5cm]
\displaystyle  -\beta(s) & \displaystyle  -\gamma(s)  & \displaystyle 1
\end{array}
\right )
 \left (
\begin{array}{ll}
\rho_{R^\prime}\\[0.5cm]
\omg_{R^\prime}\\[0.5cm]
\phi_{R^\prime}\\[0.5cm]
\end{array}
\right )~~~, ~~ {\rm also~ denoted}~~~ {\cal V}_R=R(Loop) \cdot {\cal V}_{R^\prime}
\label{eq1-33}
\ee
so that the full renormalization procedure is obtained by~:
  \be
 {\cal V}_I \Rightarrow \left[R(U_3) \cdot R(Loop) \right]  \cdot {\cal V}_{R^\prime} ~.
\label{eq1-34}
\ee
The $R^\prime$ fields are the physical vector meson fields. When working within the
Basic Solution variant,
${\cal V}_{R} \equiv {\cal V}_{I}$ and $R(U_3)$ can be dropped out from Equation (\ref{eq1-34});
otherwise, within the RS framework,  $R(U_3)$ is given in Equation (\ref{eq1-23}).

The complex "angles" in Equation (\ref{eq1-33}) can be derived from the
$\delta M^2(s)$ matrix elements~:
  \be
\begin{array}{lll}
\displaystyle
\alpha(s)= \frac{\varepsilon_{\rho^0 \omg}(s)}{\lambda_\rho(s) -\lambda_\omg(s)}~~,&
\displaystyle
\beta(s)= -\frac{\varepsilon_{\rho^0 \phi}(s)}{\lambda_\rho(s) -\lambda_\phi(s)}~~,&
\displaystyle
\gamma(s)= -\frac{\varepsilon_{\omg \phi}(s)}{\lambda_\omg(s)-\lambda_\phi(s)}\,,
\end{array}
\label{eq1-35}
\ee
where the numerators manifestly depend on the kaon loops only
and the $\lambda$'s are the eigenvalues of $M^2(s)$ matrix.
At leading order in breaking parameters, these write \cite{ExtMod3}~:
  \be
\begin{array}{lll}
\displaystyle
\lambda_\rho(s)=m_{\rho^0}^2 + \Pi_{\pi\pi}(s)+\varepsilon_{\rho^0 \rho^0}(s),&
\displaystyle
\lambda_\omg(s)=m_\omg^2+\varepsilon_{\omg \omg}(s),&
\displaystyle
\lambda_\phi(s)=m_\phi^2+\varepsilon_{\phi \phi}(s),
\end{array}
\label{eq1-36}
\ee
the vector meson masses being those displayed in Equations (\ref{eq1-26}). As the
 three $m_V^2$'s  occurring here are different, the three
 angles defined above vanish at $s=0$; in contrast,  within
the former BHLS context \cite{ExtMod3},  because the HK masses for $\rho^0$
and $\omg$ were equal, $\alpha(0)$ has a non-zero
limit at  $s=0$, compromising a proper analytic continuation of the form
factors downwards into the spacelike region\footnote{See, in particular,
the comments in Section 6.3 of \cite{ExtMod3}.}.

In summary the full transform from ideal to fully renormalized $\rho_0$,
$\omg$, $\phi$ fields is given either by (BS solution)~:
\be
\displaystyle R(I \Rightarrow R^\prime)  \equiv  R(Loop)
\label{eq1-37a}
\ee
or by (RS solution)~:
\be
\displaystyle R(I \Rightarrow R^\prime)  \equiv \left[R(U_3) \cdot R(Loop) \right] =
  \left (
\begin{array}{ccc}
\displaystyle 1     &\displaystyle -[\psi_\omg+\alpha(s)] & [ \psi_\phi+\beta(s)] \\[0.5cm]
\displaystyle [\psi_\omg+\alpha(s)] & \displaystyle 1 & \displaystyle [\psi_0 +\gamma(s)] \\[0.5cm]
\displaystyle  -[ \psi_\phi+\beta(s)] & \displaystyle  -[\psi_0 +\gamma(s)]  & \displaystyle 1
\end{array}
\right )\,,
\label{eq1-37}
\ee
where breaking terms of order greater than 1 have been discarded. We will refer from now on to
Equation (\ref{eq1-37}) for both the BS and RS variants, being understood that in the former case, the
$\psi_i$'s are zero. When relevant, the entries in the  matrix just above will be denoted
$\widetilde{\alpha}(s)$, $\widetilde{\beta}(s)$ and $\widetilde{\gamma}((s)$, using obvious notations.
Moreover, in the following, the fully
renormalized fields can be either  indexed by $R^\prime$ -- if useful -- or deprived of
indexation to lighten expressions.
For  clarity, in all displayed Lagrangian pieces,
one always writes the couplings for the {\it ideal} $\rho_0$, $\omg$, $\phi$ fields.
 To go to physical fields, one has to apply
the appropriate transformation $ R(I \Rightarrow R^\prime)$ and collects the
contributions in order to yield
the {\it physical} $\rho_0 $, $\omg $, $\phi $ couplings.

\subsection{The Effects of Anomalous Loops}
\label{pi0g_loop}
\indent \indent When defining the loops contributing to $\delta M^2(s)$, we have only considered
those generated by the non-anomalous HLS Lagrangian. However, the anomalous (FKTUY)
HLS Lagrangian pieces
\cite{FKTUY,HLSRef} allow for (anomalous) couplings also generating  loop
contributions\footnote{The Yang-Mills part of the full Lagrangian
may also contribute with other loops like  $K^* \overline{K}^*$.
The $VPPP$ Lagrangian  generate 2-loop contributions with coefficients of ${\cal O}(\epsilon)$
or ${\cal O}(\epsilon^2)$ to the transitions among the vector mesons; they are presently ignored.
}
to $\delta M^2(s)$ as, for instance, $K^* \overline{K}$ or 3-pion loops.
As all the two particle channels have thresholds above our fitting region, their loops --
of order ${\cal O} (g^2)$ -- are real  in our
fitting range and, thus, can be discarded, assuming  their  effects
numerically absorbed by the subtraction polynomials of the other
(pion and kaon) loops involved \cite{ExtMod3}.

However, one should also note that the anomalous (FKTUY) sector of the HLS Lagrangian
generates couplings of the neutral vector mesons to the $\pi^0 \gamma$, $\eta \gamma$
and  $\etp \gamma$  final states, even
when setting $c_3 = c_4$ \cite{ExtMod3,ExtMod5}. Then the FKTUY sector generates the corresponding loops, which have
couplings of order ${\cal O}(g^2 e^2)$ and develop tiny imaginary parts far inside our fitting region for the first two
and close to its upper border for the third one.
 These contributions, and their imaginary parts, are higher order  and can generally be discarded;
nevertheless, their tiny imaginary parts were accounted for in our
previous studies \cite{ExtMod3,ExtMod4,ExtMod5}
by simply adding some fixed $i \varepsilon$ to the eigenvalue differences
$\lambda_i(s)-\lambda_j(s)$ which occur in the denominators of the mixing
"angles" shown in  Equations (\ref{eq1-35}).

So, these (complex) mixing "angles" exhibit a dependence upon differences of the $M^2(s)$
eigenvalues. However, these eigenvalues -- see  Equations (\ref{eq1-36}) --
being $s$-dependent, their differences may exhibit zeros for real values of $s$
which, accordingly, generate  real poles for the mixing "angles". Poles occurring at real negative
$s$ are not a real issue as they can be dropped out by means of customary dispersive
 techniques.
For the others, the  $i \varepsilon$ trick emphasized just above, permits to avoid unwanted
singularities on the physical region $s \ge m_{\pi^0}$.
\begin{figure}[!b!]
\hspace{-0.0cm}
\begin{minipage}{0.5\textwidth}
\begin{center}
\resizebox{1.0\textwidth}{!}
{\includegraphics*{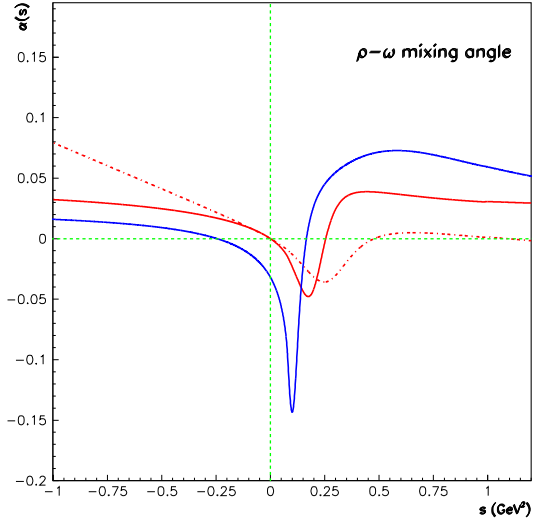}}
\end{center}
\vspace{1.5cm}\end{minipage} \hfill
\begin{minipage}[b]{0.44\textwidth}
\captionof{figure}{
Real part of the $\alpha(s)$ mixing "angle"; the full blue curve
shows the case for the (former) BHLS, the full red curve refers to the Reference Solution
of BHLS$_2$ and the dashed  red curve to its Basic Solution.
See Subsection \ref{pi0g_loop}  for more comments.}
\label{Fig:alpha_angle}
\end{minipage}
\end{figure}

In the previous BHLS version\cite{ExtMod3}, the main purpose of this {\it ad hoc} $i \varepsilon$  was  to
force\footnote{As clear from the first Equations (\ref{eq1-35}), the numerator of $\alpha(s)$
behaves as $s$ close to the origin and, as $m_{\rho^0}=m_\omg$ in  the previous BHLS
\cite{ExtMod3}, $\lambda_\rho(s) -\lambda_\omg(s)$ exhibits a similar $s$-behavior.}
 $\alpha(s=0)=0$, which
is no longer  necessary in the present version of BHLS.
A more natural way to proceed -- giving the same results -- is, however,  to take into account
the $\pi^0 \gamma$ loop which introduces a small imaginary part to the eigenvalue differences,
for $s_0 \ge m_{\pi^0}^2$
upwards. For this purpose, we use the loop expression derived in  \cite{mixing} (see  its Appendix C)
 which can be written ($s_0=m_{\pi^0}^2$)~:
\be
\displaystyle
\epsilon_{\pi^0}^V(s)= \frac{[g_{V\pi^0\gamma}^R]^2}{96 \pi^2}
\left [ G(s)-s_0^2 \right] \equiv [g_{V\pi^0\gamma}^R]^2 \epsilon_{\pi^0}(s)\epo
\label{eq1_37.1}
\ee
Along the real  axis in the complex $s$-plane, one has~:
\be
\left \{
\begin{array}{lll}
\displaystyle
G(s)=-\frac{(s_0-s)^3}{s} \ln {\frac{(s_0-s)}{s_0}}~~~, & s\le s_0 \\[0.5cm]
\displaystyle
G(s)=\frac{(s-s_0)^3}{s}  \left [
\ln {\frac{(s-s_0)}{s_0}} -i \pi \right ]~~~, & s\ge s_0\epo
\end{array}
\right.
\label{eq1-37.2}
\ee
The derivation of the $g_{V\pi^0\gamma}^R$ couplings is given in Subsection \ref{AVPcouplings}.
The constant term inside the
bracket in  Equation (\ref{eq1_37.1}) ensures that $\epsilon_{\pi^0}(0)=0$.
Full consistency would impose to add (using obvious notations)~:
\be
\displaystyle
\delta_2 M^2_{i,j}(s)=\left [ g_{V_i \pi^0\gamma}^R g_{V_j \pi^0\gamma}^R \right ] \epsilon_{\pi^0}(s)
+\left [ g_{V_i \eta\gamma}^R g_{V_j \eta\gamma}^R \right ] \epsilon_{\eta}(s)
+\left [ g_{V_i \etp \gamma}^R g_{V_j \etp\gamma}^R \right ] \epsilon_{\etp}(s)
\label{eq1_37.3}
\ee
to $\delta M^2(s)$ as expressed above; however, because of the $e^2$ factor, they are  higher order
and can be neglected, except in the diagonal where they avoid possible real poles in the physical region. Additionally,
the $\epsilon_{\eta}(s)$ and $\epsilon_{\etp}(s)$ contributions displayed for completeness
will also be discarded.

In order to substantiate the effect of the $\pi^0 \gamma$ loops,  let us anticipate on the
global fit information.
The main effect of these loops is to prevent naturally the occurrence of poles on the physical
region.
This effect is prominent for the  "angle" $\alpha(s)$  which produces a $\rho^0-\omg$ mixing;
 Figure \ref{Fig:alpha_angle} displays the behavior of Re$[\alpha(s)]$ for the
 BHLS$_2$  BS and RS variants
 and also, for comparison, its behavior within  the former BHLS, where a $i\epsilon$
 trick prevents a real pole close to the 2-pion threshold \cite{ExtMod3}.
 Even if the overall lineshapes are similar, one observes significant differences between the
 various solutions; for instance, $\alpha(s)$ vanishes at $s=0$ within the two variants of BHLS2,
 as the denominator is non-zero at the chiral point, allowing for a smooth connection between the
 spacelike and timelike regions.

 Over the range  $[-3.5 \div 1.2]$ GeV$^2$, the behavior observed
for the two other mixing angles  $\beta(s)$ and $\gamma(s)$ is much smoother
(see also \cite{ExtMod1});
nevertheless, specific
parametrizations may  lead to a real pole\footnote{One should note that a pole at such
a location is exhibited by the usual Gounaris-Sakurai formula \cite{Gounaris}.}
for $\beta(s)$ at negative $s$
below $\simeq -1.5$ GeV$^2$. As noted above, such poles can be eliminated by means of usual dispersive
methods.
\section{The Model Pion and Kaon Form Factors}
\label{form_factors}
\indent \indent Using the non-anomalous sectors of the HLS model, broken
as pointed out in the previous Sections (see Appendix \ref{AA}),
one can already derive some of the form
factors expressions needed for our global fit framework.

\subsection{The Pion Form Factor in the $\tau$ Decay}
\label{FFtau}
\indent \indent The $\tau$ sector of BHLS$_2$ is almost identical to that of BHLS
\cite{ExtMod3}; indeed the main ${\cal L}_\tau$ piece reminded in Appendix \ref{AA-2}
is unchanged, but
the newly introduced  ${\cal L}_{z_3}$  contributes. This yields~:
\be
\displaystyle
F_\pi^\tau(s) = \left[
1-\frac{a}{2}(1+\Sigma_V) \right]- \frac{ag}{2}(1+\Sigma_V)F_\rho^\tau(s)
\frac{1}{D_\rho(s)}\,,
\label{eq1-38}
\ee
with~:
\be
\left \{
\begin{array}{lll}
\displaystyle
D_\rho(s)=s-m^2_\rho -\Pi_{\rho \rho}^\tau(s)\,,\\[0.5cm]
\displaystyle
F_\rho^\tau(s) =f_\rho^\tau -g z_3 s - \Pi_{W}(s)\,,\\[0.5cm]
\displaystyle
f_\rho^\tau =a g  f_\pi^2 (1+\Sigma_V) ~~~,~~ m^2_\rho=a g^2 f_\pi^2(1+\Sigma_V)\,,
\end{array}
\right .
\label{eq1-39}
\ee
where $z_3$ is a new HLS parameter \cite{HLSRef} introduced by the ${\cal O}(p^4)$ terms of the
HLS Lagrangian (see Appendix \ref{AA-3}).  $\Pi_{\rho \rho}^\tau(s)$ is the loop correction to
the $\rho^\pm$ propagator and $\Pi_{W}(s)$ the loop correction to the $W^\pm-\rho^\pm$ transition amplitude;
both  are defined just below.

Using the following short-hand notations~:
\be
\begin{array}{lll}
\displaystyle  G_\pi= \frac{ag}{2}~~~,& \displaystyle  G_K=\frac{ag}{4z_A}\,,
\end{array}
\label{eq1-39b}
\ee
one  first defines the pion and kaon loop contributions to the $\rho^\pm$ self-energy
 $\Pi_{\rho \rho}^\tau(s)$~:
\be
\begin{array}{lll}
\displaystyle \Pi_\pi^\tau(s) =
 \left [\ell_\pi^\tau(s) + P^\tau_\pi(s)\right ]~~~,&
\displaystyle \Pi_K^\tau(s) =
\left [\ell_{K}^\tau(s)~ + P^\tau_K(s)\right ]\,,
\end{array}
\label{eq1-40}
\ee
where $\ell_\pi^\tau(s)$ and $\ell_{K}^\tau(s)$ denote  the amputated loop functions for resp.
$\pi^\pm \pi^0$ and $K^\pm K^0$, each having absorbed {\it only} a factor of resp. $G_\pi^2$ and  $G_K^2$;
in this way, only the breaking parameters affecting the $V$ sector of BHLS$_2$ are factored out
and displayed.
$P^\tau_\pi(s)$ and $P^\tau_K(s)$ are subtraction polynomials chosen to vanish at $s=0$ because of
current conservation.
Then one has\footnote{Here and in the following, one can be led to keep
breaking parameter dependencies higher than first order to simplify the expression writing.}~:
\be
\left \{
\begin{array}{lll}
\displaystyle \Pi_{\rho \rho}^\tau(s)= \left [(1+\Sigma_V) \right ]^2 \Pi_\pi^\tau(s)
+\left [ \sqrt{2}(1+\Sigma_V)\right ]^2\Pi_K^\tau(s) \,,\\[0.5cm]

\displaystyle \Pi_W(s)= \frac{(1+\Sigma_V)}{G_\pi}
\left [ 1-\frac{a}{2}(1+\Sigma_V)\right ] \Pi_\pi^\tau(s) +
\frac{\sqrt{2}(1+\Sigma_V)}{ z_A  G_K} \left [z_A-\frac{a}{2}(1+\Sigma_V)\right ]
\Pi_K^\tau(s)\epo
 \end{array}
\right .
\label{eq1-41}
\ee
As the loop functions  vanish at $s=0$, one clearly has $F_\pi^\tau(0)=1$.
One should note that the same subtraction polynomials occur in the $W-\rho$
transition amplitude  $F_\rho^\tau(s)$ and in the inverse propagator $D_\rho(s)$ but
 differently weighted.

 The expression for the loop functions $\ell_\pi^\tau(s)$ and $\ell_K^\tau(s)$
 along the real $s$-axis can be found in Subsection F.2 of the Appendix to \cite{taupaper};
 the subtraction polynomial needed to ensure that they vanish at $s=0$ has coefficients
 given in Equations (122) of this reference. The coefficients for the constant term  and for
 the second degree one given therein are correct, but the coefficient for the term linear in $s$ should
 be corrected to~:
 $$\displaystyle
 c_1= \frac{1}{\pi} \left[ 1 - \frac{m_0^2+m_c^2}{2 m_0 m_c}
 \ln{ \frac{m_0-m_c}{m_0+m_c} }\right]
$$
using the notations\footnote{$m_0^2$ and $m_c^2$ are the $s$ values
of the direct and crossed thresholds energies.} of \cite{taupaper}.

\subsection{Parametrization of Loops in $e^+e^-$ Annihilations and $\tau$ Decays}
\label{loopfc}
\indent \indent The $\rho^0$ self-energy and the
$\gam  \Rightarrow (\rho^0/\omg/\phi)$ transition amplitudes involve important
loop corrections; these already play a central role in the dynamical mixing
of vector mesons as shown in Section \ref{DynMix}. Let us define a parametrization
common to the computation of the "angles" $\alpha(s)$, $\beta(s)$ and $\gamma(s)$,
 to the self-energies and to the $A \Rightarrow V$ transition amplitudes as these are
involved in all form factors addressed in our global fitting code.

One first defines the following loops ~:
\be
\begin{array}{lll}
\displaystyle \Pi_\pi^e(s) =
 \left [\ell_\pi^e(s) + P_\pi^e(s)\right ]~,&
\displaystyle \Pi_{K^\pm} ^e(s) =
\left [\ell_{K^\pm}^e(s)~ + P^e_{K^\pm}(s)\right ]~,&
\displaystyle \Pi_{K^0} ^e(s) =
\left [\ell_{K^0}^e(s)~ + P^e_{K^0}(s)\right ]\,,
\end{array}
\label{eq1-42}
\ee
where $\ell_\pi^e(s)$, $\ell_{K^\pm}^e(s)$ and $\ell_{K^0}^e(s)$ are resp. the
$\pi^+ \pi^-$, $K^+ K^-$ and $K^0 \overline{K}^0$ loops. These are defined
as the amputated loops, including only a factor of $G_\pi^2$ for the former loop and
of $G_K^2$ for the two others (see Equations (\ref{eq1-39b})). The expression for these loops
along the real $s$-axis  can be found in Section F.1 of the Appendix to \cite{taupaper}.

The loop parts being defined this way, $P_\pi^e(s)$, $P^e_{K^\pm}(s)$
and $P^e_{K^0}(s)$ are the subtraction polynomials, chosen of the second degree
and vanishing at $s=0$.

However, as already observed in Section \ref{DynMix}, besides the pion loop, it is rather
the sum and difference of the kaon loops which are relevant~:
\be
\left \{
\begin{array}{lll}
\displaystyle  \Pi_{K_{\rm sum}} ^e(s) =
\left [
\ell_{K^\pm}^e(s)+ \ell_{K^0}^e(s)+ P^e_{S}(s)
\right ] \equiv  \Pi_S ^e(s) \,,\\[0.5cm]
 \displaystyle  \Pi_{K_{\rm diff}} ^e(s) =
\left [\ell_{K^\pm}^e(s)-\ell_{K^0}^e(s)+P^e_{D}(s)\right ]
 \equiv  \Pi_D ^e(s)\,,
\end{array}
\right .
\label{eq1-43}
\ee
where $P^e_{S}(s)$ and $P^e_{D}(s)$ are denoting their respective subtraction polynomials.
As already noted, one can consider $ \Pi_{K_{\rm diff}} ^e(s)$ and $P^e_{D}(s)$
to be  first order in breaking (${\cal O}(\epsilon)$). So, it is certainly more appropriate
to parametrize $P^e_{S}(s)$ and $P^e_{D}(s)$ for fits and propagate them into
the subtraction polynomials of $ \Pi_{K^\pm} ^e(s)$ and $ \Pi_{K^0}^e(s)$ by
stating~:
 \be
\begin{array}{lll}
\displaystyle
 P^e_{K^\pm}(s) =\frac{1}{2} \left [  P^e_{S}(s)+ P^e_{D}(s) \right ]~~,&
 \displaystyle
 P^e_{K^0}(s) =\frac{1}{2} \left [  P^e_{S}(s)- P^e_{D}(s) \right ]\,,
\end{array}
\label{eq1-44}
\ee
keeping in mind that
$P^e_{D}(s)$  is ${\cal O}(\epsilon)$, while $P^e_{S}(s)$ is ${\cal O}(1)$.

\vspace{0.5cm}

On the other hand, as one certainly has~:
 \be
\left \{
\begin{array}{lll}
\displaystyle
\Pi_\pi^\tau(s) \simeq \Pi_\pi ^e(s) + {\cal O}(\epsilon)\,,\\[0.5cm]
\displaystyle
\Pi_K^\tau(s) \simeq \Pi_{K_{\rm sum}} ^e(s) + {\cal O}(\epsilon)\,,
\end{array}
\right .
\label{eq1-45}
\ee
and it is appropriate to impose~:
 \be
\left \{
\begin{array}{lll}
\displaystyle  P_\pi^\tau(s) =P^e_\pi(s)+ \delta P^\tau(s)\,,\\[0.5cm]
\displaystyle  P_K^\tau(s) =P^e_{S}(s)\,,
\end{array}
\right .
\label{eq1-46}
\ee
and submit to fit the coefficients of the $\delta  P^\tau(s)$ polynomial
rather than those for $P_\pi^\tau(s) $ directly. This turns out to attribute
the full breaking in $e^+e^-$ versus $\tau$ to the pion loop subtraction
term; as our fitting region is mostly below the $K\overline{K}$ thresholds, this  looks
a safe assumption. The second relation, additionally, is an assumption which allows to reduce
the number of free parameters in the minimization procedure by two units without
any degradation of the fit quality.

As a general statement, all our subtraction polynomials  have been
chosen of second degree and vanishing at $s=0$. An exception is made, however, by
choosing the third degree for
$ \delta P^\tau(s)$ which is found to provide a significantly better description
of the $\tau$ spectra, especially for the Belle dipion spectrum \cite{Belle}.

\subsection{The Pion Form Factor in the $e^+e^-$ Annihilations}
\label{pionFF}
\indent \indent
Because of the vector meson mixing, deriving the pion form factor in the $e^+e^-$
annihilation is slightly more involved than in the $\tau$ decay.
 To start, one needs
to propagate the transformation in  Equation (\ref{eq1-37}) into the Lagrangian in Appendix
\ref{AA-1} and collect all contributions to the fully renormalized $\rho$,
$\omg$ and $\phi$ field couplings.
The derived pion form factor in $e^+e^-$ annihilations $F_\pi^e(s)$  writes~:
\be
\displaystyle
F_\pi^e(s) = \left [ 1-\frac{a}{2}(1+\Sigma_V )\right] -
F_{\rho \gamma}^e(s) \frac{g_{\rho \pi \pi}}{D_\rho(s)}
- F_{\omega \gamma}^e(s) \frac{g_{\omega \pi \pi}}{D_\omega(s)}
- F_{\phi \gamma}^e(s) \frac{g_{\phi \pi \pi}}{D_\phi(s)}\epo
\label{eq1-47}
\ee

Using the $\rho^I \pi^+ \pi^-$ coupling in the Lagrangian  displayed
in Appendix \ref{AA-1} and the $I \Rightarrow R$ transform given
by Equation (\ref{eq1-37}), one readily gets~:
\be
 \begin{array}{llll}
\displaystyle g_{\rho \pi \pi} = \frac{a g}{2}(1+\Sigma_V+\xi_3) ~,&
\displaystyle g_{\omega \pi \pi} = -\frac{a g}{2}
[\psi_\omg+\alpha(s)]~,&
\displaystyle g_{\phi \pi \pi} = \frac{a g}{2}  [\psi_\phi+ \beta(s)]\,,
  \end{array}
\label{eq1-48}
\ee
keeping only the leading ${\cal O}(\epsilon)$ terms in breaking. So, BHLS$_2$ generates
$\omg$ and $\phi$ direct couplings to $\pi \pi$. The inverse $\rho^0$
propagator writes~:
 \be
\displaystyle
D_ {\rho^0}(s) =s-m_{\rho_0}^2 -\Pi_{\rho \rho}^e(s)\,,
\label{eq1-49}
\ee
where $m_{\rho_0}^2$ is defined in Equations (\ref{eq1-26}) and $\Pi_{\rho \rho}^e(s)$
is the self-energy of the physical ({\it i.e.} fully renormalized) $\rho^0$.
This can be defined in terms of the loop functions constructed in the Subsection just
above.  $\Pi_{\rho \rho}^e(s) $  is made up of two pieces~:
 \be
 \left \{
 \begin{array}{llll}
\displaystyle \Pi_\pi^e(s) =  \left [\widetilde{g}^\rho_\pi \right ]^2
 \left [\ell_\pi^e(s) + P^e_\pi(s)\right ]~~~,~~~
\displaystyle \widetilde{g}^\rho_\pi =(1+\Sigma_V+\xi_3)\,,\\[0.5cm]
\displaystyle \Pi_K^e(s) =
 \left [ \widetilde{g}^\rho_{K\pm}\right ]^2
\ell_{K^\pm}^e(s) + \left [\widetilde{g}^\rho_{K0} \right ]^2
\ell_{K^0}^e(s)+
\frac{1}{2}\left [ \left [\widetilde{g}^\rho_{K\pm}\right ]^2+
\left [\widetilde{g}^\rho_{K0}\right ]^2
\right ]  P^e_S(s)+ {\cal O}(\epsilon^2)\,,
\end{array}
\right.
\label{eq1-50}
\ee
where the couplings $\widetilde{g}^\rho_{K\pm}$ and
$\widetilde{g}^\rho_{K0}$ are given in Appendix \ref{AA-4}.
A term$\left [ \left [\widetilde{g}^\rho_{K\pm}\right ]^2-
\left [\widetilde{g}^\rho_{K0}\right ]^2 \right ] P^e_D(s)/2$ has
been omitted in the expression for $\Pi_K^e(s) $  as being ${\cal O}(\epsilon^2)$.

 As for the inverse propagators $D_ {\omg}(s)$ and $D_ {\phi}(s)$,
 one uses the expression given by Equation (\ref{bw1})
with the appropriate  values for HK masses $m_V^2$, and having
([$\widetilde{m}_V$, $\widetilde{\Gamma}_V$], $V=\omg,~\phi$) as parameters to be determined
by fit.

\vspace{0.5cm}

\indent \indent
The loop corrected $V-\gamma$ transitions amplitudes $F_{V\gamma}^e(s)$ are given by~:
\be
\displaystyle F_{V\gamma}^e(s)= f_{V\gamma} -c_{V\gamma}~ z_3~ s - \Pi_{V\gamma}(s)~~,~~
(V=\rho^0_R,~\omg_R,~\phi_R)~~,
\label{eq43q}
\ee
where the new-comer $z_3 s$ terms occur as in $F_{\rho}^\tau(s)$ above.
Using together the ideal couplings Equations (\ref{AA-3}), the renormalization matrix
Equation (\ref{eq1-37}) and the  RS conditions (see Section \ref{RefSol}),  one derives~:
 \be
 \left \{
 \begin{array}{llll}
\displaystyle
f_{\rho \gamma}=a g  f_\pi^2 \left [ 1+\Sigma_V+\xi_3+\frac{[\psi_\omg+\alpha(s)]}{3}
+z_V\frac{\sqrt{2}[\psi_\phi+\beta(s)]}{3} \right ]\,,\\[0.5cm]
\displaystyle
f_{\omg \gamma }=\frac{a g  f_\pi^2}{3} \left [ 1+\Sigma_V+\xi_8-3[\psi_\omg +\alpha(s)]
+z_V\sqrt{2}[\psi_0+\gamma(s)] \right ]\,,\\[0.5cm]
\displaystyle
f_{\phi\gamma}=\frac{a g  f_\pi^2}{3} \left [-z_V\sqrt{2}(1+\xi_8) +
3[\psi_\phi+\beta(s)] +[\psi_0+\gamma(s)] \right ]\,,
\end{array}
\right.
\label{eq1-51}
\ee
which become $s$-dependent thanks to the dynamical vector mixing. Similarly, using
the couplings which can be read off  ${\cal L}_{z_3}$  in Equation (\ref{eqAA-5}) and
the transformation matrix given by Equation (\ref{eq1-37}), one also derives~:
 \be
 \left \{
 \begin{array}{llll}
\displaystyle
c_{\rho \gamma}=g \left [ 1+ \xi_3+\frac{[\psi_\omg+\alpha(s)]}{3}
+\frac{\sqrt{2}[\psi_\phi+\beta(s)]}{3} \right ]\,,\\[0.5cm]
\displaystyle
c_{\omg \gamma}=\frac{g }{3} \left [ 1+\xi_8-3[\psi_\omg +\alpha(s)]
+\sqrt{2}[\psi_0+\gamma(s)] \right ]\,,\\[0.5cm]
\displaystyle
c_{\phi \gamma}=\frac{g}{3} \left [-\sqrt{2}(1+\xi_8) +
3[\psi_\phi+\beta(s)] +[\psi_0+\gamma(s)] \right ]\,,
\end{array}
\right.
\label{eq1-51b}
\ee
neglecting terms of breaking order greater than 1.
 Finally, the  relevant loop transition terms $\Pi_{V\gamma}(s)$ are collected in Appendix
\ref{AA-5}.

\subsection{The Charged and Neutral Kaon Form Factors}
\label{kaonFF}
\indent \indent  The charged and neutral kaon electromagnetic form-factors can
also be derived from the BHLS$_2$ Lagrangian given in Appendix \ref{AA-1}~:
 \be
\begin{array}{ll}
\displaystyle
F^e_{K_c}(s)=&\displaystyle \left[1 -\frac{a}{6z_A}
(3+2\Sigma_V-\frac{3\Delta_A}{2}) \right]\crn[0.5cm] & \hspace*{1cm}\displaystyle
-F_{\rho\gamma}(s) \frac{g_{\rho K^+ K^-} }{D_\rho(s)}
-F_{\omg\gamma}(s)\frac{g_{\omg K^+ K^-} }{D_\omg(s)}
-F_{\phi\gamma}(s)\frac{g_{\phi K^+ K^-} }{D_\phi(s)}\,,\\[0.5cm]
\displaystyle
F^e_{K_0}(s)=&\displaystyle -\frac{a\Sigma_V}{6z_A}
- F_{\rho\gamma}(s) \frac{g_{\rho K^0  \overline{K}^0}}{D_\rho(s)}
-F_{\omg\gamma}(s)\frac{g_{\omg K^0  \overline{K}^0} }{D_\omg(s)}
-F_{\phi\gamma}(s)\frac{g_{\phi K^0  \overline{K}^0} }{D_\phi(s)}\,,
\end{array}
\label{eq1-52}
\ee
where the $\gamma - V$ transition amplitudes $F_{V\gamma}$ and the propagators have
been already defined and the $VK\overline{K}$ couplings are displayed in the Appendix
\ref{AA-4}.

The experimental spectra  measured in the spacelike region, in particular by the NA7 Collaboration
\cite{NA7,NA7_Kc},  being  squared form factors, the modulus squared of the form factors
given in Equation (\ref{eq1-47}) and in  Equations (\ref{eq1-52}) apply directly.
In the timelike region, experimentalists rather publish cross-sections~:
\be
\displaystyle
\sigma(e^+e^- \ra P \overline{P})=\frac{8 \pi \alpha_{\rm em}^2}{3 s^{5/2}} q_P^3 |F^e_P(s)|^2\,,
\label{eq1-53}
\ee
where  $q_P=\sqrt{s-4 m_P^2}/2$ is the meson
momentum in the center-of-mass system ($P=K^+, K^0$). In the case of the
$K^+ K^-$ final state, the cross-section  should take into account the significant
interaction between the charged
kaons emerging from the $e^+e^-$ annihilation. Traditionally, this is done
by multiplying the $e^+e^- \ra K^+ K^-$ cross-section by the leading term describing
the Coulomb interaction as formulated by \cite{Gourdin,BGPter}~:
\be
\displaystyle
Z(s)= \left [ 1 + \pi \alpha_{\rm em} \frac{1+v^2}{2 v}
\right ]~~~,~~ v=\frac{\sqrt{s-4 m_{K^\pm}^2}}{s}\epo
\label{eq1-54}
\ee

As the data have become more and more precise, it looks more convenient
to use the exponentiated expression derived in \cite{IFSVP2} which includes also
the Final State Radiation (FSR) correction factor; indeed, the $K^+ K^-$ spectra
provided by the various experiments are not amputated from FSR effects.

\subsection{Form Factors in BHLS versus BHLS$_2$}
\indent \indent It looks worth pointing out the difference between the form factor
expressions in   BHLS and in BHLS$_2$.   As already noted,
the ${\cal L}_A$ Lagrangian is  common to both  frameworks.
In contrast, the ${\cal L}_V$ Lagrangian is significantly different.
Obvious differences for ${\cal L}_V$  have
already been noted.  For instance, within BHLS$_2$, $\Delta_V=0$ and its sharing parameter $h_V$
drop out whereas they are a key ingredient in the diagonalization procedure
of BHLS.

Three parameters\footnote{Additional constraints arising in the diagonalization  procedure of the vector
mass term, have reduced this number to two as $\xi_0=\xi_8$ is requested. }
 arise from the Covariant Derivative breaking ($\xi_3$,~$\xi_0$,~$\xi_8$).
 Moreover, a key role is played by the $\psi$ rotation angles within the RS solution
 of BHLS$_2$. None of these parameters occur in the BHLS framework of \cite{ExtMod3}.

The effect of a significant  $\xi_3$  value is obviously equivalent to
 shifting apart the Higgs-Kibble masses of both $\rho$'s and their couplings
 to $\pi\pi$ in a correlated way. As this kind of exercise performed
 within BHLS concluded to a small effect \cite{ExtMod3,ExtMod4},
 one expects a small value for $\xi_3$. Also, BHLS$_2$
 deals with different subtraction polynomials -- the $\delta P^\tau(s)$ function
 already referred to above --
for the $\pi^\pm \pi^0$ and $\pi^+ \pi^-$ loops, a new (loop) mechanism able alone to
contribute to physical mass and width differences between the $\rho^0$ and $\rho^\pm$
 mesons.

One should also note another difference~: In the present framework,
as already noted in Subsection \ref{FFtau}, the pion and kaon loops entering
the amplitudes $F_\rho^\tau(s)$ and $F_\rho^e(s)$ carry fitted subtraction
polynomials already involved in the charged and neutral $\rho$ inverse propagators
$D_\rho(s)$. In BHLS, one allowed  the subtraction polynomial in
$F_\rho^{e/\tau}(s)$  to carry a piece independent of both
$D_\rho(s)$'s. Because of the ${\cal O}(p^4)$ term still introduced, it
means that a second-degree term in $s$ has been removed from
$F_\rho^{e/\tau}(s)$ in the BHLS$_2$ framework. We should see
below the effect of this on the $\pi \pi$
$P$-wave phase-shift above the $\phi$ mass.

\section{The Anomalous Sector of BHLS$_2$}
\label{anomalous}
\indent \indent As in \cite{ExtMod3}, the anomalous FKTUY
Lagrangian \cite{FKTUY,HLSRef}  is used to address
 the $e^+e^- \ra (\pi^0/\eta) \gamma$ and
$e^+e^- \ra  \pi^0\pi^+\pi^- $ processes, and also the radiative decays  with
couplings of the form $VP\gamma$ and $P \gamma \gamma$ which are parts of
our global fitting framework. One can display
its various pieces~:
\be
\begin{array}{ll}
\displaystyle
{\cal L}_{\rm anomalous}&=
(1-c_4)~{\cal L}_{AAP}+c_3{\cal L}_{VVP}+
\left [ 1-\frac{3}{4}(c_1-c_2+c_4)  \right ]{\cal L}_{APPP}\\[0.5cm]
~~&+(c_1-c_2-c_3)~{\cal L}_{VPPP}+
(c_4-c_3)~{\cal L}_{AVP}\,,
\end{array}
\label{eq2-1}
\ee
where, temporarily,  the weight of each piece has been
extracted out  to exhibit its
dependence upon the unconstrained  $c_i$ constants of the anomalous HLS Lagrangian
\cite{FKTUY}.
If one imposes to yield the Wess-Zumino-Witten terms \cite{WZ,Witten}
at the chiral point -- the so-called triangle and box anomalies --
 the condition $c_3=c_4$ is known to be
mandatory \cite{HLSOrigin,HLSRef}. Reference \cite{ExtMod3} having also shown that
this condition is indeed favored by fits to the annihilation data,
one endorses the $c_3=c_4$ constraint and, thus, cancels out the last term in
Equation (\ref{eq2-1}). Hence, the $VP\gamma$ couplings become  entirely generated
by the combination of $VVP$ couplings with the  $V-\gamma$ transitions provided by
 the non-anomalous HLS Lagrangian.

The surviving Lagrangian pieces displayed in Equation (\ref{eq2-1}) are given by~:
\be
\left \{
\begin{array}{lll}
{\cal L}_{VVP}=& \displaystyle - \frac{N_c g^2}{4 \pi^2 f_\pi} ~c_3
 \epsilon^{\mu \nu \alpha \beta}{\rm Tr}[ \partial_\mu V_\nu \partial_\alpha V_\beta P] \,,\\[0.5cm]
 {\cal L}_{AAP}=& \displaystyle - \frac{N_c e^2 }{4 \pi^2 f_\pi} ~(1- c_4)
 \epsilon^{\mu \nu \alpha \beta}\partial_\mu A_\nu \partial_\alpha A_\beta{\rm Tr}[Q^2 P]\,,\\[0.5cm]
 {\cal L}_{VPPP}=& \displaystyle - i \frac{N_c g }{4 \pi^2 f_\pi^3} (c_1-c_2-c_3)
   \epsilon^{\mu \nu \alpha \beta}{\rm Tr}[V_\mu \partial_\nu P \partial_\alpha P \partial_\beta P]\,,\\[0.5cm]
  {\cal L}_{APPP}=& \displaystyle - i \frac{N_c e}{3 \pi^2 f_\pi^3} [1- \frac{3}{4}(c_1-c_2+c_4)]
  \epsilon^{\mu \nu \alpha \beta} A_\mu{\rm Tr}[ Q \partial_\nu P \partial_\alpha P \partial_\beta P]\,,
\end{array}
\right .
\label{eq2-2}
\ee
where $N_c$ is the number of colors and the $c_i$  factors have been  reabsorbed
 in the normalization of the various Lagrangian pieces.
$P$ and $V$ denote the  {\it bare} pseudoscalar  and vector field matrices. In the present
context, one has, however~:
\be
\displaystyle   V_\mu= V_\mu^I +\delta V_\mu \,,
\label{eq2-3}
\ee
where $V_\mu^I$ is the (usual) $U(3)$ symmetric vector field matrix and
$\delta V_\mu=\delta V_\mu^3+\delta V_\mu^0+\delta V_\mu^8$ which has been
 expressed in term of the ideal field combinations in Equations (\ref{eq1-10}).

For the reader's convenience, Appendix \ref{BB} reminds  briefly the renormalization
procedure which leads from the bare PS fields to their renormalized partners \cite{ExtMod3}.

\subsection{ Renormalization of the $VVP$ Couplings}
\indent \indent As displayed in Equation (\ref{eq1-34}), referring to~:
$$\widetilde{{\cal V}}=(\rho_0,~\omg,~\phi)~~,$$
the relation between the fully renormalized vector fields (denoted $R$ here and from now on)
 and their ideal combination  is given by~:
$$
{\cal V}^I \Rightarrow \left[R(U_3) \cdot R(Loop) \right]  \cdot {\cal V}^{R}\,,
$$
where the product $R(U_3) \cdot R(Loop)$,   given by Equation (\ref{eq1-37}),  will be denoted from now on~:
\be
\displaystyle  {\cal R}(s) = R(U_3) \cdot R(Loop)\epo
\label{eq2-4}
\ee
 Let us point out that ${\cal R}(s)$ is a real analytic $3\times 3$ matrix function and
  fulfills at  order ${\cal O}(\epsilon)$~:
\be
\displaystyle   \widetilde{ {\cal R}}(s+i\epsilon) \cdot {\cal R}(s+i\epsilon)=
{\cal R}(s+i\epsilon)  \cdot \widetilde{ {\cal R}} ( s+i\epsilon)=1
\label{eq2-5}
\ee
along the right-hand cut on the physical sheet \cite{taupaper}. The $VVP$ Lagrangian
displayed in Appendix \ref{CC} can
be symbolically written~:
\be
\displaystyle
{\cal L}_{VVP}^{\rm ideal}(P)=
\displaystyle
\frac{C}{2} \epsilon^{\mu \nu \alpha \beta} \left \{ P_0~~
  \partial_\mu {\cal V}^I_{i~\nu}  G_{i,j}^I (P_0) \partial_\alpha {\cal V}_{j~\beta}^I
 +G_{i}^I (\pi^\pm) [
 \pi^- ~\partial_\mu \rho^+_{\nu} \partial_\alpha {\cal V}_{i~\beta}^I
 + \pi^+  ~\partial_\mu \rho^-_{\nu}\partial_\alpha {\cal V}_{i~\beta}^I
 ]
 \right \}\,,
\label{eq2-6}
\ee
in terms of ideal field combinations, $P_0$ being any of the $\pi^0$, $\eta$ or $\etp$ fields;
summation over greek and latin indices is understood.
The 3-vector $G^I (\pi^\pm)$ and the 3$\times$3 matrix $G^I (P_0)$ can easily be constructed using
the relevant pieces of information given in Appendix \ref{CC}. One thus has~:
\be
\displaystyle
\widetilde{G^I} (\pi^\pm)=\left ( \widetilde{g}^\pm_{\rho^0 \rho^\pm},
\widetilde{g}^\pm_{\omg \rho^\pm},
\widetilde{g}^\pm_{\phi \rho^\pm} \right )\,,
\label{eq2-7}
\ee
with the obvious  $(\rho,~\omg,~ \phi)$ component indexing. On the other hand, one has defined~:
\be
G^I (P_0) =
 \left (
\begin{array}{lll}
\displaystyle \widetilde{g}_{\rho \rho}& \displaystyle\frac{1}{2}\widetilde{g}_{\rho \omg}
& \displaystyle\frac{1}{2}\widetilde{g}_{\rho \phi}\\[0.5cm]
\displaystyle\frac{1}{2} \widetilde{g}_{\omg \rho}& \displaystyle\widetilde{g}_{\omg \omg}
& \displaystyle\frac{1}{2}\widetilde{g}_{\omg \phi}\\[0.5cm]
\displaystyle \frac{1}{2} \widetilde{g}_{\phi \rho}& \displaystyle\frac{1}{2}\widetilde{g}_{\phi \omg}
& \displaystyle \widetilde{g}_{\phi \phi}
\end{array}
\right)\,,
\label{eq2-8}
\ee
where, for each $P_0$,   $\widetilde{g}_{i j}$ is the coupling of the form $P_0 {\cal V}_i {\cal V}_j$
which can be read off the relevant expressions given  in  Appendix \ref{CC}.

Performing  the replacement ${\cal V}^I={\cal R}(s) {\cal V}^R$ in  Equation (\ref{eq2-6}),  one can derive the fully
renormalized $VVP$ Lagrangian~:
\ba
\displaystyle
{\cal L}_{VVP}^{\rm ren}(P)&=& \displaystyle   \frac{C}{2} \epsilon^{\mu \nu \alpha \beta} \left [
P_0~~ \partial_\mu {\cal V}^R_{i~\nu}  G_{i,j}^R (P_0) \partial_\alpha
{\cal V}_{j~\beta}^R  \right. \crn[0.2cm] && \left.\displaystyle + \pi^- ~\partial_\mu
\rho^+_{\nu}G_{i}^R (\pi^\pm)\partial_\alpha {\cal V}_{i~\beta}^R
+ \pi^+  ~\partial_\mu \rho^-_{\nu}G_{i}^R
(\pi^\pm)\partial_\alpha {\cal V}_{i~\beta}^R \right ]\,,
\label{eq2-9}
\ea
where, the coupling vector $G^R (\pi^\pm)$ and the coupling
matrix $G^R(P_0)$ inherit from ${\cal R}$  a $s$-dependence;
they are related to their ideal partners by~:
\be \begin{array}{lll}
\displaystyle   G^R (\pi^\pm) =\widetilde{ {\cal R}}(s) G^I (\pi^\pm)~~~,& \displaystyle G^R (P_0)=\widetilde{ {\cal R}}(s)G^I
(P_0){\cal R}(s)\,,
\end{array}
\label{eq2-10}
\ee
which is a concise way to express the renormalized couplings.

\subsection{The Renormalized AVP Effective Lagrangian}
\label{AVPcouplings}
\indent \indent  As already noted,   setting  $c_3=c_4$ turns out to cancel
the direct FKTUY $AVP$ Lagrangian piece;  hence, all  $AVP$ couplings inside
BHLS$_2$  are
generated by $VVP$ couplings followed by one  $V \ra \gamma$ transition.
Let us consider the $V-\gamma$ transition term of the non-anomalous BHLS$_2$ Lagrangian (see
Equation (\ref{eqAA-2})). Having defined the transposed vector
$\widetilde{f_{V \gamma}^I}=(f_{\rho \gamma}^I,~f_{\omg \gamma}^I,~f_{\phi \gamma}^I)$,
one can rewrite it~:
\be
\displaystyle
{\cal L}_{AV} = -e \left [ \widetilde{f_{V \gamma}^I} \cdot {\cal V}^I_\mu \right ] A^\mu
=-e \left [ \widetilde{f_{V \gamma}^R}(s) \cdot {\cal V}^R_\mu \right
] A^\mu \,,
\label{eq2-11}
\ee
which defines the 3-vector $f_{V \gamma}^R$ as~:
\be
\displaystyle
f_{V \gamma}^R= \widetilde{{\cal R}}(s)f_{V \gamma}^I\epo
\label{eq2-12}
\ee

The effective ${\cal L}_{AVP} $ Lagrangian can be derived from Equation (\ref{eq2-9})
by replacing one neutral vector meson, say ${\cal V}_i^R$, using the rule~:
\be
\displaystyle
{\cal V}^R_{i~\mu}\Rightarrow (-ief_{\gamma V_i}^R) \left( \frac{-i}{D_{V_i}(0)} \right ) A_\mu
= \frac{ef_{\gamma v_i}^R}{m_{i}^2}  A_\mu \equiv e H_{i~\gamma} A_\mu\,,  ~~~(i=\rho^0,~\omg,~\phi)
 \label{eq2-13}
\ee
where  $m_{i}^2$  is the renormalized squared mass of the ${\cal V}_i$ vector
meson as given in Equations (\ref{eq1-26}); let us note that each
$H_{i~\gamma}$ carries a hidden factor of $1/g$. Starting with the charged pion part of
Equation (\ref{eq2-9}), one gets~:
\be
 \displaystyle
{\cal L}_{AVP}^{\rm ren}(\pi^\pm)=
\displaystyle
\frac{eC}{2} [G^R(\pi^\pm) \cdot H_{\gamma V}] ~\epsilon^{\mu \nu \alpha \beta}  \partial_\mu A_\nu
\left[ \partial_\alpha \rho^+_\beta \pi^- +
\partial_\alpha \rho^-_\beta \pi^+\right]\,,
\label{eq2-14}
\ee
where $H_{\gamma V}$ is the 3-vector constructed from the $H_{i~\gamma}$
defined just above~:
\be
 \displaystyle
 \widetilde{H}_{\gamma V}=\left ( \frac{f_{\gamma \rho_0}^R}{m_{\rho_0}^2},~
\frac{f_{\gamma \omg}^R}{m_{\omg}^2},~
\frac{f_{\gamma \phi}^R}{m_{\phi}^2} \right )\epo
\label{eq2-15}
\ee

For the $P_0$ part of the Lagrangian  Equation (\ref{eq2-9}), the algebra is
slightly more involved and gives~:
\be
\displaystyle
{\cal L}_{AVP}^{\rm ren}(P_0)=
\displaystyle
eC  \sum_{i=\rho,\omg,\phi}
 [\widetilde{H}_{\gamma V} G^R(P_0)]_i ~~\epsilon^{\mu \nu \alpha \beta}
  \partial_\mu V_{i~\nu}^R \partial_\alpha A_{\beta}~~ P_0~~~,~~~P_0=(\pi^0,~\eta,~\etp)\,,
\label{eq2-16}
\ee
with~:
\be
 \displaystyle
C=-\frac{N_c g^2}{4\pi^2 f_\pi}~c_3\epo
\label{eq2-17}
\ee
This effective Lagrangian is the tool to parametrize the $e^+e^- \ra (\pi^0/\eta) \gamma$
cross-sections and the $VP\gamma$ radiative decays.

\subsection{The $e^+e^- \ra (\pi^0/\eta) \gamma$ Cross-Sections}
\label{Pg-cs}
\indent \indent Using the $AAP$ and $AVP$ Lagrangians derived above in terms
of renormalized vector and pseudoscalar fields, one can construct the amplitudes
for $\gamma^* \ra (\pi^0/\eta) \gamma$ transitions. These can be written~:
\be
\displaystyle
T(\gamma^* \ra P_0 \gamma)=
-i\frac{3 \alpha_{\rm em}}{\pi f_\pi} \left [g^2 c_3 \widetilde{H}_{\gamma V} G^R(P_0) K_{\gamma V}
-(1-c_3) L_{P_0} \right ]= -i\frac{3 \alpha_{\rm em}}{\pi f_\pi} F_{P_0 \gamma}(s)\,,
\label{eq2-18}
\ee
where the (co)vector $ \widetilde{H}_{\gamma V}$ is defined in Equation (\ref{eq2-15}), the
$G^R(P_0)$ matrix in Equation (\ref{eq2-10}) (and  Equation (\ref{eq2-8})). One has also defined~:
 \be
 \displaystyle
 \widetilde{K}_{\gamma V}=\left ( \frac{F_{\gamma \rho^0}^R(s)}{D_{\rho^0}(s)},~
\frac{F_{\gamma \omg}^R(s)}{D_{\omg}(s)},~
\frac{F_{\gamma \phi}^R(s)}{D_{\phi}(s)} \right )~~,
\label{eq2-19}
\ee
the functions $F_{\gamma V}^R(s)$ and $D_{V}(s)$ having been  defined
in Subsection \ref{pionFF}. Finally, one has~:
\be
\begin{array}{ll}
 \displaystyle
L_{\pi^0}= \frac{g_{\pi^0 \gamma \gamma}}{3}~~~,& {\rm and~~}
 \displaystyle
L_{\eta}= \frac{g_{\eta \gamma \gamma}}{\sqrt{3}}~~~,
\end{array}
\label{eq2-20}
\ee
 with $g_{P_0 \gamma \gamma}$, derived in \cite{ExtMod3},
 and reminded in Appendix \ref{DD}.

 In the case of the $e^+e^- \ra (\pi^0/\eta) \gamma$ annihilations, the available data
 are always cross-sections which are related to $F_{P_0 \gamma}(s)$ just defined by~:
 \be
 \displaystyle
 \sigma(e^+e^- \ra P_0 \gamma,s)= \frac{3 \alpha_{\rm em}^3}{8 \pi f_\pi^2}
\left [  \frac{s-m_{P_0}^3}{s} \right ]^3 |F_{P_0 \gamma}(s)|^2\epo
\label{eq2-21}
\ee

\subsection{The $e^+e^- \ra \pi^0\pi^+\pi^-$ Cross-section}
\indent \indent  The amplitude for the  $\gamma^* \ra \pi^0\pi^+\pi^-$ transition
involves most of the FKTUY Lagrangian pieces; it can be written~:
\be
 \displaystyle
 T(\gamma^* \ra \pi^0\pi^+\pi^-)=T_{APPP}+T_{VPPP}+T_{VVP}~~,
\label{eq2-22}
\ee
 labeling each term  by the particular piece of the FKTUY Lagrangian
from which it originates. As already noted, because  $c_3=c_4$ is assumed,
there is no $T_{AVP}$ piece.

The $T_{APPP}$ contribution to the full  $T(\gamma^* \ra \pi^0\pi^+\pi^-)$
amplitude can be derived from the information given  in Appendix \ref{DD}; it
writes~:
\be
\hspace{-1cm}
\begin{array}{ll}
\displaystyle T_{APPP}= C_{APPP}
\left [1  -G(\delta_P) \right ]
\epsilon_{\mu \nu \alpha \beta} ~~\epsilon_\mu(\gamma) p_0^\nu
p_-^\alpha p_+^\beta \,,
\end{array}
\label{eq2-23}
\ee
where one has defined~:
\be
\begin{array}{llll}
\displaystyle G(\delta_P)=  \frac{\Delta_A}{2}+
\epsilon_1 \sin{\delta_P} -\epsilon_2 \cos{\delta_P}~~,& \displaystyle
C_{APPP}= -\frac{ie}{4 \pi^2 f_\pi^3}\left [ 1-\frac{3}{4} (c_1-c_2+c_4) \right ]\epo
\end{array}
\label{eq2-24}
\ee

Three pieces are coming from the $VPPP$ Lagrangian also displayed in Appendix  \ref{EE};
they are collected in~:
\be
\hspace{-0.5cm}
\begin{array}{ll}
\displaystyle T_{VPPP}=& \displaystyle C_{VPPP}
\left [ \sum_{V=\rho,\omg,\phi} \frac{F^e_{V\gamma}(s)}{D_V(s)}  g_{V\pi}^R (s) \right ]
\epsilon_{\mu \nu \alpha \beta} ~~\epsilon_\mu(\gamma) p_0^\nu p_-^\alpha p_+^\beta\,,
\end{array}
\label{eq2-25}
\ee
where the renormalized vector couplings   $g_{V\pi}^R (s)$ to 3 pions have to be derived
from using Equations (\ref{EE2}) and Equation (\ref{EE5}),  ${\cal R}(s)$ being given by
Equation (\ref{eq1-37}). The $V-\gamma$ amplitudes $F^e_{V\gamma}(s)$ have been constructed
in Subsection \ref{pionFF} and the inverse propagators in  Subsection \ref{pionFF} for
the $\rho^0$ meson. The inverse propagators for the  $\omg$ and $\phi$ mesons have
been discussed and defined in  Section \ref{BasicSol}. One has also
defined~:
\be
\displaystyle
C_{VPPP}= -\frac{3ige}{4 \pi^2 f_\pi^3} \left [
c_1-c_2-c_3 \right ]\epo
\label{eq2-26}
\ee
\vspace{0.3cm}

 The $VVP$ Lagrangian piece in Equation (\ref{CC7}) has
been rewritten in terms of renormalized fields in Equation (\ref{eq2-9}) with
(renormalized)  couplings derivable by means of Equation (\ref{eq2-10}).
The simplest way to write down $T(\gamma^* \ra \pi^0 \pi^+ \pi^-)$  in a way
easy to code within our global fit procedure is displayed just below.

One first defines the $H_i(s)$ functions~:
 \be
\hspace{-1cm}
\left \{
\begin{array}{llll}
\displaystyle  H_0(s) =1  \,, & \displaystyle H_1(s) =  \frac{1}{D_\rho(s_{+-})} +
\frac{1}{D_\rho(s_{0+})} + \frac{1}{D_\rho(s_{0-})}\,,~\\[0.5cm]
\displaystyle H_2(s)=\displaystyle \frac{1}{D_\rho(s_{+-})}~~,&
\displaystyle H_3(s)=\displaystyle
\widetilde{\alpha} (s_{+-}) 
\left [ \frac{1}{D_\rho(s_{+-})} - \frac{1}{D_\omg(s_{+-})} \right ]\,, 
\end{array}
\right.
\label{eq2-27}
\ee
where $s$ is the incoming squared energy and the $s_{ij}$'s indicate the invariant mass squared of the
corresponding outgoing $(i,j)$ pairs; the primed mixing angles have been defined in Subsection \ref{angle}. 
$T_{VVP}$ depends on the 3 functions $(H_i(s), ~i=1~\cdots 3  $
with $s$-dependent coefficients $F_i(s)$ given below.

Collecting all terms, the full amplitude writes~:
 \be
\displaystyle T(\gamma^* \ra \pi^0 \pi^+ \pi^-)=
\left [ F_0(s) H_0(s)+ C_{VVP}\sum_{i=1\cdots 3}  F_i(s) H_i(s) \right ]
\epsilon_{\mu \nu \alpha \beta} ~~\epsilon_\mu(\gamma) p_0^\nu p_-^\alpha p_+^\beta\,,
\label{eq2-28}
\ee
with~:
 \be
 \displaystyle C_{VVP}= -i \frac{3 e g m^2 }{8 \pi^2 f_\pi^3} (1+\Sigma_V) c_3\epo
\label{eq2-29}
\ee
In this way,  to write down the full amplitude, the various $F_i(s)$ functions only depend on
the incoming off-shell photon energy squared $s$; the dependence upon the various sub-energies $s_{ij}$
is, instead,  only carried by the $H_i(s)$ functions as clear from Equations (\ref{eq2-27}). One has~:
\be
\left \{
\begin{array}{llll}
\displaystyle F_0(s)= C_{APPP} \left [1-G(\delta_P)\right ]
+C_{VPPP}\left [\frac{F^R_{\rho\gamma }(s)}{D_\rho(s)}  g_{\rho\pi}^0 (s)
+\frac{F^R_{\omg\gamma}(s)}{D_\omg(s)}  g_{\omg\pi}^0 (s)+
\frac{F^R_{\phi\gamma}(s)}{D_\phi(s)}  g_{\phi\pi}^0 (s)
\right ]
\,,\\[0.5cm]
\displaystyle F_1(s) = \displaystyle
\widetilde{\alpha}(s)\frac{F^R_{\rho\gamma}(s)}{D_\rho(s)}
+ \left [ 1+\frac{2\xi_0+\xi_8}{3} \right ]\frac{F^R_{\omg\gamma}(s)}{D_\omg(s)}
+\left [\frac{\sqrt{2}}{3} (\xi_0-\xi_8)  +\widetilde{\gamma}(s)\right ]
\frac{F^R_{\phi\gamma}(s)}{D_\phi(s)}
\,,\\[0.5cm]
\displaystyle F_2(s) = \displaystyle
\left [ \epsilon_2 \cos{\delta_P}-
\epsilon_1 \sin{\delta_P} - \frac{\Delta_A}{2} \right ] \frac{F^R_{\rho\gamma}(s)}{D_\rho(s)}
+ \frac{3\xi_3}{2}  \frac{F^R_{\omg\gamma}(s)}{D_\omg(s)}
\,, \\[0.5cm]
 \displaystyle F_3(s) = \displaystyle
\frac{F^R_{\rho\gamma}(s)}{D_\rho(s)}
 \epo
\end{array}
\right.
\label{eq2-30}
\ee
where $\widetilde{\alpha}(s)=\alpha(s)+\psi_\omg$ and $\widetilde{\gamma}(s)=\gamma(s)+\psi_0$.

One should note, as in BHLS \cite{ExtMod3} and earlier in \cite{ExtMod1},  that all parameters
already fitted with the (five) $e^+e^- \ra (\pi \pi/K \overline{K}/\pi^0 \gamma / \eta \gamma)$
processes fully determine all  the $(F_i(s), ~i=1~\cdots 3)$. The only new parameter
coming with $e^+e^- \ra \pi^0 \pi^+\pi^-$ is $c_1-c_2$,  only affecting $F_0(s)$; for practical purposes,
it is handled in a specific manner, as in the previous versions of the (broken) HLS model just quoted.

Indeed, a global fit to all cross-sections
but $e^+e^- \ra \pi^0 \pi^+\pi^-$ allows to yield the relevant  parameters
with a good approximation;  thus,  having at hand all ingredients  defining the $F_i(s)$'s,
a first minimization run including the $e^+e^- \ra \pi^0 \pi^+\pi^-$ cross-section
can be performed to also derive a first estimate for $c_1-c_2$. The output of this {\sc minuit}
minimization run is then used as input for a next minimization step. This initiates an iteration
procedure involving all cross-sections which is carried on up to convergence --
when some criterion is met.

This method has been proven to converge in a very few minimization steps \cite{ExtMod3}. What makes such a
 minimization procedure unavoidable is that the $e^+e^- \ra \pi^0 \pi^+\pi^-$ cross-section expression implies
to integrate over the $s_{ij}$'s at each step. This is obviously  prohibitively computer time consuming for a negligible gain.
Hence, at each step, one starts by tabulating  coefficient functions,
exhibited in the next expression between curly brackets~:
\be
\begin{array}{ll}
\displaystyle
\sigma(e^+e^- \ra \pi^0 \pi^+\pi^-,s)=&  \displaystyle
\frac{\alpha_{\rm em} ~s^2}{192 \pi^2}
\displaystyle
\times
\left [
\left \{ \int G dx dy \right \} |F_0(s)|^2 \right. \crn[0.3cm]&
\left.\hspace*{-2.5cm}
+ C_{VVP}^2 \sum_{i,j=1\cdots 3}  F_i(s) F_j^*(s)
\left  \{ \int G H_i H_j^* dx dy \right  \}
\right .
 \\[0.3cm]
~~~&  \displaystyle  \left . \hspace*{-2.5cm}
+C_{VVP} \sum_{i=1\cdots 3} \left ( F_0(s) F_i^*(s) \left  \{ \int G H_i^* dx dy\right  \}
\displaystyle
+ F_0^*(s) F_i(s) \left  \{ \int G H_i dx dy\right  \}  \right )
\right ]
\end{array}
\label{eq2-31}
\ee
and these tables are used all along this step. Equation (\ref{eq2-31}) uses
the Kuraev-Silagadze parametrization \cite{Kuraev}
and its kernel $G(x,y)$ function; these are reminded in Appendix H of \cite{ExtMod3}.

\section{The Data Samples Submitted to Global Fits}
\label{DataSamples}
\indent \indent Basically, 48 experimental data samples are presently available which enter
the scope of the HLS Model.  Relying on the global fits performed in
\cite{ExtMod3,ExtMod4}, one has been led to discard a few of them
from our fitting procedure; this situation happens again within the present framework.
Before going on, let us remind  the most important part of the data samples and list
the newcomers. Newcomers encompass either newly collected samples or
 existing form factor spectra covering spacelike momenta not treated within the former BHLS framework
 \cite{ExtMod3}.

 Substantially, the set of available  data samples includes~:
  \begin{itemize}
  \item
  {\bf i/} {\it For the  $\pi^+\pi^-$ annihilation  channel}~:

  The earliest
  data samples date back to the beginning of the eighties \cite{DM1,Barkov}. These have been
  followed by quite precise data samples collected in scan mode by
  CMD-2 \cite{CMD2-1995corr,CMD2-1998-1,CMD2-1998-2}   and SND \cite{SND-1998}  on
  the VEPP-2M  collider at Novosibirsk. The published spectra cover the   energy region from 370 MeV
  to 970  MeV; presently,  there is still no published spectrum covering the $\phi(1020)$ energy region.

  These measurements, referred to globally as NSK, have been followed by large statistics data collected
  using the so-called initial state radiation (ISR)  mechanism (see \cite{Benayoun_isr}, for instance).
  In this way, three data samples \cite{KLOE08,KLOE10,KLOE12} have been collected with
  the KLOE detector running on   the DAPHNE collider up to  $\simeq 1$ GeV, just below the $\phi$ meson mass.
   In the same period of time, another $\pi^+\pi^-$  sample
   has  been collected by BaBar \cite{BaBar, BaBar2} covering the energy range up to 3 GeV. By 2015,
   BESIII has also published a  $\pi^+\pi^-$ sample \cite{BESS-III} and, by end
   of 2017, a group from the CLEO-c Collaboration has published the latest to date $\pi^+\pi^-$  sample \cite{CESR}
   collected with the CESR detector.

   As a whole, the most precise data samples represent now 8 spectra. One has been led to discard
   from our global framework  KLOE08 \cite{KLOE08} and BaBar  \cite{BaBar,BaBar2}
   for different reasons\footnote{As for the BaBar spectrum, a recent comprehensive study \cite{Kubis2016}
   analyzing the $\omega \rightarrow \pi ^+\pi ^-$ branching ratio reaches conclusions similar to ours.
   One also has to keep in mind that the BaBar $\pi\pi$ spectrum is in conflict with the $\tau$ data corrected 
   for the isospin breakings. This is what we learn from Figure 1 of Ref. \cite{ExtMod5}.}
   discussed in \cite{ExtMod4,ExtMod5}. A reanalysis  unifying the three KLOE data
   samples has been recently performed \cite{KLOEcomb,Teubner3} and an 85 data-point spectrum has been
   derived which will be commented at the appropriate place. Moreover, the full error covariance
   matrix including the correlations between KLOE08, KLOE10 and KLOE12 is now available; this is used
   here  to account for the (weak) correlations between KLOE10 and KLOE12.
   \item
 {\bf ii/}  {\it For the  $K^+K^-$ annihilation  channel}~:

 	Up to very recently, the available data samples covering this channel were the three
	scans collected  by CMD-2   \cite{CMD2KKb-1,CMD2KKb-3} and the two scans of SND
	 \cite{SNDKKb}. They were included within the global fits performed in the (previous) BHLS framework
	 (see \cite{ExtMod3,ExtMod5}).  Quite recently, the CMD-3 Collaboration has found that the two CMD-2
	 scans  \cite{CMD2KKb-3} should undergo a rescaling by \cite{Mainz2018}  $1.094 \pm 0.04$ which represents
	 an important correction of the spectrum absolute normalization and of its (correlated)
	 systematics ($2.2\% \rightarrow 4.45\%$). The influence of this correction on the
	 physics information previously
	 derived within BHLS in \cite{ExtMod3,ExtMod4,ExtMod5} imposes the quite significant update
	   performed in the present study.

	BaBar has published in 2013 the first measurement \cite{BaBarKK} of this cross-section performed
	in the ISR mode. This high statistics spectrum covers the full energy range from
	threshold  to 3 GeV. Moreover,
	very recently, CMD-3 has also produced a new high statistics measurement of this
	spectrum   in scan mode \cite{CMD3_KpKm}. However, as emphasized  in the CMD-3 publication,
	this measurement is inconsistent with the BaBar data sample \cite{BaBarKK}. This issue
	is addressed  below.
    \item
 {\bf iii/}  {\it For the $K_LK_S$ annihilation  channel}~:

 	To our knowledge, the first reported measurement of the $e^+e^-\ra K_LK_S$ cross section has
	been provided by CMD-2 in 1995
	\cite{CMD2KKb-1} and has been followed in 2004 by the 4 scans reported in \cite{CMD2KKb-2}. In the meantime
	SND also produced 4 data samples, 2 in the charged decay mode of the $K_S$ meson, 2 in
	its neutral mode \cite{SNDKKb}. These were the data encompassed in our former studies. The present work
	also includes the CMD-3 data sample \cite{CMD3_K0K0b} recently published.

     \item
 {\bf iv/}  {\it For the $(\pi^0/\eta)\gamma$ annihilation  channels}~:

 		As for the $e^+e^-\ra \pi^0 \gamma$ cross section, in our energy range,  the
	 available data samples are scarce~: One sample has been provided by CMD-2  \cite{CMD2Pg2005}
	 and another one by SND \cite{sndPi0g2016} -- which supersedes the former
	\cite{sndPg2003} used in our previous studies\cite{ExtMod3,ExtMod4,ExtMod5};
	the older SND $\pi^0 \gamma$ spectrum given in \cite{sndPg2000} is also considered.

	Regarding the $e^+e^-\ra \eta \gamma$ cross section, the situation is more favorable as
	six data samples corresponding to different $\eta$ decay modes have been collected
	by CMD-2  \cite{CMD2Pg1999,CMD2Pg2001,CMD2Pg2005} and SND \cite{sndPg2000,sndPg2007}.
	Nevertheless, no newly collected data sample for this channel  has been reported.

 	  \item
  	{\bf v/} {\it For the  $\pi^+\pi^-\pi^0$ annihilation  channel}~:

	As no newly collected  sample covering the energy region up to the $\phi$
	mass has been reported, we are left with only
	the samples already examined in our previous analyses \cite{ExtMod3,ExtMod4,ExtMod5}.
	These  have been collected by CMD-2  \cite{CMD2KKb-1,CMD2KKb-2,CMD2-2006,CMD2-1998}
	and SND \cite{SND3pionLow,SND3pionHigh} and cover either the $\omg$ peak region
	or the $\phi$ peak region. We also consider the  3-pion spectrum given in \cite{ND3pion-1991} as it
	mostly
	deals with the region in between the $\omg$ and $\phi$ peaks, allowing for a valuable constraint
	on the $\rho^0$ meson physics background.

	However,  preliminary fits showing that the sample
	in  \cite{CMD2-2006} exhibits an average $\chi^2$ per data point much above 2 -- as also yielded by
	 \cite{Kubis_3pi} in a different theoretical context -- we have been led to discard it from our fits.
	 Overall, one is left with nine data samples, including an old sample from the former CMD
	 Collaboration.

 	  \item
  	{\bf vi/} {\it The pion and kaon form factors in the spacelike region}~:

	One of the motivations to develop a new breaking procedure for the HLS Model is to
	design it in such a way that the resulting broken model (BHLS$_2$) could encompass
	accurately the close spacelike region
	of the pion and kaon form factors.
	Indeed, a fair  account of $F_\pi(s)$ in a region which stretches on both sides of
	$s=0$ is the best way to
	ascertain the low energy behavior of the $e^+e^- \ra \pi^+\pi^-$  cross section
	 which provides an enhanced contribution to $(g-2)_\mu$.

	The NA7 Collaboration has measured the pion form factor \cite{NA7} in the spacelike range
	$Q^2 \in [-0.253,-0.015]$ GeV$^2$  with good statistics and a small
	reported normalization error (0.9\%); two years later, NA7 has also published
	a measurement of the charged kaon  form factor \cite{NA7_Kc} over the range
	$Q^2 \in [-0.0950,-0.0175]$ GeV$^2$ with also a small reported normalization uncertainty
	of 1\%. On the other hand, an experiment at Fermilab had previously published less
	precise spectra on the pion and kaon form  factors \cite{fermilab2,fermilab2_Kc}
	over resp. $Q^2 \in [-0.092, -0.039]$ GeV$^2$ and  $Q^2 \in [-0.1145,-0.0409]$ GeV$^2$.
	Unlike  former experiments\footnote{ Reference
	 \cite{Bebek_all} has collected  the results of several of them.}  at higher
	virtualities and, more modern experiments\footnote{Former references can be found
	in the quoted papers.}  \cite{Volmer,Tadevosyan,Blok,Huber,Desy1,Desy2},
	Refs.  \cite{NA7,fermilab2} -- and  \cite{NA7_Kc,fermilab2_Kc} as well -- report on
	results derived in
	resp. $\pi$ and $K$  scattering on atomic electrons and are, thus, model independent.

	As a matter of fact, one should also note that there is no $K^0$ form factor spectrum
	measurement in the spacelike region, which would be a real challenge; nevertheless,
	 measurements  of $<r_{K^0}^2>$ exist  \cite{Lai_NA48,Abouzaid_KTEV} which  report negative values.

\end{itemize}

	A common feature of the data samples we have to deal with is that their systematic	
	errors are quite generally reported as correlated; this implies that the corresponding
	error covariance matrices $W$ can be of large dimension (especially for the BaBar
	and KLOE samples). As the input to the $\chi^2$ expressions which feed our fitting code
	are just their inverses $W^{-1}$, the issue has been to ascertain the quality of these. For this
	purpose, we have chosen using the subroutines of the {\sc NAG} library. Numerical checks have 
	nevertheless been performed on the eigenvalues, the diagonal and non--diagonal elements of the products 
	$W \cdot W^{-1}$ to verify that numerical departures from unit matrix properties 
	were indeed negligible.

\section{Comments on the Model Parameters}
\label{comments}
\indent \indent Before going on, it is worthwhile commenting
on the fit parameters of the broken HLS Models.
Basically, the number of parameters of the original (unbroken)  HLS Model
 \cite{HLSOrigin,HLSRef,FKTUY}
to be fixed from data is small. As for its non-anomalous sector, besides $f_\pi$, these are
the universal vector coupling $g$ and the specific  HLS parameter
$a$, generally found to slightly depart from 2, the value expected
from standard VMD approaches. Its anomalous sector \cite{FKTUY} involves a
few more parameters already
displayed in the Lagrangian pieces presented in Section \ref{anomalous}~:
$c_1-c_2$, $c_3$ and $c_4$. There
are strong motivations \cite{HLSOrigin,HLSRef} to impose $c_3=c_4$.
So, when unbroken, the HLS model depends on only four unconstrained parameters.

\vspace{0.5cm}

There are two parameter categories introduced in our approach: the coefficients of
the subtraction polynomials which
supplement the corresponding loops and, on the other hand, the breaking parameters inherently
affecting the present Lagrangian -- and those defined in \cite{ExtMod3}.

Traditionally, as illustrated by the Gounaris-Sakurai formula \cite{Gounaris}, the coefficients of the
subtraction polynomials are reexpressed in terms of the resonance mass and width, which are determined by
a fit to the data. In a global approach like ours, the subtraction polynomials affect loops which
come {\it simultaneously}  in  the various amplitudes associated with the various annihilation processes
{\it simultaneously} under examination. In this case, as there is no reason to favor a specific channel
to fix the meaning and the value of parameters which come in all channels, the strategy adopted is to let the data, {\it as a whole},
 determine the subtraction parameter values via a {\it global} fit. Even if the parameter
content may look less intuitive, this approach -- also adopted in BHLS \cite{ExtMod3} --
looks the most motivated. This does not prevent to extract usual physics quantities,
relying on the fit results, as done since \cite{ExtMod2, ExtMod3}, for instance.
\begin{figure}[!phtb]
\begin{minipage}{0.50\textwidth}
{\includegraphics[width=\textwidth]{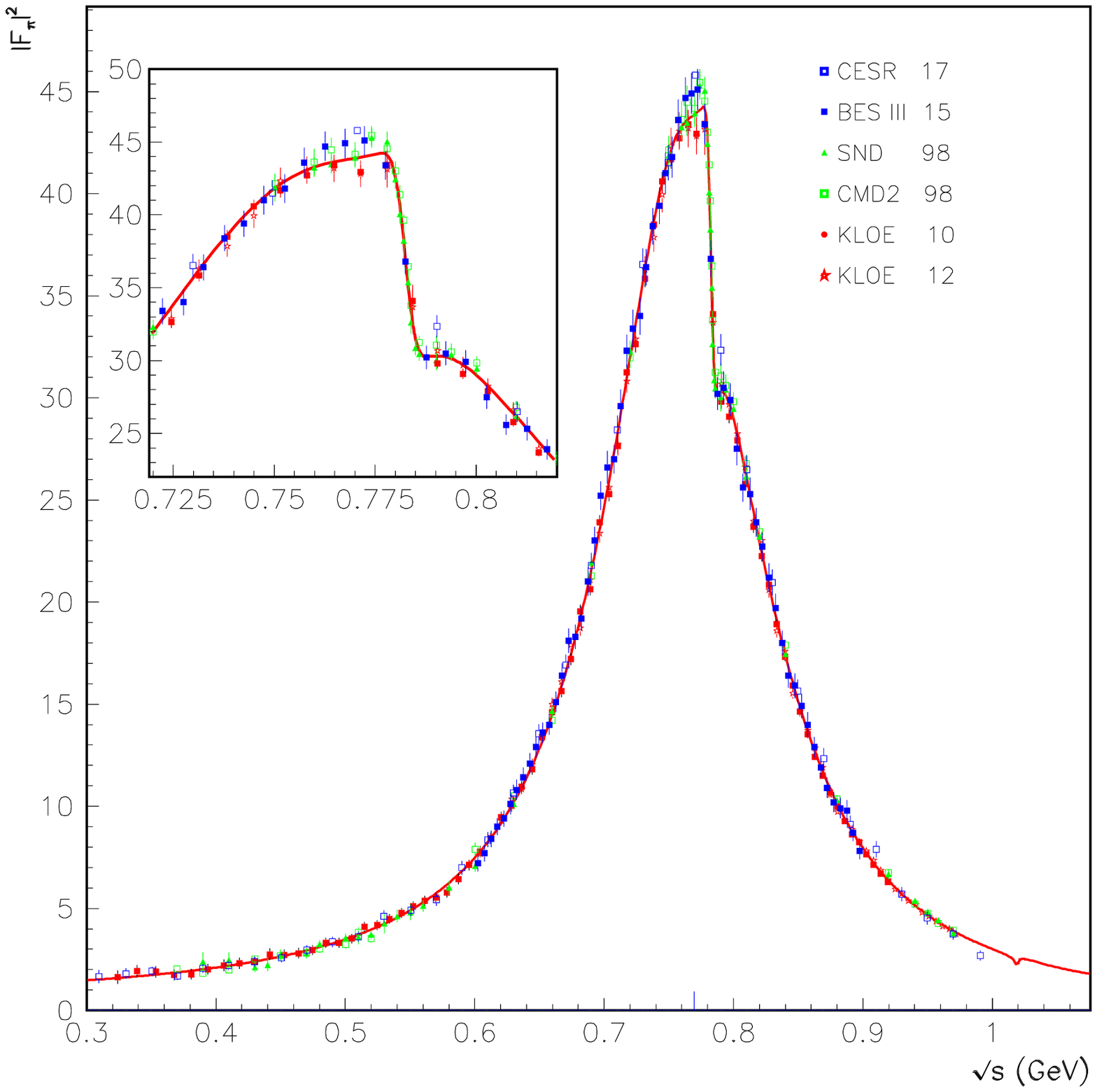}}
\end{minipage}
\hspace {0.01\textwidth}
\begin{minipage}{0.50\textwidth}
\includegraphics[width=\textwidth]{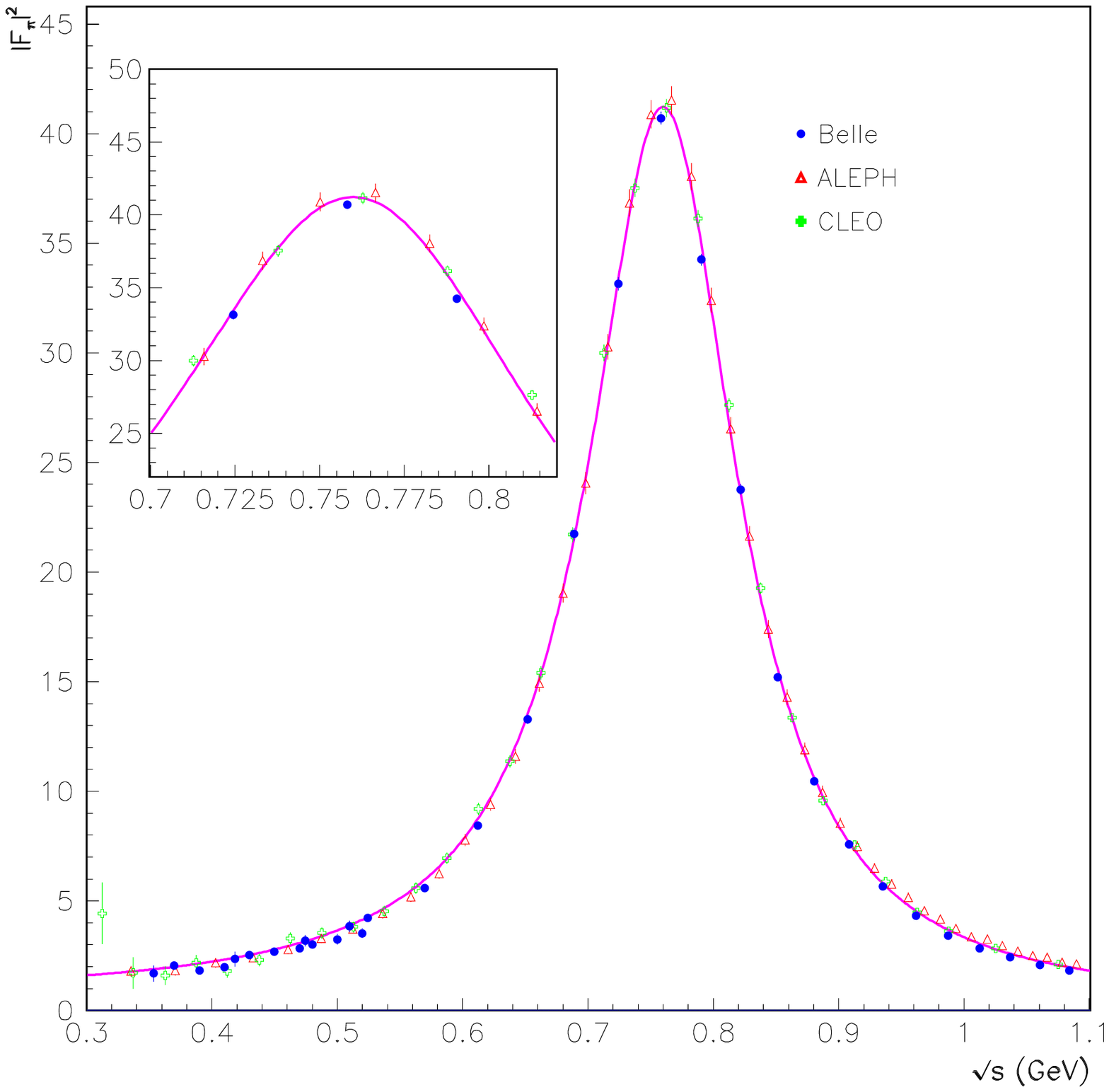}
\end{minipage}
\begin{center}
\vspace{-0.3cm}
\caption{\label{Fig:PIPI} BHLS$_2$ fit to the  $\pi \pi $ data, the RS solution~:
The leftmost panel shows the pion form factor squared in the
$e^+e^-$ annihilation and the rightmost one displays the same
spectrum in the $\tau$ decay. The fitted regions extend up to $s=1.0$ GeV$^2$.}
\end{center}
\end{figure}

In the second category of parameters, besides the parameters of the unbroken HLS model reminded
just above,
one finds all those introduced by the various aspects of the breaking procedure at work.
As for the parameters generated by the BKY mechanism and reminded in Section \ref{BKYbrk},
six in total, the condition $\Sigma_A=0$ is already stated since the original BHLS Model
\cite{ExtMod3}. It has been shown in Section \ref{BasicSol} that the
diagonalization procedure of BHLS$_2$ imposes $\Delta_V=0$. Therefore, four BKY parameters only
($\Sigma_V$, $\Delta_A$, $z_A$ and $z_V$) remain free.

In Sections \ref{BasicSol} and \ref{propagators}, it was also shown that
$\xi_0=\xi_8$ comes as a natural constraint on the parameters of the CD
breaking defined in Section \ref{Vbrk}. Therefore the CD breaking involves
two more parameters\footnote{Rigorously speaking, as already noted, two more breaking parameters
can be defined addressing the charged and neutral $K^*$ mesons.}
 $\xi_0(=\xi_8)$ and $\xi_3$. Finally, switching on the Primordial
Mixing  introduces three more parameters ($\psi_0$, $\psi_\omg$ and $\psi_\phi$) into the fit
procedure.

\section{The BHLS$_2$ Global Fits~: General Features}
\label{global_fits}
\indent \indent Various kinds of global fits have been performed using the data
samples and channels listed in Section \ref{DataSamples}, with or without the
spacelike data, with or without the $\tau$ data, fixing some of the model
parameters (the $\psi_\alpha$'s angles and $z_V$) or leaving all of them free.

When mixing as many data samples carrying important --  sometimes dominant --
overall normalization uncertainties, the effect of the so-called Peelle's
Pertinent Puzzle (PPP) \cite{Peelle} cannot be ignored to avoid getting  biased
estimates for physics quantities. The use of $\chi^2$ minimization methods,
 first questioned, has been finally justified  \cite{Fruehwirth,Chiba}; however,
when dealing with distributions or form factors, the pertinent method turns out
to invoke the  underlying (theoretical) model function which is just what
fits are supposed to provide. Following \cite{Ball},  \cite{ExtMod5}
has defined an iterative procedure proved to
cancel out biases (see the Appendix in \cite{ExtMod5}).
\begin{figure}[!phtb]
\vspace{-1.0cm}
\begin{center}
\begin{minipage}{0.80\textwidth}
{\includegraphics[width=\textwidth]{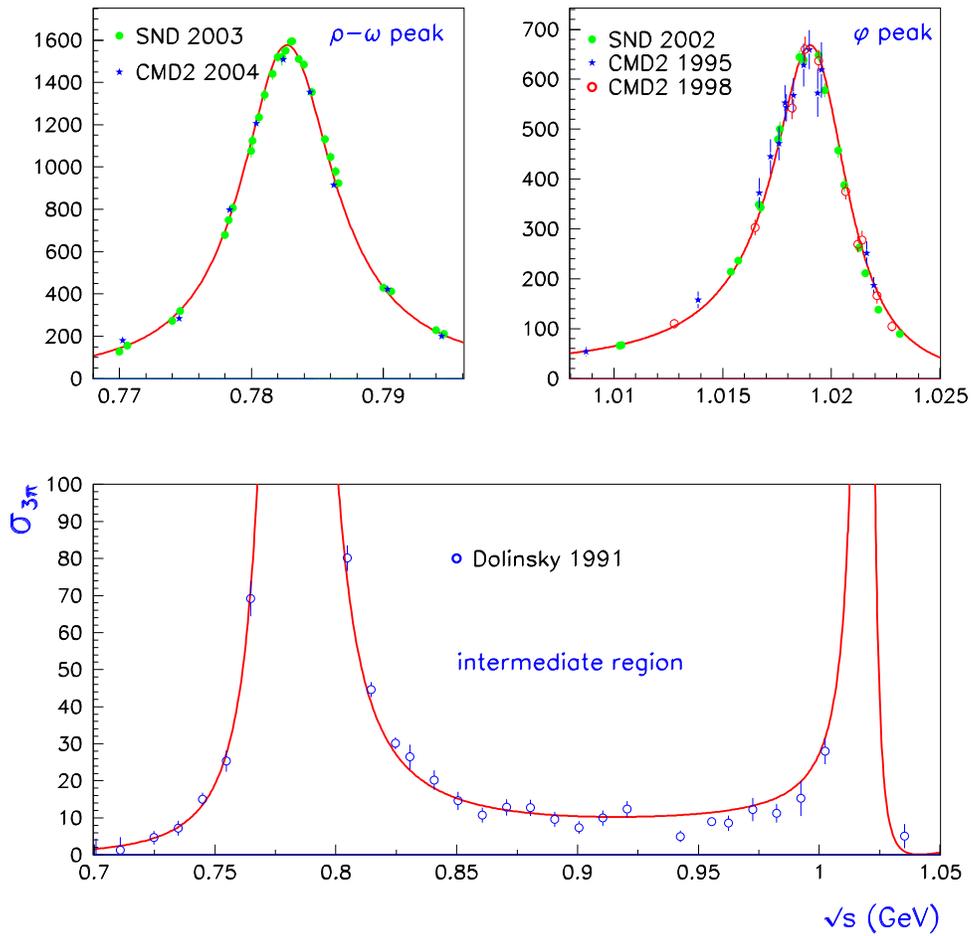}}
\end{minipage}
\caption{\label{Fig:PIPIPI} BHLS$_2$ fit to the  $\pi^+ \pi^- \pi^0$  spectra, the RS solution~:
Top panels show the fit and data in, resp. the $\omg$ and $\phi$ regions, the bottom
one focuses on the intermediate region.}
\end{center}
\end{figure}

It is worthwhile briefly sketching the conclusions of\footnote{See, in particular,
its Section 4. }  \cite{ExtMod5}  concerning
the issue just reminded~: {\bf 1/} The normalization uncertainty (denoted $\sigma$,
possibly $s$-dependent) of any spectrum $T$ is absorbed within the total covariance
matrix contributing to the general $\chi^2$,
{\bf 2/} The correction which affects the normalization  of $T$ is
{\it derived} from the data, the theoretical function provided by the fit, $\sigma$
and the {\it statistical} covariance matrix -- which may absorb the uncorrelated systematics.

Let us make a few general statements about fit results and data before focusing on specific
 topics, mainly the $K \overline{K}$ and $\tau$ sectors and the spacelike region behavior of the
 pion and kaon form factors. The detailed properties will be emphasized in the next Sections.

\begin{figure}[!phtb]
\vspace{-1.0cm}
\begin{center}
\begin{minipage}{0.80\textwidth}
{\includegraphics[width=\textwidth]{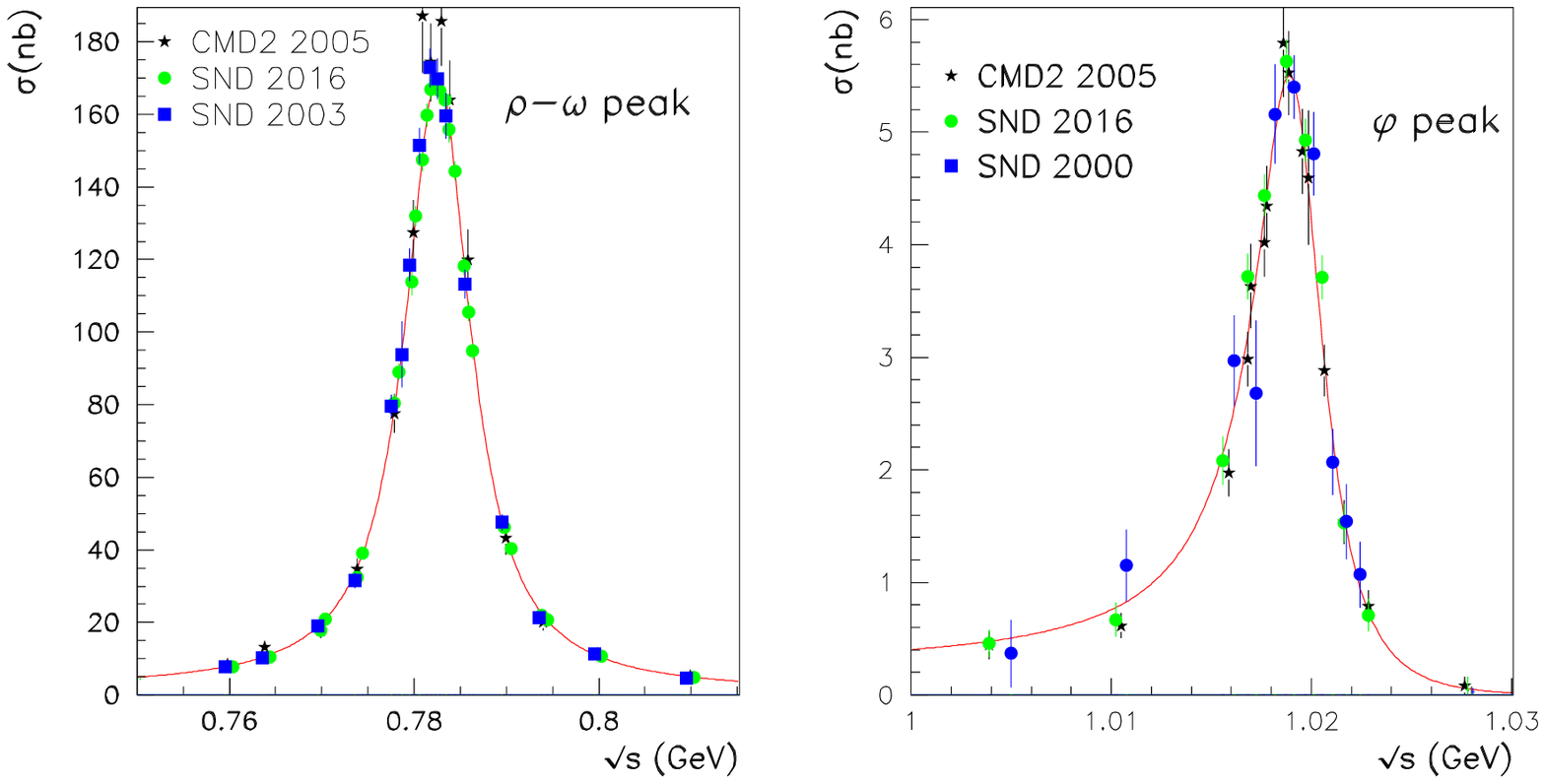}}
\end{minipage}
\begin{minipage}{0.80\textwidth}
\vspace{-1.5cm}
{\includegraphics[width=\textwidth]{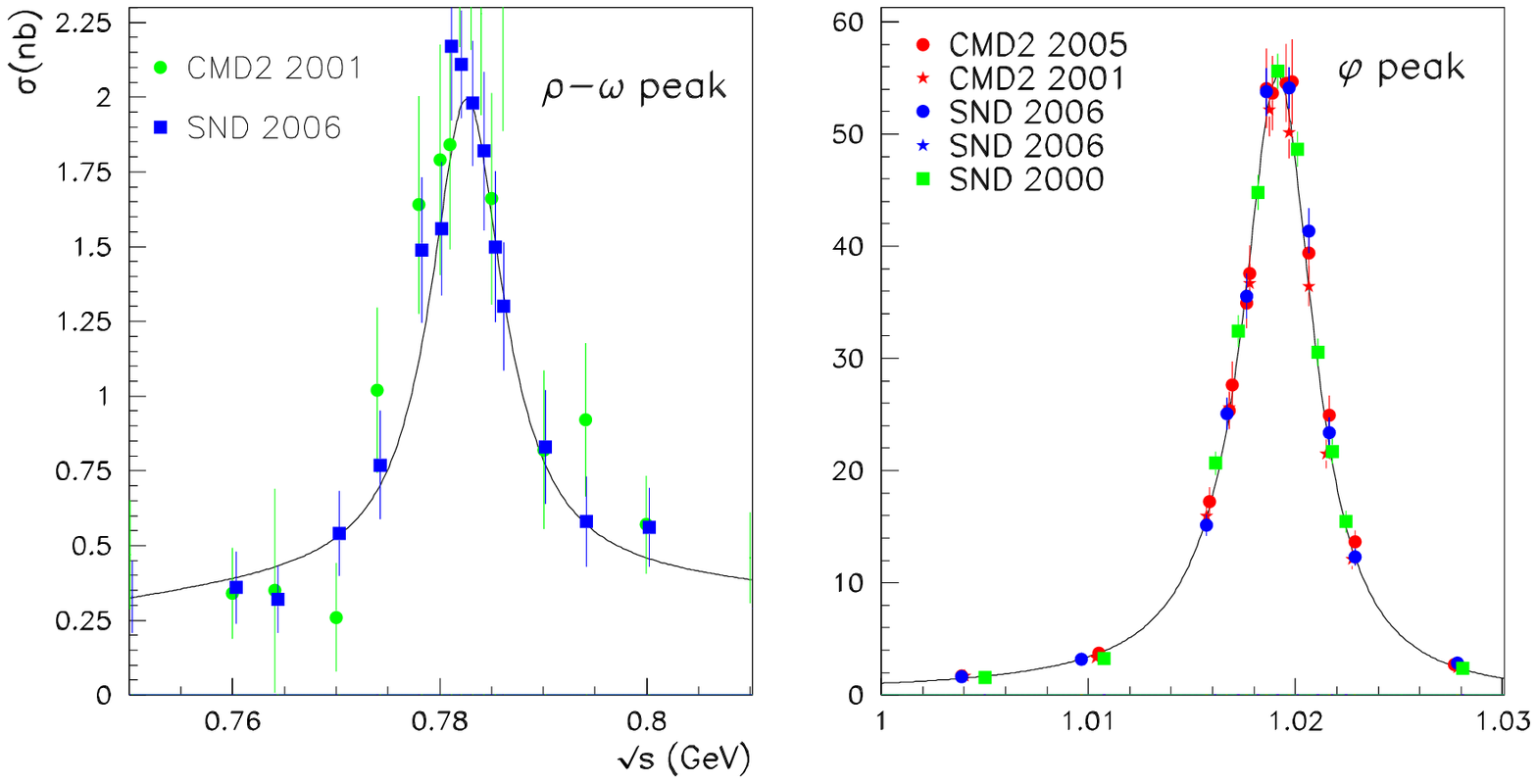}}
\end{minipage}
\vspace{-0.5cm}
\caption{\label{Fig:PSG} BHLS$_2$ fit to the  $(\pi^0/\eta) \gam$  spectra, the RS solution~:
Top panels show the case for the $\pi^0 \gam$ spectra in the $\omg$ and $\phi$
regions, the bottom panels display the corresponding plots for the $\eta \gam$ spectra.
}
\end{center}
\end{figure}

 \begin{itemize}
 \item
 As for the behavior of the data samples within our global fits, one can state
 that the issues met with the dipion spectra from BaBar \cite{BaBar,BaBar2}
 and KLOE08  \cite{KLOE08} within the
 BHLS framework \cite{ExtMod3} are confirmed within the BS and RS variants of
 the BHLS$_2$ framework studied here; they are also discarded in the
 present study.

Whatever the specific conditions of the various fits performed,
the  $\pi^+\pi^-$  data samples from NSK ({\it i.e.} CMD-2 and SND), KLOE10/12,
 BESIII  \cite{BESS-III} and  CLEO-c \cite{CESR} are as well described within the
 BHLS$_2$ framework as they were already within BHLS  \cite{ExtMod3}; more precisely,
 NSK and KLOE10/12 contribute an average $\chi^2$ per data point of $\simeq 1$, while
 the BESIII and CLEO-c samples yield $\chi^2/N_{\rm pts} \simeq 0.6$.
 Figure \ref{Fig:PIPI} displays the
spectra and fits for  the pion form factor in the $e^+e^-$ annihilation and
in the $\tau$ decay.

\item
In order to account for the difficulty to fully address the
$\pi^+ \pi^- \pi^0$ channel in the $\phi$ mass region within
BHLS, a so-called  B model was defined which simply turns out to discard
from fit this mass region for (only) this channel \cite{ExtMod3}.
In contrast, BHLS$_2$  does not meet any issue\footnote{As for
 the  discarded data sample \cite{CMD2-2006}, see the above Section \ref{DataSamples}
and also  \cite{Kubis_3pi}.}  and achieves
a fair description of the $\pi^+ \pi^- \pi^0$
 data up to, and including, the $\phi$ mass region.
  The spectra are shown in Figure \ref{Fig:PIPIPI}.
\item
Finally, the $e^+e^- \ra (\pi^0/\eta) \gam$ spectra are also nicely fitted within BHLS$_2$ -- as well as
in the former BHLS -- as illustrated by Figure \ref{Fig:PSG}. Quite generally,
one obtains $\chi^2/N_{\rm pts} \simeq 90/112$ and $\chi^2/N_{\rm pts} \simeq 120/182$ for resp.
the $ \pi^0 \gam$ and $\eta \gam$ channels.
 \end{itemize}

In order to perform global fits, one obviously should make an assumption on the energy
calibration of each of the data samples considered. For the dipion spectra, one
has gathered into the fit data coming from
the VEPP-2M machine at Novosibirsk (CMD-2 and SND), data from DAPHNE (KLOE10 and KLOE12),
from BESIII and from CESR.
A significant mismatch of the relative energy calibrations of these would have produced
 a failure of their common fit because of the
 $\rho-\omg$ drop off region; this is clearly not seen neither in the $\chi^2$
contributions of the various samples, nor in  Figure \ref{Fig:PIPI} (see the inset in
its leftmost panel). Moreover, comparing both panels in this Figure excludes a significant
energy calibration  mismatch between $e^+ e^-$ and $\tau$ data.

Moreover, the data collected by CMD-2 and SND around the $\omg$ and $\phi$ peaks
in the 3-pion channel as well as in the $(\pi^0/\eta) \gam$ final states indicate a
good consistency
 between the energy calibrations of the various data samples collected
 in different  runs distributed over several years. This is visible
 in the fit properties as well as in the
 Figures where the fit function and the various data are displayed.
\section{The BHLS$_2$ Global Fits~: The $e^+e^- \ra K \overline{K}$ Channels}
\label{kkb_data}
\indent \indent The available data covering the  $ K^+K^-$
 channel\footnote{In the following,
as already noted in Section \ref{DataSamples}, the two CMD-2 scans \cite{CMD2KKb-3}
are corrected as  stated in \cite{Mainz2018}.
As for the BHLS predictions in \cite{ExtMod3,ExtMod4,ExtMod5},
they are updated correspondingly.}
come from SND, CMD-2 and BaBar. Recently CMD-3 has published a new $ K^+K^-$ measurement
 \cite{CMD3_KpKm}.
As for the neutral channel, most data come from CMD-2 and SND and have been listed in
Section \ref{DataSamples}. A new data sample covering our fitting range has also been
produced by CMD-3 \cite{CMD3_K0K0b}; unfortunately, the published $K_L K_S$ spectrum from BaBar
starts at 1.06 GeV, just above our fitting range \cite{BaBarklks}.

\subsection{Preliminary Remarks}
\indent \indent
As for the $K^+K^-$ channel~:  In its \cite{CMD3_KpKm}, the  CMD-3 Collaboration points
toward disagreements with both the BaBar \cite{BaBarKK} and the original CMD-2 data samples
\cite{CMD2KKb-3}.  Once the correction factor ($1.094 \pm 0.040$) for the trigger
efficiency of CMD-2 is applied, the discrepancy between  the CMD-3 and CMD-2 $K^+K^-$
samples looks  much reduced \cite{Mainz2018} -- possibly washed out.
Anyway, a full examination of
the various $e^+e^- \ra K \overline{K}$ data samples\footnote{As for the contribution
to the muon HVP 
of the energy region from $s \in [4m_K^2, 1.06$ GeV$^2$] to $a_\mu(K^+K^-)$, BaBar
obtains $18.64 \pm0.16_{\rm stat} \pm 0.13_{\rm syst} \pm 0.03_{VP}$ while CMD-3 gets
$19.33 \pm 0.040$ in units of 10$^{-10}$, a difference of $\simeq 0.7 \times10^{-10}$
between  the central values, a $3.25 \sigma$ effect.} is worth   addressing.

On the other hand, preliminary BHLS$_2$ analyses based on the CMD-3 and BaBar data,
made us aware of a visible
degradation of  both  $K \overline{K}$ channel data description above
$1.025 \div 1.030$  GeV;
therefore, in the present work, one limits the fit region for the  $K \overline{K}$ data
to the $\sqrt{s} \in [2m_K, 1.025$ GeV] range. The explanation for this effect
is unclear. It could be
an intrinsic weakness of the BHLS/BHLS$_2$ models\footnote{The standard
HLS framework \cite{HLSRef}, which underlies the BHLS/BHLS$_2$ models, does not account
for vector particles  not belonging  to the fundamental vector nonet.} which, as it stands,
does address  a possible  onset of the high mass vector meson contributions; it could
also reflect an experimental issue. For the sake of consistency, we have chosen to impose
the same fitting range also to the  former CMD-2
 and SND  data samples in both the charged and neutral modes,

\subsection{A Specific Issue of the $K \overline{K}$ Data}
 \indent \indent As already pointed out, one does not detect any indication for a mismatch
 in the energy calibration of the various data samples covering
 the 3-pion and $(\pi^0/\eta)\gam$ channels, all collected with the CMD-2 and SND
 detectors on the same VEPP-2M collider. One also does not  observe any mismatch
 between the energy calibration of the 3-pion and $(\pi^0/\eta)\gam$ data and those
 collected by the same collaborations in the $K \overline{K}$ channels.
 Nevertheless\footnote{E. Solodov, Budker Institute, Novosibirsk, private communication.},
 CMD-2 and SND  carry a (correlated) absolute energy uncertainty in the range
of $30\div 40$ keV, while for
CMD-3, the  absolute energy uncertainty is $\simeq 60$ keV, statistically
independent of that of CMD-2 \& SND. Moreover, BaBar also reports
a 50 keV correlated energy scale uncertainty\cite{BaBarKK}.
When dealing with an object as narrow as the $\phi$ meson within a global framework,
these uncertainties should be
accounted for otherwise, the fits cannot succeed.

As the bulk of data samples for both $K\overline{K}$ channels has been provided by CMD-2
and SND, it looks quite natural to choose their common energy
calibration as a reference; this will
be denoted  $E_{\rm NSK}$ in the following. For this purpose, one
defines  $E_{\rm CMD3}=E_{\rm NSK}+\delta E_{\rm CMD3}$ and $E_{\rm
BaBar}=E_{\rm NSK}+\delta E_{\rm BaBar}$ to
recalibrate the CMD-3 and BaBar data point energies in a fully correlated way and derive
their $E_{\rm NSK}$ equivalent energy.

Therefore, in order to deal with the  $K \overline{K}$ data, we are faced with
a case where two kinds of correlated uncertainties have to be accounted for: Energy
calibration and absolute normalization of the cross-sections. Hence,  one merges
in the minimization procedure the method developed in \cite{ExtMod5} and the
two calibration parameters $\delta E_{\rm CMD3}$ and $\delta E_{\rm BaBar}$ to be also determined
 within the same fits.

 For this purpose, one assumes that there is no {\it functional}  relation
 between   the normalization uncertainty $\lambda_i$ -- of an experiment labeled $i$ --
 and the energy scale uncertainty  $\delta E_i$; then,
  as $\partial \lambda_i/\partial [\delta E_i] =0$ for each experiment,
 substituting, as emphasized in \cite{ExtMod5}, the value  for $\lambda_i$ derived from
 $\partial \chi^2 /\partial\lambda_i =0$ into the expression for $\chi^2$ is unchanged
 compared to  \cite{ExtMod5}.

\subsection{Fitting the $K \overline{K}$ Channels~: NSK, BaBar and CMD-3 Separately}
\indent \indent
In the following, the non-kaon sector data and channels used in our fits
include, unless otherwise stated, the $\pi^+\pi^-$, $\pi^+\pi^-\pi^0$, $\pi^0 \gam$, $\eta\gam$ annihilations,
the $\tau$ decay to $\pi^\pm \pi^0$ and a few decay partial widths (see \cite{ExtMod3}).
One starts by discussing the  fits where the CMD-2 \& SND (NSK),
BaBar and CMD-3 kaon data are not influencing each other within the minimization procedure.
Table \ref{Table:T1} gathers most of our results in this case; these are
commented right now.

\begin{itemize}
\item
With regard to  NSK~:  $<\chi^2>$, the average (partial) $\chi^2$ per NSK data points, is of order 1
in both the neutral and charged kaon channels; thanks to the fair description of the
other channels involved in the fit, the global fit probability exceeds the 90\% level.

As reminded above, the procedure defined in \cite{ExtMod5} allows us to derive from
the fit procedure  the normalization (absolute scale) correction to be applied to each
spectrum\footnote{One should stress that such a "renormalization" is the
natural outcome to the scale uncertainty problem and that the main advantage of
any kind of global fit is to allow determining such a correction in consistency
with a (large) set of data samples coming from different sources.}.
For the charged channel, one gets +5.0\% for SND (expect. value 7.1\%), a $0.7
\sigma$ effect, and  --1.6\% for CMD-2 (updated expectation 4.4\%),
a $0.3\sigma$ effect; in the neutral channels, one correspondingly
obtains --1.8\%  (expect. value 4.2\%) for SND and +0.6\% (expect. 1.7\%) for CMD-2, resp.
$0.4 \sigma$ and $0.3 \sigma$ effects.

Therefore the CMD-2 and SND data samples successfully fit  the BHLS$_2$ framework and,
moreover, the derived scale corrections meet the reported scale uncertainty expectations.

 \item
With regard to  BaBar~: One has chosen to supply the NSK neutral data to the fits
in order to populate each dikaon channel as the BaBar $K_L K_S$ data \cite{BaBarklks}
do not cover our fitting energy range.
As the BaBar charged spectrum \cite{BaBarKK} has been
collected in the ISR mode, it looks appropriate to compare, within the fit,
each spectrum datum in a bin
to an average value of the model spectrum over the bin width.
The second data column in Table \ref{Table:T1} shows
that the BaBar sample yields $<\chi^2>=1.7$, somewhat large, and a global fit probability also
above  the 90\% level.

The minimization procedure returns as energy shift vs NSK $\delta
E_{\rm BaBar}=-125.8\pm 19.0$
keV; this looks fair when compared to the independent $30\div 40$ keV (NSK) and 50 keV (BaBar)
energy calibration uncertainties. Moreover, the spectrum scale correction
is at the $10^{-4}$ level -- so that the absolute scale of the BaBar spectrum has not to
be corrected. On the other hand, the scale corrections for the
neutral NSK data included in the fit  are --2.1\% (SND) and +0.7\% (CMD-2), almost identical to
what has been obtained in the fit of the NSK data in isolation mode.

 Interestingly
one can also derive the scale correction for the charged data of
SND (+5.2\%) and CMD-2 (--3.7 \%) even if they are not submitted to the fit; these predictions
are clearly in the expected  ballpark. This indicates, beforehand, that NSK and BaBar data
in the kaon sectors are  consistent with each other.

\item
As for CMD-3~: This Collaboration has provided data samples in both dikaon final states,
When fitting these data, the correspondence is made -- as with NSK --
between each datum and the theoretical model computed at the nominal energy. As for NSK and
BaBar,
the systematic uncertainties are treated as fully correlated and serve to build up the
systematic  covariance matrices to be added up  to the statistical ones; these are
used in the fitting of the neutral and charged CMD-3 data within the global BHLS$_2$ framework.
The fit results are displayed in Table \ref{Table:T1}.

The  $<\chi^2>$ for both CMD-3 data samples are clearly very large and the global fit
probability poor. The scale corrections found are $+1.2$\% (charged channel)
 and $-1.2$\%  (neutral channel) to be compared to the reported r.m.s., resp. 2.0\%
 and 1.8\%. The possibility of treating the bin information as in ISR
 experiments\footnote{Comparing the datum information to the fit function averaged over
 the bin width, as just sketched  for BaBar.} has been examined -- specifically, only in the neutral
 channel -- and in  view of the outcome, we gave up.

 For the sake of completeness, we have nevertheless also performed the fit of the CMD-3 data in isolation
 by discarding the {\it non-diagonal part}  of the total covariance
 matrix\footnote{This turns out to consider that the systematic uncertainties are
 fully uncorrelated; some configurations intermediate between full correlation
 and no correlation can lead to similar results.}, as seemingly done in
 \cite{BaBarKK} for the BaBar data; the fit results change dramatically as one reaches
 $<\chi^2>=14/16$ and $<\chi^2>=17/17$ for resp. the charged and neutral modes
 of the CMD-3 data and  with a fit probability exceeding 90\%.
 Moreover, in this case (CMD-3 data in isolation) the corrections
 to the original  normalization of the cross-sections become negligible ($\simeq 10^{-4}$)
 for both the charged and neutral channels\footnote{The $<\chi^2>$'s  ~of the {\it fit prediction}
 to NSK data remain in reasonable correspondence with the NSK fit in isolation (55/49 and 105/92)
 while the distance of this prediction to BaBar data degrades more severely (76/27).}.
 As for the energy shift vs NSK it moves from $-51.4 \pm 13.7$ keV (see Table \ref{Table:T1})
 to  $-44.6 \pm 14.8$ keV, still in the expected ballpark.
   \end{itemize}
\begin{table}[!phtb!]
\begin{center}
\begin{tabular}{|| c  || c  | c | c ||}
\hline
\hline
\hhhd   $\chi^2$ ($K\overline{K}$ Sample Set) & \hhhv NSK  &  \hhhv BaBar &  \hhhv CMD3  \\
\hline
\hline
 \hhhv NSK $K_L K_S$ (92) &      $107 $   &  $100$          & $-$       	  \\
\hline
 \hhhv NSK $K^+ K^- $(49) &      $51$     &  $-$	    & $-$	          \\
\hline
 \hhhv BaBar $K^+ K^- $ (27) &   $-$      &  $46$      	    & $-$	                 \\
\hline
 \hhhv CMD3 $K_L K_S$ (17) &     $- $     &  $-$            & $120$     		  \\
\hline
 \hhhv CMD3 $K^+ K^- $(16) &     $-$      &  $-$	    & $88$			 \\
\hline
 \hhhv $\chi^2/N_{\rm pts}$ & $1087/1210$  &  $1078/1188$    & 1155/1102 	         \\
 \hhhw Probability      &     97.5\%      &  95.2 \%        & 3.8 \%    	         \\
\hline
\hline
 \hhhv $\delta E$ (keV) &     $-$         &  $-125.8\pm 19.0$   & $-51.4\pm 13.7$		\\
\hline
\hline
\end{tabular}
\end{center}\vspace{-0.3cm}
\caption {
BHLS$_2$ fit properties in the kaon sectors.  The kaon data from NSK, BaBar and CMD-3
are fitted in isolation. Running BaBar data "alone" is complemented by the  NSK data for the
$K_L K_S$ channel. The data and channels other than kaon are identical in all cases;
the  full  $\chi^2$ value and the total number of data points ($N_{\rm
pts}$) refer to the global fit.
The last line displays the shift values $\delta E_{\rm CMD3}$ or
$\delta E_{\rm BaBar}$ relative
to the NSK data energy calibration.
\label{Table:T1}
}
\end{table}

 \subsection{Fitting the $K \overline{K}$ Channels~: NSK, BaBar and CMD-3 Combined}
\indent \indent This Subsection is devoted to analyzing the consistency of the
three available groups of data. Quite generally, the covariance matrices used within the fit
procedure assume full correlation for the systematics, as in the previous paragraph.
Nevertheless, one  also briefly reports on additional information about a fit performed assuming
the systematics uncorrelated for CMD-3 data.
It is worthwhile proceeding with pairwise combinations.
\begin{itemize}
\item As clear from the first data column in Table \ref{Table:T2} -- and as could be
inferred from the results already reported --  one observes a fair consistency between
NSK and BaBar samples within the global BHLS$_2$ framework. Comparing the present NSK and BaBar
contributions to the (total) $\chi^2$ with their analogs  in their  fits in isolation clearly shows
that there is no significant tension between them. The only noticeable difference is
the central value for $\delta E_{\rm BaBar}$ found closer to 100 keV by $1\sigma$.

Other pieces of information concerning the absolute spectrum normalization\footnote{
For SND and CMD-2, as their corrections are computed run by run and sample by sample,
here we only give average values, for simplicity.}
are worth to be mentioned. The fit outcome indicates that
 the BaBar data normalization should be downscaled by $\simeq 0.4 \%$,
 the  (updated) CMD-2 charged spectrum by 1\% while the SND charged data normalization
 should be increased by 8.1\%. These corrections are always  consistent
with the respective expectations reminded above.
 \begin{table}[ph!]
\hspace{0.0cm}
\begin{tabular}{|| c  ||  c| c | c|| c||}
\hline
\hline
\hhhd   $K\overline{K}$ Sample Set  &  \hhhv NSK+BaBar &  \hhhv NSK+CMD3 &  \hhhv BaBar+CMD3 &\hhhv NSK  +BaBar   \\
\hline
\hhhd    (\# data points)           &  ~~              &  (CMD3 corr.)        & (CMD3 corr.)             & \hhhv + CMD3 uncorr. \\
\hline
\hline
 \hhhv NSK $K_L K_S$ (92)           & $106$	  & $105$  	& $(108)$  & $113$	\\
\hline
 \hhhv NSK $K^+ K^- $(49)           & $50$	  & $52$	& $(50)$   & $50$	\\
\hline
 \hhhv BaBar $K^+ K^- $ (27)        & $46$	  &$-$		&$45$	   & $56$		\\
\hline
 \hhhv CMD3 $K_L K_S$ (17)          &$-$	  &$121$	&$122$	   &$18$		\\
\hline
 \hhhv CMD3 $K^+ K^- $(16)          & $-$	  &$98$		&$97$  	   &$23$\\
\hline
\hline
 \hhhv $\chi^2/N_{\rm pts}$          & 1133/1237     &$1331/1243$  &$1212/1129$	&$1193/1270$\\
 \hhhw Probability                & 93.3\%	    &  1.1 \%     &  0.8 \%	&  81.5 \%\\
\hline
\hline
 \hhhv $\delta E_{\rm CMD3}$ (keV)       	& $-$  &$-46.5\pm 6.7$ &$-45.9\pm 3.9$    &$-38.2\pm 11.7$\\[-0.2cm]
 \hhhv $\delta E_{\rm BaBar}$ (keV) & $-105.8\pm 17.2$  &$-$	       &$-105.5\pm 16.1$  &$-103.2\pm 16.7$\\
\hline
\hline
\end{tabular}
\caption {BHLS$_2$ fit properties in the kaon sectors.
The kaon data from NSK, BaBar and CMD-3 are fitted pairwise combined.
The $\chi^2$ values are displayed for individual
channels or for the total number of data points $N_{\rm pts}$. In the two
rightmost data columns, the top  entry  is  $\delta E_{\rm CMD3}$ and
the bottom one  $\delta E_{\rm BaBar}$. The rightmost column displays the fit results when
treating the CMD-3 systematics as uncorrelated.
\label{Table:T2}
}
\end{table}
\item The second data column in Table  \ref{Table:T2} displays results from fits
combining NSK and CMD-3. Here also one observes a striking resemblance between the individual
partial $\chi^2$'s in the combined fit and in their fits in isolation, {\it i.e.} 
the NSK
data samples go on yielding  $<\chi^2>_{{\rm NSK}} \simeq 1 $
while  one gets $<\chi^2>_{{\rm CMD3}} > 6 $; the global fit
probability is also poor.

Besides the energy shift $ \delta E_{\rm CMD3}=-46.5\pm 6.7$ keV, the spectrum rescalings derived from
the fit are interesting pieces of information. In this way,  the fit returns an increase of the
charged CMD-3 data normalization by only 0.2\% while the neutral CMD-3 ones should be downscaled by 2.6\% (a $1.5 \sigma$ effect).
One also finds that the SND charged data normalization should undergo an increase by 9.2\%,
while for CMD-2  the increase is only 1.5\%.
As for the neutral kaon normalizations, SND should be downscaled by 2.1\%,
while CMD-2 is increased by 0.5\%. So, one does not observe major changes in the NSK normalizations.

\item The third data column in Table  \ref{Table:T2} displays results from fits
combining BaBar and CMD-3 kaon data, excluding their NSK analogs. The global fit probability
clearly indicates that such a combination is not really favored; one should remark, nevertheless,
that all properties exhibited by CMD-2 and SND are in fair correspondence with their
analogs yielded in their fits in isolation; additionally, even the $\chi^2$ distances
of the CMD-3+BaBar fit function to the NSK data (not under fit) -- given within parentheses --
are analog to their values derived in  NSK alone or NSK+BaBar fits.

\begin{figure}
\centering
\includegraphics[width=0.495\textwidth]{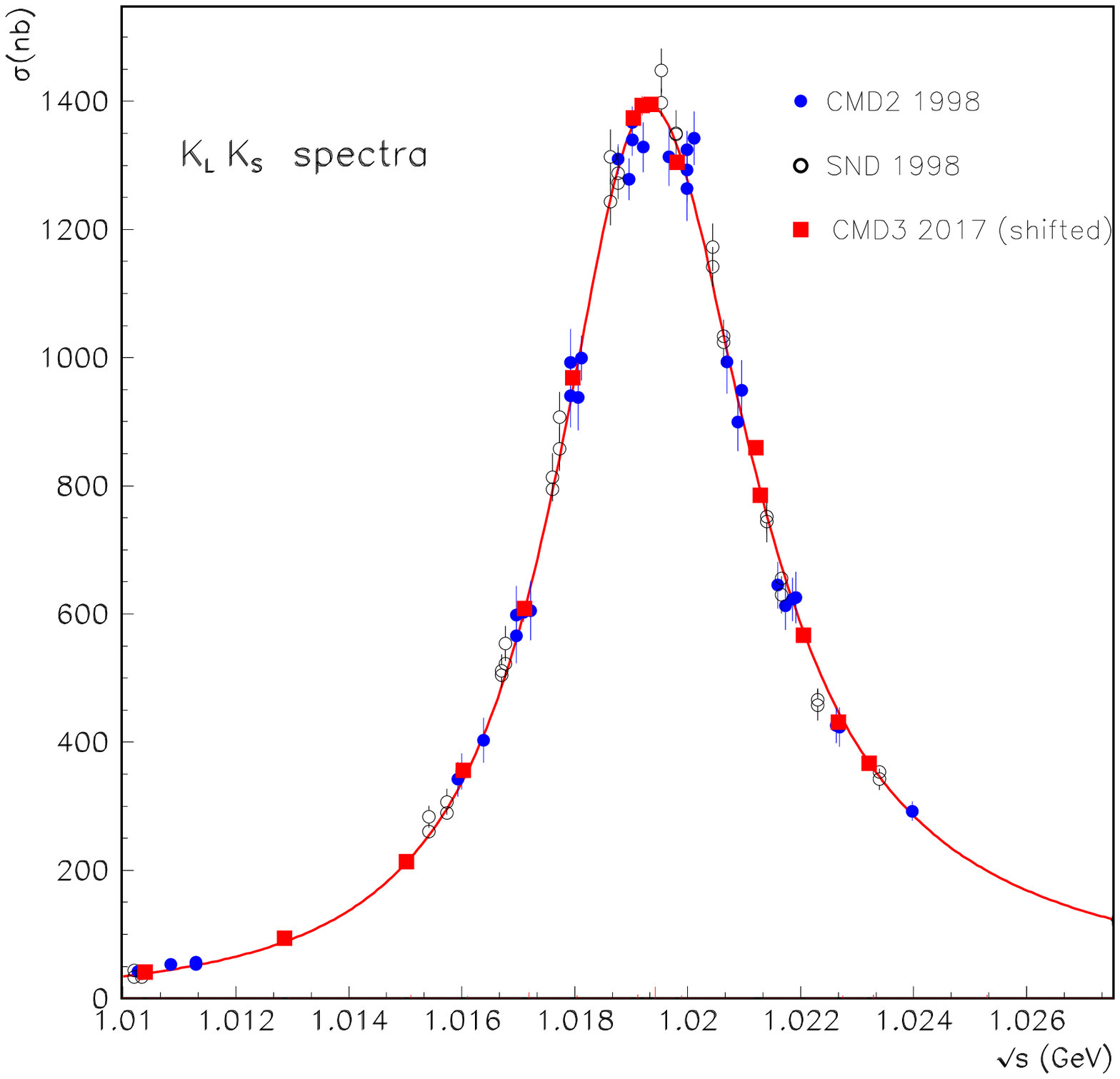}
\includegraphics[width=0.495\textwidth]{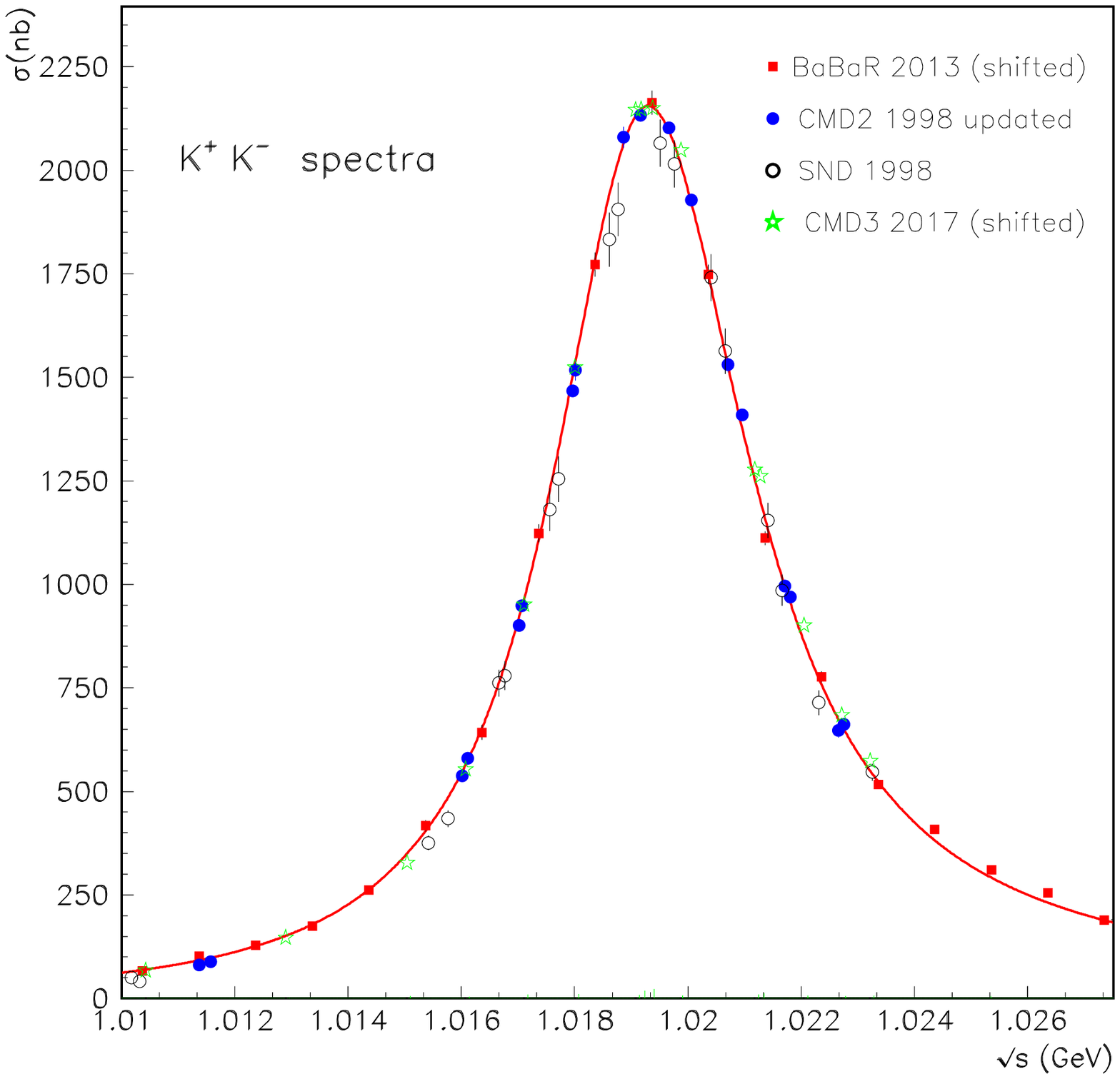}
\caption{\label{Fig:KLKS} BHLS$_2$ global fit to the $K^+K^-$ and $K_L K_S$ spectra
(RS solution)~: The plotted CMD-3 data are {\it not}  fitted
but are shifted by a  $\delta E_{\rm CMD3}$ determined elsewhere. The value for
$\delta E_{\rm BaBar}$ is derived from this fit. All experimental spectra are
rescaled as emphasized in the text. }
\end{figure}

Quite interestingly, the tension observed between CMD-3 on the one hand, and BaBar and NSK on the other hand,
is not manifest in plots, as soon as the energy calibration of CMD-3 vs NSK is applied. For instance
Figure (\ref{Fig:KLKS}), shows the $K\overline{K}$ cross sections
derived from the fit
to NSK + BaBar; the corresponding data are rescaled and appropriately shifted as emphasized
in the previous item; one has {\it superimposed} the (unfitted) data from CMD-3 shifted by $-45.9$ keV
and rescaled as {\it predicted} by the  NSK + BaBar fit~: $-4.3$\% and -2.4\% for resp. the
charged and neutral CMD-3 data samples.

On the other hand, one has found worth to also perform the fit sketched
in the rightmost data column in Table \ref{Table:T2},  {\it i.e.} treating the NSK and BaBar
uncertainties as fully correlated and both CMD-3 data samples as carrying uncorrelated
uncertainties. Using  the corrections to the absolute normalization
of each sample as derived from this fit\footnote{
This amounts to the following rescalings w.r.t. the original normalizations~:
--1.3\% (CMD-3 $K^+K^-$), --0.5\% (CMD-3 $K_L K_S$), +8.1\%  and --0.3\% ($K^+K^-$ from
resp. SND \& CMD-2), +0.4\%  and --0.5\% ($K_L K_S$ from resp. SND \& CMD-2) and
$\simeq -0.3 \% $ for BaBar $K^+K^-$.}, one gets a figure which
 superimposes exactly to  Figure
(\ref{Fig:KLKS}), in the smallest detail.

This clearly indicates that  the observed tension  is essentially due to
the non-diagonal parts of
the CMD-3 error covariance matrices constructed from experimental information,
as also done for the other data samples, CMD-2 in particular.
\end{itemize}

\section{Fit Properties of the Various Broken HLS Models}
\label{genfitall}
\begin{table}[!phtb!]
\begin{center}
\begin{minipage}{0.9\textwidth}
\begin{tabular}{|| c  || c  || c | c || c ||}
\hline
\hline
\hhhd ~~~~ Channels 	& \hhhv BHLS \cite{ExtMod3}   &  \hhhv BHLS$_2$ (BS)  	 &  \hhhv BHLS$_2$ (BS)&  \hhhv BHLS$_2$ (RS)\\
\hhhd ~~~~  		&  ~~                         &  space, NO $\tau$          & space, $\tau$  & space, $\tau$\\
\hline
\hline
 \hhhv NSK  $\pi^+\pi^-$ (127)&    $137$    &  $140$          & $139$    & $137$\\
\hline
 \hhhv KLOE $\pi^+\pi^-$ (135)&    $140$    &  $138$          & $154$    & $137$\\
\hline
 \hhhv	$\tau$ (ABC) (85) &        $85$     &  ${\cal -}$       & $98$    & $92$ \\
\hline
 \hhhv $\pi^0 \gam$ (112)   &      $91$     &  $88$	          & $89$     &  $86$\\
\hline
 \hhhv $\eta \gam$( 182) &        $120$     &  $121$	          &$122$     & $120$\\
\hline
\hline
 \hhhv $\pi^+\pi^-\pi^0$ (158)&   $79/96$   &  $148 $           & $153$     & $141$ \\
\hline
 \hhhv NSK $K_L K_S$ (92) &       $97$     &   $107$           & $106$     &$106$\\
\hline
 \hhhv NSK $K^+ K^- $(49) &       $51$      &  $49$	     	  & $48$	& $50$ \\
\hline
 \hhhv BaBar $K^+ K^- $ (27) &    $49$      &  $45$      	  & $44 $	& $46 $\\
\hline
 \hhhv space $K^+ K^- $ (25)&     $-$         &  $19$      	  &$19 $	& $18$\\
\hline
 \hhhv space $\pi^+ \pi^- $(59) &    $-$      &  $58$           & $63$      & $55$\\
\hline
 \hhhv $\chi^2/N_{\rm pts}$ &    $949/1056$  &  $1054/1152$      & 1185/1237 & 1133/1237 \\
 \hhhw Probability      &         96.7\%      &  93.9 \%          & 67.6\%    & 93.2\%\\
\hline
\hline
\end{tabular}
\end{minipage}
\end{center}
\caption {
\label{Table:T3}
Global fit properties of BHLS and of the BS and RS variants of BHLS$_2$~:
$\chi^2$ values yielded by each group of data samples for the
indicated channel. Numbers within parentheses in the first column
indicate the numbers of data points
submitted to  fits; for BHLS, the actual number of fitted data points in the
3$\pi$ channel (model B)  is shown in the entry as $\chi^2/N_{\rm pts}$.}
\end{table}

 \indent \indent All fits referred to in this  Section involve  the collection of data samples
 and channels  already described  and listed in Section \ref{DataSamples}; as for
 the $K \overline{K}$ channels,
one limits oneself to using only the  samples from CMD-2 (updated), SND and BaBar
 which do not show any obvious tension. The motivation for this choice is to avoid
 disentangling  model dependence effects, on the one hand, and identified tensions
 among data samples\footnote{The issue between BaBar and CMD-3 charged kaon
 data has already been noted in \cite{CMD3_KpKm}.} on the other hand. There is indeed
 no gain in increasing the statistics if this turns out to increase the systematics
 in an uncontrollable way.

 The BHLS$_2$ fits presented up to now have been derived using the so-called
 Reference Solution (RS); this solution
 combines the so-called Basic Solution (BS) defined in Section \ref{BasicSol} and
 the Primordial Mixing defined in Section \ref{RefSol}.

 Nevertheless, in order to explore model dependence effects, analyzing also the
 fit results derived using BS is certainly relevant. On the other hand, for the same concern,
 we have rerun the BHLS model \cite{ExtMod5} using the SND together with the {\it updated}
 CMD-2 data; it then looked appropriate to also include the BaBar dikaon sample.

 Therefore, with the BS and RS variants of BHLS$_2$ and with BHLS, one
  actually has at disposal  three approaches
 to the same data; as for BHLS, one concentrates on the so-called model B
 which circumvents the issue met at the $\phi$ mass in the three pion annihilation channel
 which is one of
 the motivations having led to BHLS$_2$. The other one was to ascertain the description
 of the dipion threshold region and to improve the accountability of the spacelike region.
 The main fit properties are sketched in Table \ref{Table:T3} and deserve some comments.

 The leftmost data column -- dealing with BHLS -- shows the  various $\chi^2$ for the
 displayed  channels and the sum of the contributions of all data submitted
 to the fit. In all channels the  average $\chi^2$ is of the order 1 or
 better\footnote{One should note the smaller number of data points considered in the fit
 procedure for the $\pi^+\pi^-\pi^0$ annihilation channel, because of excluding the
 data points from the $\phi$ region.}  and the (global) fit probability is fairly good.

 The rightmost data column displays the corresponding information derived with the RS
 variant of BHLS$_2$. Here also, the fit quality is good and one yields a fair probability.
 One should note the quite favorable $<\chi^2>$'s for the pion and kaon form factors in
 the spacelike energy region in conjunction with good $\chi^2$'s in their timelike regions;
 the spacelike data samples -- coming from NA7 and Fermilab -- are discussed
 in detail in the next Section. As far as we know, the fits presented here are the
 first ones involving the pion and kaon form factors simultaneously in their
 timelike and spacelike regions with such quality.

 \vspace{0.5cm}

 The third data column displays the fit properties derived from the BS variant of BHLS$_2$
 using the same data as those just discussed for the RS variant. The fit probability is  
  lower than usual in the HLS context. Scrutinizing the various items displayed in Table \ref{Table:T3}, 
  one clearly observes a significant 
 tension in  the BS fit  absent from the RS fit results {\it when  the
 $\tau$ data are involved}.  Indeed, when going from RS to BS  $\chi^2_\tau$ increases by 6 units, but
 $\chi^2_{3\pi}$ increases by 12 and  $\chi^2_{{\rm space} \pi}$ by 8; moreover\footnote{The different behavior
 of the NSK and KLOE dipion data under BS global fits  reflects the different  magnitude 
 of the uncertainties
 which makes the NSK dipion data -- and likewise for those from  BESIII and CESR -- more permissive than their 
 KLOE analogs.},  $\chi^2_{KLOE~2\pi}$ increases by 17 units whereas
 $\chi^2_{NSK~2\pi}$ is marginally affected (a 2 unit increase). 
  

Instead,  the BS fit performed {\it when  discarding the $\tau$ data} leads to a picture
in good correspondence with the RS fit results already commented; indeed, 
 the partial $\chi^2$'s reported in the second data column of Table  \ref{Table:T3}  are similar 
 to  those obtained by fitting with the RS variant of the BHLS$_2$ model.
 Nevertheless,  some tension survives in the 3-pion sector while the dipion spectra 
 and the spacelike data almost recover their optimum fit quality; so, the BS
variant of BHLS$_2$   exhibits a quite acceptable behavior once the $\tau$ sector is discarded.

 In conclusion, despite their different structures -- and taking into account the peculiarities 
 of the BS variant of BHLS$_2$ --  these models/variants open the possibility to
 examine  model dependence effects in the evaluation of  some reconstructed physical quantities; this may 
 allow to compare these with the systematics generated by dealing with contradictory samples.

\section{The Pion and Kaon Form Factors in the Spacelike Region}
\label{spacelike}
\indent \indent
By allowing  different HK masses for the  $\rho^0$ and $\omg$ mesons,
BHLS$_2$ naturally permits a smooth connection
between the spacelike and timelike sectors of the pion  and kaon
form factors.  Moreover, applying in this context the vector meson dynamical mixing
(see Section \ref{DynMix}), does not break this smooth connection because
$\alpha(s\!=\!0)\!=\!0$ becomes automatic and supplements  the (already) vanishing properties
of the other angles~: $\beta(s\!=\!0) \!= \! \gamma(s\!=\!0)\! =\! 0$.
\begin{table}[phtb!]
\begin{center}
\begin{tabular}{|| c  || c  | c ||}
\hline
\hline
\hhhd  BHLS$_2$~: BS fit		& \hhhv incl. spacelike data    &  \hhhv   excl. spacelike data	**  \\
\hline
\hline
 \hhhv space $\pi^+ \pi^-$ (59)  &    $57.6$    &  $58.6$		\\
\hline
 \hhhv space  $K^+ K^-$  (25)    &    $18.8$   &  $19.0$ 		\\
\hline
 \hline
 \hhhv $\chi^2/N_{\rm dof}$ &    $1054/1152$  &  $977/1068$	  \\
 \hhhw Probability &    93.9\%  &  92.8 \% \\
\hline
\hline
\end{tabular}
\end{center}
\caption {
\label{Table:T4} BS variant global  fits excluding $\tau$ data: The leftmost data
column shows the partial  $\chi^2$'s in the fit  including both kinds
of spacelike data, the rightmost one shows the corresponding results in the fit
excluding all spacelike data. Global fit information is displayed in the
last lines.
The numbers of fitted data points in each  channel are given within parentheses.
}
\end{table}
\subsection{The Pion and Kaon Form Factors in the Close Spacelike Region}
\begin{figure}[!t!]
\centering
\includegraphics[width=0.86\textwidth]{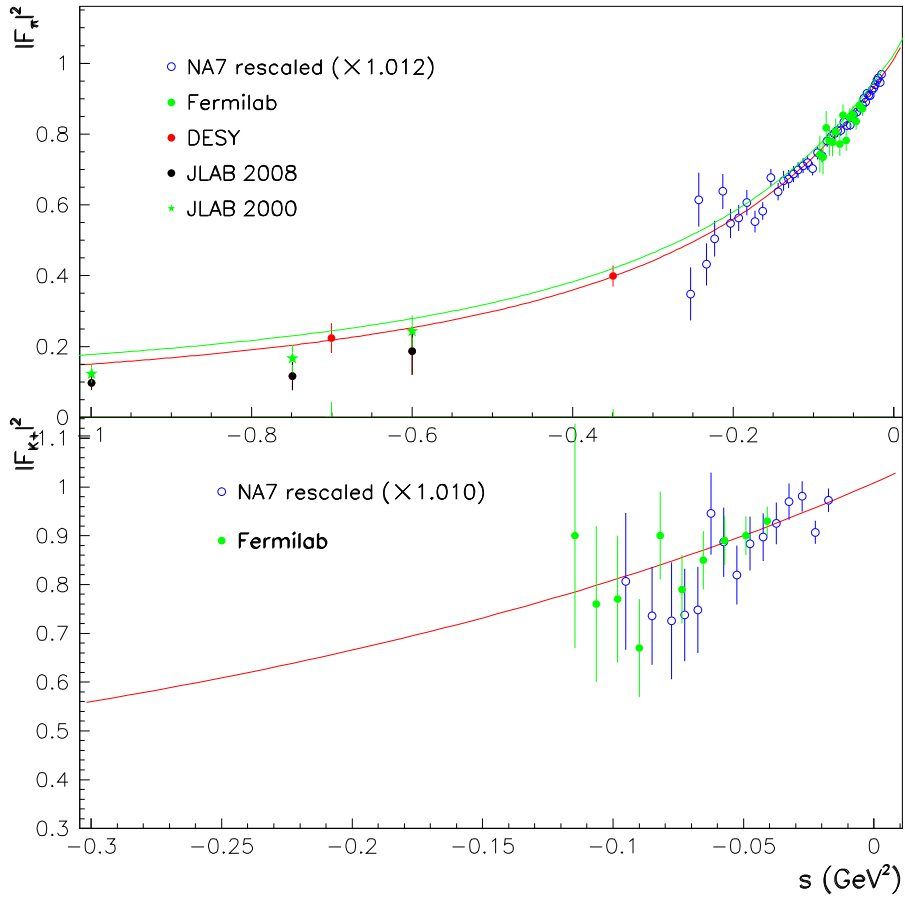}
\caption{\label{Fig:spacelike_1} Form factors squared in the spacelike region.
The top panel displays the BHLS$_2$ prediction
(green curve) and the fit (red curve) together with
the rescaled data for $|F_\pi(s)|^2$. The bottom  panel displays the fit and
the rescaled data for $|F_{K^\pm}(s)|^2$; the prediction  from the fit
excluding the spacelike data exactly superimposes to the fit (red curve).
The unfitted points from DESY and JLab are only superimposed.
See Section \ref{spacelike}  for details.}
\end{figure}

\indent \indent
Figure \ref{Fig:spacelike_1} displays the pion and kaon form factors in the spacelike region
coming out from the fit. The spacelike data supplied to the global fit procedure
are only the (model-independent) data samples  collected by the NA7 \cite{NA7,NA7_Kc}
and  Fermilab experiments
 \cite{fermilab2,fermilab2_Kc}. Altogether, these experiments cover photon
  virtualities  down to $\simeq -0.25$ GeV$^2$  and  $\simeq -0.12$ GeV$^2$
  for resp. the pion and kaon form factors.

  The pion and kaon form factors derived from the global fit including
 the spacelike data are the red curves in Figure \ref{Fig:spacelike_1}.
 Global fits have also been performed using the BS and RS variants
 of BHLS$_2$,  {\it excluding the spacelike data}. The relevant fit results
 obtained using, for instance, the BS  variant
 are displayed in Table \ref{Table:T4}. They correspond to the green curve in
 Figure \ref{Fig:spacelike_1}; as for the kaon form factor
  the predicted (green) curve
 superimposes exactly to the fitted (red) one.

Table \ref{Table:T4} and Figure \ref{Fig:spacelike_1} obviously
prove that  the spacelike data are not a real constraint
in the BHLS$_2$ global fit approach; stated otherwise, the BHLS$_2$
model supplied
with {\it only}  timelike data fairly well predicts the spacelike behaviors
far inside the spacelike region for both the pion and kaon form factors.
It is worthwhile to  note  that these analytic continuations are performed
across energy squared distances which can be as large as  1 GeV$^2$.
For instance, the kaon form factor displayed here is  the analytic
continuation of the charged kaon form factor fitted in the $\phi$ mass region, {\it i.e.}
the curve shown on the rightmost panel of Figure \ref{Fig:KLKS}. Such a property
is rare enough to be underlined.

 Figure \ref{Fig:spacelike_1} gives   the opportunity to comment on
 the representation of spectra affected by a normalization (scale) uncertainty;
the NA7  spacelike data are a quite simple illustration of handling the rescaling effects.
 When a  scale uncertainty affects a spectrum, there are obviously two possible
 ways to display the result~: Either rescaling the data, keeping the fit/prediction untouched
or performing the converse.

So, in Figure \ref{Fig:spacelike_1} one has chosen to
display the NA7 data rescaled, keeping unmodified the fit (and prediction) function(s),
the plotted uncertainties being the statistical
ones. As for the Fermilab data samples, taking into account their low accuracy,
no rescaling  has been performed
and the plotted uncertainties are just the reported total errors.

One can, for once, go into  a few plotting details in simple wording.
Following the method given in  \cite{ExtMod5}, the scale correction is given by~:
 \be
\displaystyle
\lambda= \frac{f_i V^{-1}_{ij} [m_j-f_j]}
{\displaystyle f_k V^{-1}_{kl} f_l +\frac{1}{\sigma^2}}\,,
\label{eq4-1}
\ee
where (NA7) $\sigma=0.9\%$ or 1.0\%, $f_i=  |F_P(s_i)|^2$  ($P=\pi^\pm,~K^\pm$),
$V$ is the statistical error covariance matrix
and $m_i$ is the measured datum at $s=s_i$.  The plotted NA7 data  $m^\prime$ are
related with the original ones $m$ by~:
\vspace{-0.3cm}
$$ m_i \ra m_i^\prime=m_i -\lambda f_i \left (\simeq [1-\lambda] m_i \right ).$$
Looking at  Figure \ref{Fig:spacelike_1},  the precise data (NA7 pion form factor) are
indeed observed
to tightly follow the fit function, {\it once the normalization correction is applied}.

It is relevant to provide the precise values for the $\lambda$'s derived from the
fitted parameters. One thus finds~:
$$ {\rm NA~pion~form~factor~:} \hspace{0.5cm} \lambda_\pi= (1.18 \pm 0.02_{syst}  \pm 0.03_{fit}) \%$$
\vspace{-0.5cm}
$$ {\rm NA~kaon~form~factor~:} \hspace{0.5cm} \lambda_K= (1.05 \pm 0.03_{syst}  \pm 0.02_{fit}) \%$$
which exhibit noticeable accuracies. Subsection \ref{LQCD_FF} below -- and  Figure
\ref{Fig:LQCD} -- illustrates  that the predicted scale correction for the pion form factor
is supported by the LQCD data which should be intrinsically free of scale uncertainties.

\subsection{The Pion Form Factor~: The Model-Dependent Data}
\indent \indent As  one cannot measure
the pion form factor in a model-independent way for photon virtualities
$Q^2 =-s > 0.3$ GeV$^2$, this physics is performed using the scattering
process $p e \ra e \pi^+ n$, referred to in the literature as
$^1$H$(e,e^\prime\pi^+)n$. In this approach, two questions should be addressed~:

{\bf i/} Separating out the longitudinal cross section $\sigma_L$,

{\bf ii/} Extracting $F_\pi$ from $\sigma_L$.
\begin{figure}[!ht]
\hspace{1.cm}
\begin{minipage}{0.9\textwidth}
\begin{center}
\resizebox{1.0\textwidth}{!}
{\includegraphics*{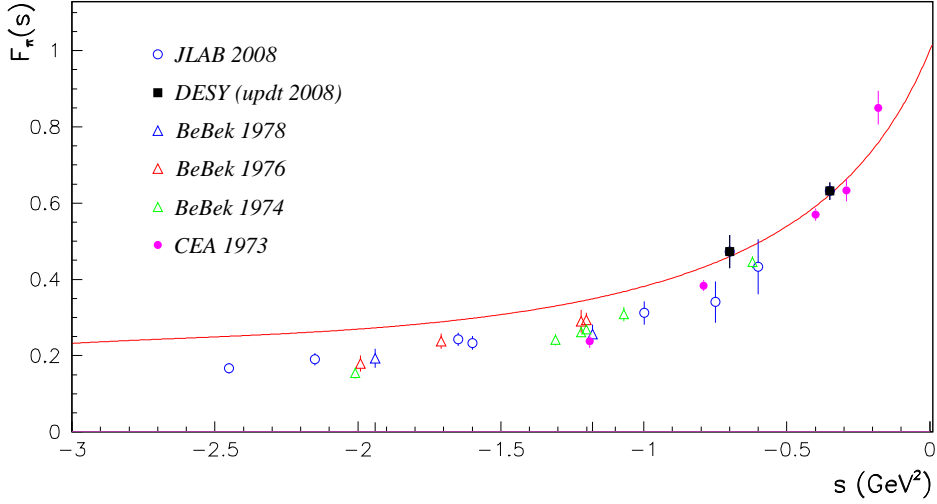}}
\end{center}
\end{minipage}
\begin{center}
\vspace{-1.cm}
\caption{\label{Fig:jlab} The pion form factor in the spacelike region;
the curve displays the function $F_\pi(s)$ coming from our global fit
of the timelike data and of the spacelike data from  NA7 \cite{NA7} and Fermilab
\cite{fermilab2}.
The (unfitted) data flagged by JLAB and DESY are extracted from  \cite{Huber},
the other can be found in \cite{Bebek_all}. See also Figure \ref{Fig:spacelike_1}.
}
\end{center}
\end{figure}

One can refer to \cite{Tadevosyan,Blok,Huber} for  comprehensive
accounts of the issues. As for item {\bf i/}, the method has evolved and
improved in the course of time  from the early experiments (also reported in
\cite{Bebek_all}) to the more recent ones \cite{Desy1,Desy2, Volmer}. Item
{\bf ii/} addresses more deeply the nuclear physics content of the process
in order to account at best for the $\pi^+ p n$ vertex and for the off-shell
character of the intermediate $\pi^+$. Presently, the favored approach
is the Regge Model  developed by Vanderhaeghen, Guidal and Laget \cite{VGL97,VGL98}.
The data extracted for $F_\pi(s<0)$ have thus improved in time but remain
unavoidably model-dependent. These data samples have not been included in our fitting approach,
but simply  compared with the extrapolation of our fit function  at $s<0$,
 a pure prediction below  $s\simeq -0.25$ GeV$^2$.

In Figure  \ref{Fig:spacelike_1}, besides the fitted data  from NA7 and Fermilab,
one has displayed the ({\it unfitted}) data from the DESY
experiments \cite{Desy1,Desy2} and the lowest photon virtuality data
collected at Jefferson Lab by the $F_\pi$ Collaboration. The data
points flagged by JLAB 2008 and DESY have been extracted from the
latest -- to our knowledge -- publication of the $F_\pi$ Collaboration
\cite{Huber}. One can observe that the updated \cite{Huber}  DESY data points fall
exactly on the  $F_\pi(s)$ function derived from a fit using also
the NA7 \cite{NA7} and Fermilab \cite{fermilab2} data.
 The three JLab plotted data points are found systematically
below the fit function;  however,  most of these JLab data points can also be found in
a previous JLab publication \cite{Volmer}. The difference between the
values in \cite{Huber} and \cite{Volmer}
seems mostly coming from using different extraction methods of $\sigma_L(Q^2)$
\cite{Tadevosyan}. The data points from \cite{Volmer} seem in closer agreement with
the continuation of our  $F_\pi(s)$ than the estimates given in \cite{Huber}. They
are also in better agreement with what can be expected from the DESY data points
as updated in  \cite{Huber}; this updating seems  to deal with the extraction of
$F_\pi$ from the longitudinal cross section $\sigma_L$.
The  behavior of (the extrapolated) $F_\pi(s)$
beyond $-1$  GeV$^2$ is shown in Figure \ref{Fig:jlab} together with the other
reported data.

\vspace{0.5cm}

The content of  Figure \ref{Fig:spacelike_1}
 gives a hint that the extrapolation of the BHLS$_2$  $F_\pi(s)$ could well be valid down to
$s \simeq -1$ GeV$^2$. This  also reflects a $\rho$ dominance well inside
the non-perturbative spacelike region. This dominance is illustrated by
fit properties~: When fitting the pion spacelike data, beside the $\rho$ contribution,
the fit function contains also  those from the $\omg$ and $\phi$ mesons; the fit
returns $<\chi^2>/N = 55/59$; however, computing the $\chi^2$ distance
of the fit function {\it amputated} from the  $\omg$ and $\phi$ contributions to
the {\it pion} data from NA7 and Fermilab, one obtains  $<\chi^2>/N = 56/59$.
This indicates that the (tiny) contributions of the $\omg$ and $\phi$ mesons
introduce a (tiny) noise and that nothing else than the $\rho$ contribution
is actually required.

This $\rho$ dominance phenomenon does not happen for the kaon form factor as,
truncating the prediction for $F_K(s)$ from its $\omg$ and $\phi$ contributions
returns a  disastrous $\chi^2$ distance; for instance, the full
$\chi^2$ distance of the NA7 kaon data to $F_K(s)$ which is $\simeq 14$, becomes $\simeq 2300$
when $F_K(s)$ is amputated from its $\omg$ and $\phi$ contributions. Such
property is quite unexpected as the parametrization for the  $\omg$ and $\phi$
\cite{ExtMod3}, reminded in Section \ref{propagators}, is nothing but a Breit-Wigner lineshape
modified in such a way that the inverse propagators fulfill $D_V(0)=-m_V^2$, these HK
masses being those in  Equations (\ref{eq1-26}).
 That the analytic continuation,  over more than $\simeq 1 $ GeV$^2$,
 of such a simple parametrization, with parameter values fixed
 at the $\phi$ mass, can provide a good   description of the existing kaon spacelike data
may  look quite amazing.

 \begin{figure}[!ph]
\hspace{0.5cm}
\begin{minipage}{0.9\textwidth}
\begin{center}
\resizebox{1.0\textwidth}{!}
{\includegraphics*{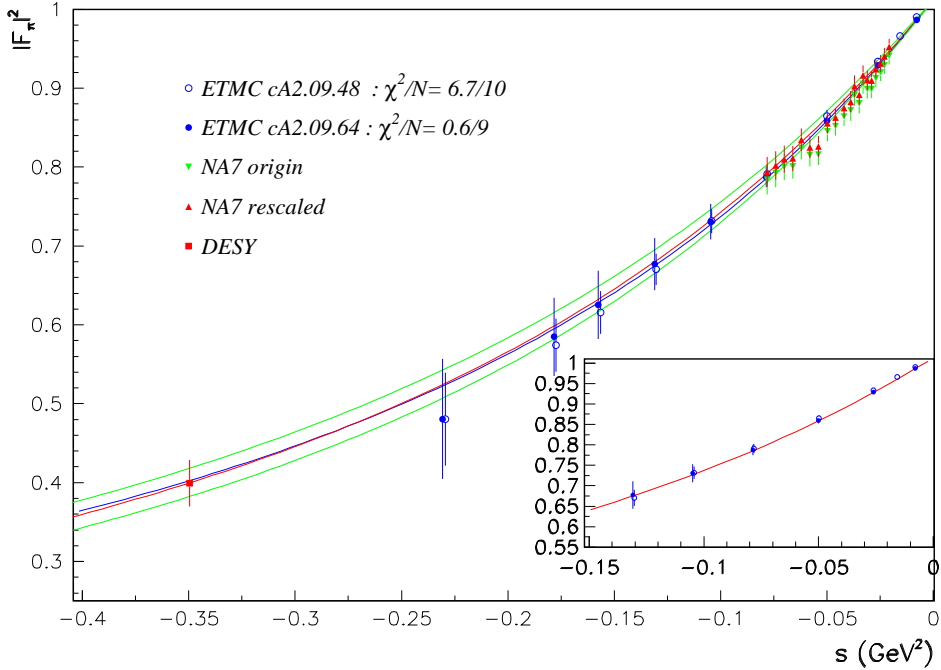}}
\end{center}
\end{minipage}
\begin{center}
\vspace{-1.cm}
\caption{\label{Fig:LQCD}  The red curve is  $|F_\pi(s)|^2$ derived from our global fit,
the blue curve is the HPQCD parametrization and the green curves display this parametrization
at  $\pm 1\sigma$ of the central value for their  $<r_\pi^2>$. The LQCD spectra from
ETMC and the experimental data shown  are commented in Subsection \ref{LQCD_FF}; the
LQCD spectra and the data point from DESY is not fitted. The inset magnifies the
threshold behavior of the predicted form factor with the ETMC data {\it superimposed}
(not fitted).
}
\end{center}
\end{figure}

\subsection{The Pion Form Factor~: The Lattice QCD Data}
\label{LQCD_FF}
\indent \indent Some Collaborations have published information on
the pion form factor derived from Lattice QCD (LQCD) data  at pion masses close
to the physical pion mass value \cite{RPP2016}. We compare these LQCD
form factors to our fit function $F_\pi(s<0)$ which, already, fairly well
describes the  accurate NA7 experimental data as obvious
 from Figure \ref{Fig:spacelike_1} and Tables \ref{Table:T3} or \ref{Table:T4}.

 The HPQCD Collaboration has thus produced  a parametrization\footnote{We thank L.
 Lellouch for having drawn our attention to this paper.}
of the pion form factor \cite{HPQCD-FF1}
$F_\pi(Q^2)$ for virtualities in the range $Q^2\in [-0.1,0.0]$ GeV$^2$. In order to perform
a numerical comparison with our fit $F_\pi(s<0)$, we have plotted their pole parametrization,
setting the lattice spacing $a$ at zero and using their pion radius squared
 $<r_\pi^2>=0.403(18)(6)$ fm$^2$.
In Figure  \ref{Fig:LQCD}, our fit function for $F_\pi(s<0)$ is shown by the red curve; the blue curve
is the HPQCD parametrization at the central value for $<r_\pi^2>$. The red and
blue curves are clearly very close to each other, if not overlapping.
The green curves display the HPQCD parametrization
with $<r_\pi^2>$ at $\pm 1 \sigma$ from its central value. This indicates that the
LQCD parametrization \cite{HPQCD-FF1} is in fair agreement with the  BHLS$_2$
fit function $F_\pi(s<0)$ much beyond the expected HPQCD range of validity
and validated by the DESY data points \cite{Desy1,Desy2};
Figure \ref{Fig:LQCD} clearly shows that
the agreement between HPQCD \cite{HPQCD-FF1} and BHLS$_2$ extends to the
{\it prediction}, shown in this Figure by the green curve, derived
when fitting with discarding the spacelike data.

\vspace{0.5cm}

Six spectra for $F_\pi(Q^2)$ have been published by the ETM Collaboration \cite{ETMC-FF};
two of them correspond to a pion mass close to its physical value, namely the
gauge ensembles $cA2.09.48$ and $cA2.09.64$.
 These spectra are given by the form factor values $F_\pi(Q^2)$ and uncertainties
 as functions of $Q^2/M^2_\pi$. In order to compare with phenomenology and experimental
 data, one has to restore the values for $Q^2$   from the provided
 values for $Q^2/M^2_\pi$. 
 Ref. \cite{ETM_MPI_pi}, which sketches the generation of these ensembles,  indicates that,
 numerically, $M_\pi \simeq 134.98$ MeV, almost identical to the RPP $\pi^0$ mass.

 Figure  \ref{Fig:LQCD} shows the $|F_\pi(Q^2)|^2$ spectra derived from
 the  $cA2.09.48$ and $cA2.09.64$ ensembles in  \cite{ETMC-FF}. They are displayed
 with their reported statistical errors\footnote{It is worth noting that
 the original (unscaled) NA7 stay {\it below} the ETM data in the spectrum range
 where they are the most precise; this should be compared with Figure 6 in  \cite{ETMC-FF},
 especially its right panel, which exhibits the same trend.}. 
 The $\chi^2$ distances\footnote{Reference \cite{ETMC-FF} does not give special 
information about systematics and performs its own fits using uncorrelated $\chi^2$'s.
 So, following this line, we define the $\chi^2$ distance of the ETM data to our prediction by the sum 
 of squared terms $d_i^2$   where~: $d_i= (F_{ETM}(Q^2_i) - F_{BHLS_2}(Q^2_i))/\sigma_{ETM}(Q^2_i)$,
using obvious notations. The quantity $\sum d_i^2$ is {\it not} submitted to the minimization procedure.} 
of $cA2.09.48$ and $cA2.09.64$ to
 the BHLS$_2$  form factor squared $|F_\pi(s)|^2$ have been computed, assuming
 the systematics
 negligible compared to the statistical errors   \cite{ETMC-FF}; they are displayed
 inside  Figure  \ref{Fig:LQCD}. One can remark that the
 average $\chi^2$ distance per datum is 0.67 for $cA2.09.48$, quite a good value
 for  data
 outside the fit procedure; it is only 0.06 for the $cA2.09.64$ form factor.
 This can be considered as an independent confirmation concerning the smallness of the systematics
 versus the statistical errors for the considered ETM spectra.

 Stated otherwise,
 the data points of the $cA2.09.64$ form factor are almost exactly on the fit function.
 Out of curiosity, we have redone the fit in the RS variant of  BHLS$_2$  discarding the spacelike
 experimental data in order to get, online, the predicted form factor for any $s<0$ {\it using only
 timelike data}; the average $\chi^2$ distances become  0.59 and 0.10 for resp.
 $cA2.09.48$ and $cA2.09.64$. So,
 the $\chi^2$ distance of the ETM samples to the BHLS$_2$ prediction indicates that the form
 factor lineshape is quite consistent with LQCD expectations; the inset in Figure \ref{Fig:LQCD},
 by magnifying the energy region close to the chiral point, illustrates the tight agreement of the prediction
 and of the ETM data.   
 In other words, the ETM spectra provide strong support to the  BHLS$_2$ fits; 
  this BHLS$_2$ model spectrum is also
 intimately related to the dealing with the reported correlated systematics as promoted
 in \cite{ExtMod5}.

  For illustration, one has also displayed in Figure  \ref{Fig:LQCD}
the closest among the two data points from DESY and
 a part of the NA7 spectrum rescaled {\it and}  not rescaled. As expected
from their  $\chi^2$ distances to the fit function,
the ETM spectra are observed closer to the rescaled NA7 spectrum than
to the original one.

\subsection{The Neutral Kaon Form Factor at $s<0$}
\begin{figure}[!hp!]
\hspace{1.cm}
\begin{minipage}{0.9\textwidth}
\begin{center}
\resizebox{1.0\textwidth}{!}
{\includegraphics*{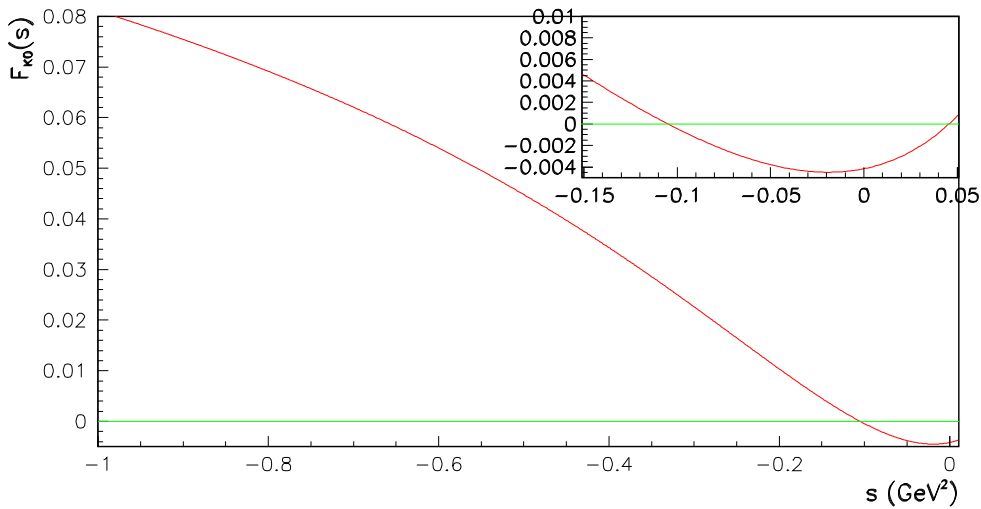}}
\end{center}
\end{minipage}
\begin{center}
\vspace{-1.cm}
\caption{\label{Fig:spacelike_k0}
The neutral kaon electromagnetic form factor predicted by BHLS$_2$~:
The function is computed at the central value of the fit parameters.
}
\end{center}
\end{figure}

\indent \indent As seen in the previous Subsections,
BHLS$_2$ predicts  the charged pion and kaon form factors
in the close spacelike region with  fair accuracy.  Likewise, BHLS$_2$  also
predicts the neutral kaon form factor $F_{K0}(s)$; it is
displayed in Figure \ref{Fig:spacelike_k0}.  This is
the analytic continuation of the form factor given in Subsection
\ref{kaonFF} and numerically determined within the fit procedure by the
$e^+e^- \ra K_L K_S$ annihilation data.
One clearly observes a behavior
quite different of those for $F_{K^\pm}(s)$, with a negative slope down to
almost $s=0$, when coming from the spacelike region.

Actually, as already stated in Section \ref{propagators} one has
$F_{K0}(s\!=0)\! =\!0$ up to contributions of second order in breaking
parameters. What is displayed in Figure \ref{Fig:spacelike_k0} is
$F_{K0}(s)$ computed at the fitted central value of the fit
parameters\footnote{This is equivalent to performing only one sampling
on the multidimensional error covariance matrix of the fitted parameters.
}.
To take  into account their uncertainties, a Gaussian Monte-Carlo   has
been run which returns the averaged central value $F_{K0}(0) \simeq 10^{-6}$
or less, depending on the fit conditions. This indicates that the
negative excursion close to zero observed in Figure \ref{Fig:spacelike_k0}
  could well be a numerical  artifact.

\begin{table}[hbpt!]
\begin{center}
\begin{tabular}{|| c  || c  | c ||}
\hline
\hline
\hhhd ~~~~            & \hhhv $<r_{\pi^\pm}^2>$    &  \hhhv  $<r_{K^\pm}^2>$	  \\
\hline
\hline
 \hhhv BHLS$_2$ fit  &    $0.430 \pm 0.002_{\rm mod} \pm 0.001_{\rm
 fit}$    &  $0.268\pm 0.004_{\rm mod} \pm 0.001_{\rm fit}$\\
\hline
 \hhhv NA7  \cite{NA7,NA7_Kc}  &    $0.439 \pm 0.008 $    &   $0.40\pm 0.11 $		\\
\hline
 \hhhv Fermilab \cite{fermilab2,fermilab2_Kc}     &    $0.439 \pm 0.030 $    &   $0.28\pm 0.05 $		\\
\hline
 \hhhv CHS   \cite{Colangelo:rpi}  &    $0.429 \pm 0.001_{\rm stat}\pm
 0.004_{\rm syst}$    &   --	\\
\hline
 \hhhv ACD   \cite{Caprini}  &    $0.432 \pm 0.004$    &   --	\\
\hline
 \hhhv ChPT  2 flavors \cite{Bijnens2}  &    $0.437 \pm 0.016$    &   --	\\
\hline
 \hhhv ChPT  3 flavors \cite{Bijnens3}  &    $0.452 \pm 0.013$    &   $0.363 \pm 0.072$	\\
\hline
 \hhhv HPQCD   \cite{HPQCD-FF1}       &    $0.403 \pm 0.018_{\rm stat} \pm
 0.006_{\rm syst}$    &   --		\\
\hline
 \hhhv ETM \cite{ETM_MPI_pi}  &    $0.443 \pm 0.021_{\rm stat} \pm
 0.020_{\rm syst}$    &   --	\\
\hline
 \hline
\end{tabular}
\end{center}
\vspace{-0.5cm}
\caption {
\label{Table:T5} Pion and kaon charged radii in units of fm$^2$. See also \cite{FLAG_2016} for
additional LQCD evaluations of $<r_{\pi^\pm}^2>$.
}
\end{table}

As for the charge radius squared of the neutral kaon, both  BS and RS variants
of BHLS$_2$ return an average value~:
\be
\displaystyle
<r^2>_{K^0} = 0.00 \pm 0.10 ~~{\rm fm}^2
\label{rk0}
\ee
consistent with all reported measurements, in particular, those of
the NA48 Collaboration \cite{Lai_NA48}
($<r^2>_{K^0} = -0.090 \pm 0.021$ fm$^2$),  or those of KTEV \cite{Abouzaid_KTEV}
($<r^2>_{K^0} = -0.077 \pm 0.007_{\rm stat} \pm 0.011_{\rm syst}$ fm$^2$). It also agrees
with the ChPT result obtained by Bijnens and Talavera \cite{Bijnens3}
($<r^2>_{K^0} = -0.042 \pm 0.012$ fm$^2$) and with the former
measurements collected in the Review of Particle Properties \cite{RPP2016}.
We don't know about predictions for $F_{K0}(s<0)$ at larger
virtualities to which one could  compare.
\subsection{Pion and Kaon Charge Radii}
\indent \indent The fit parameter central values and the parameter error covariance matrix can serve
to generate several samplings ($\alpha=1, \cdots n $) and, for each, the corresponding
pion and kaon form  factors $F_P^\alpha(s)$.
These can be used to  evaluate derived physics quantities by computing
the  average and the r.m.s. values of the various estimates obtained from the set
$\{ F_P^\alpha(s), \alpha=1, \cdots n\}$.

Among the relevant physical properties of the pion and kaon form factors,
their behavior close to $s=0$ deserves special interest. As clear from Section
 \ref{form_factors}, the charged pion and kaon vector form factors are 	analytic
 functions of $s$ and, so, can be expanded in Taylor series  around the origin.
Neglecting terms of degree higher than 2 in $s$, one can write~:
\be
\displaystyle
F_P(s) =a_P +b_P s+ c_P s^2   ~~~,~~P=\pi^\pm,K^\pm\,,
\label{rpik}
\ee
where the $a_P$'s are expected equal to 1 (up to second order terms in breakings) and the
$b_P$'s are related with the so-called charged radii by  $b_P=<r_P^2>/6$ for which several
values have been reported.

In order to have a reasonable lever arm, one has chosen a  $s$ interval
extending  on both sides of $s=0$ and
bounded\footnote{ Considering the real part of $F_P(s)$ permits to avoid (minor) issues
with the $\pi^0 \gamma$ threshold effects.}
by $\pm 0.05$ GeV$^2$. Using the values of each $F_P^\alpha(s)$ at three $s$ values
($0, \pm 0.05)$ GeV$^2$,  one constructs the sets
$\{ [a_P^\alpha, b_P^\alpha,c_P^\alpha]  , \alpha=1, \cdots n\}$,   and derives
their averages and r.m.s.'s.  Working within the RS variant of BHLS$_2$ and using
the largest set of experimental data samples, one derives~:
 \be
 \begin{array}{lll}
 \displaystyle b_\pi = [1.840 \pm 0.003] {\rm GeV}^{-2}~~,
& \displaystyle c_\pi = [4.155 \pm 0.054] {\rm GeV}^{-4}~~,
\end{array}
 \label{rpi}
\ee
and~:
 \be
 \begin{array}{lll}
\displaystyle b_K = [1.159 \pm 0.005] {\rm GeV}^{-2}~~,
& \displaystyle c_K = [1.15 \pm 0.24] {\rm GeV}^{-4}~~.
\end{array}
 \label{rK}
\ee

As it  is more customary to express the slope term in units of fm$^2$, one displays
our values together with external information  in Table \ref{Table:T5}.  Our results are given
with two uncertainties; one is derived from the sampling on the error covariance matrix
and is indexed by "fit". One has also  varied  the content of the data sample set
 submitted to the global fit and  compared the results to those derived with the maximal
 sample set. It is observed that the central values for $<r_{\pi^\pm}^2>$  and  $<r_{K^\pm}^2>$
 get shifted while the uncertainties are almost unchanged. One has also switched from the RS to the
 BS variant of BHLS$_2$; in this case, the central values get marginally shifted. Merged
 together, these shifts contribute an uncertainty denoted  "mod" for "model" which, however,
 reflects as much  the (weak)  tensions between the various data samples of the largest sample
 set -- defined in  Section \ref{genfitall} -- than the model properties by themselves.

As for $<r_{\pi^\pm}^2>$, Table \ref{Table:T5} shows an agreement of our result
with the experimental data from NA7 and Fermilab at the $1 \sigma$ level or better.
The result of G. Colangelo {\it et al.} \cite{Colangelo:rpi} (CHS) is derived
from a global fit  of the timelike and spacelike  pion form factor data  based on a
parametrization through the Omn\`es representation of $F_\pi(s)$ taking into account
isospin breaking effects ($\omg$) and prescriptions for the
asymptotics of the $\pi\pi$ $P$-wave phase shift; this estimate  is in perfect agreement
with our own fit result. Using an elaborate method mixing the low energy $\pi \pi$ phase information
and some
form factor modulus information, B. Ananthanarayan {\it et al.} \cite{Caprini}
recently derived  the value flagged by ACD in Table \ref{Table:T5} which is very close to ours.
The two flavor result of J. Bijnens {\it et al.}  \cite{Bijnens2}  flagged as "ChPT  2 flavors"
as well as the three flavor evaluation in \cite{Bijnens2} and the
most recent Lattice QCD derivations of the HPQCD   \cite{HPQCD-FF1} and ETM \cite{ETM_MPI_pi}
Collaborations are also in good correspondence with ours. Former LQCD evaluations for
$<r_{\pi^\pm}^2>$ can also be found in Table 22 of \cite{FLAG_2016}.

As for the curvature term, varying the fit conditions as just indicated for  $<r_{\pi^\pm}^2>$,
one can assess the effects observed when switching from the RS to the BS
variant of BHLS$_2$ and when modifying the set of data samples submitted to fit.
Altogether, this leads to our final result~:
 \be
\displaystyle
\displaystyle c_\pi = [4.155 \pm 0.040_{\rm mod} \pm 0.054_{\rm fit}]~~ {\rm GeV}^{-4}\epo
\label{rpik2}
\ee

The 2 flavor ChPT estimate based on all data samples available
at that time (1998), \cite{Bijnens2} reports $c_\pi= 3.85 \pm 0.60$ GeV$^2$
and the 3 flavor evaluation \cite{Bijnens3} ($c_\pi= 4.49 \pm 0.28$ GeV$^2$)
look in fair agreement with the BHLS$_2$ evaluation given in Equation (\ref{rpik2}).

\vspace{0.5cm}
As the  pion and kaon form factors, in the close spacelike
region, are frequently parametrized by a pole expression~:
 \be
\displaystyle
F_P(s) =\frac{1}{1 -b_P s}  ~~~,~~P=\pi^\pm,K^\pm\,,
\label{rpik4}
\ee
we performed alike without, however, introducing
additional normalization factors as commonly done as indeed  we believe  in
the relevance of the absolute scale of the $F_P(s)$'s returned by the BHLS$_2$ fits.
Such a parametrization turns out to impose the constraint $c_P=b_P^2$ to the
curvature term. This leads to~:
$<r_{\pi^\pm}^2> =0. 424 \pm 0.002_{\rm mod}\pm 0.001_{\rm fit}$  and  
$<r_{K^\pm}^2>=0.268 \pm 0.003_{\rm mod} \pm 0.001_{\rm fit}$.

\vspace{0.5cm}

 For what concerns  $<r_{K^\pm}^2>$, the numbers reported in Table \ref{Table:T5},
 show that our evaluation is at $1 \sigma$ from the NA7 measurement \cite{NA7_Kc}
 and coincides with  those from \cite{fermilab2_Kc}. As for the
 ChPT prediction
 from  Bijnens and Talavera \cite{Bijnens3},  it differs from the value returned by BHLS$_2$
 at the $1 \sigma$ level.
 There is no reported evaluation of the curvature term of the kaon form factor to which
 one could compare our own evaluation in Equations (\ref{rK}).

 Therefore,  concerning the low energy behavior of the pion and kaon
 form factors, the overall picture is  a fair agreement of  BHLS$_2$ with
 the corresponding experimental measurements  as well as with the existing
 LQCD and ChPT evaluations.

\section{The Isospin 1 $P$-wave $\pi \pi$ Phase Shift}
\label{phaseshift}
\indent \indent Another relevant piece of information is the $I=1$  dipion
$P$-wave phase shift $\delta_{11}$. This physics has not been included within our HLS
frameworks\footnote{A relative energy scale calibration issue between phase shift 
and annihilation data 
 may have been identified \cite{Colangelo:rpi}, probably amplified by the steep rise of
 the phase in the interesting region, combined with  a 20 MeV binning size
 for what concerns the CERN-Munich data  \cite{Ochs1}. }; 
 nevertheless, a comparison of the $P$-wave $\pi \pi$ phase shift
extracted from BHLS/BHLS$_2$ with other sources provides an additional cross check.
 
\begin{figure}[!t]
\centering
\includegraphics[width=0.86\textwidth]{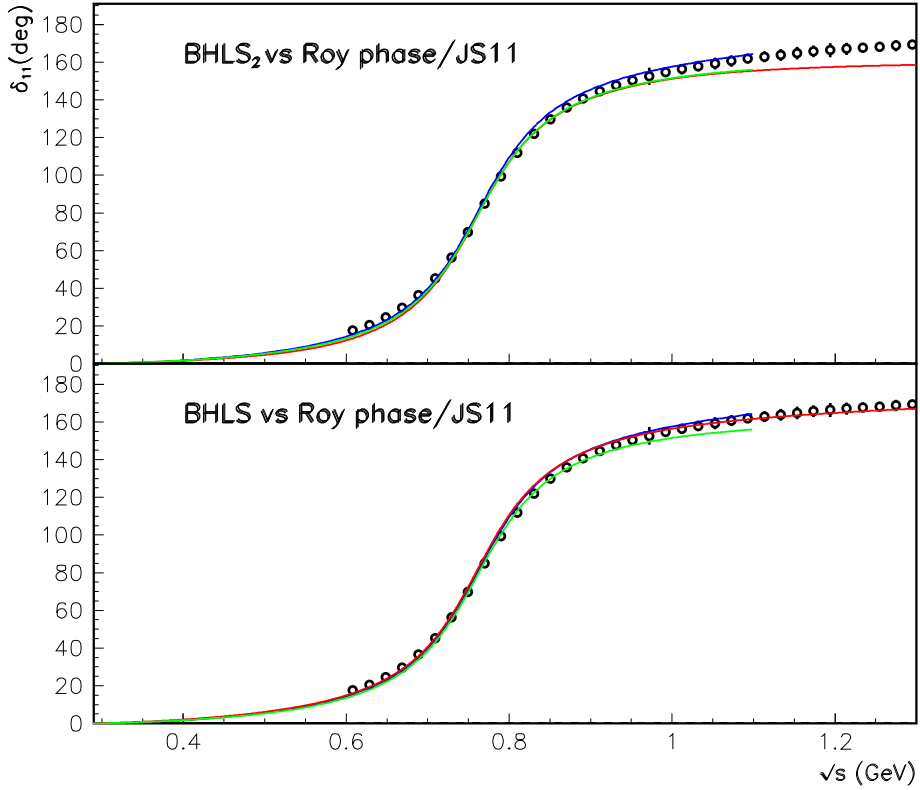}
\caption{\label{Fig:phaseShift} The $\delta_{11}$ $\pi \pi$ phase shift~:
In both panels, one displays the CERN-Munich data \cite{Ochs1}, the phase shift
from  \cite{Fred11} (JS11) is shown by the green curves and the blue curves
are the  phase shift from \cite{RoyEq} based on the Roy Equations.
The red curves display the predictions from BHLS$_2$ (top panel) and
from BHLS \cite{ExtMod3,ExtMod5} (bottom panel) using the pion form
factor amplitude amputated from its $\omg$ and $\phi$ contributions.
}
\end{figure}

Both panels of Figure \ref{Fig:phaseShift} display
the experimental data collected by the CERN-Munich Collaboration \cite{Ochs1},
the phase shift derived using the Roy Equations \cite{RoyEq}  and the phase shift
derived in  \cite{Fred11} in the context of a common fit of the $e^+e^- \ra \pi^+ \pi^-$
cross-section and of the dipion spectrum in the $\tau$ decay\footnote{Plots
including other data  \cite{Protopescu} and the ChPT predictions \cite{CGL} can be found in Figure 5 of
\cite{ExtMod5}; they are not shown here in order to avoid overburdening Figure \ref{Fig:phaseShift}.
}. The phase shift coming from BHLS$_2$  is  displayed in the top panel
while the  phase shift from the previous BHLS \cite{ExtMod3} is displayed in the bottom
panel. Both HLS based phase shifts displayed correspond to the phase of the pion form
factor amplitude amputated from the  $\omg$ and $\phi$ terms (see Equation (\ref{eq1-47}))
to meet the usual definition of the $I=1$  $P$-wave $\pi \pi$ phase shift \cite{RoyEq,Fred11}. 
The full ({\it i.e.} non amputated) BHLS/BHLS$_2$ pion form factor phases are visually
identical to the ones displayed in Figure \ref{Fig:phaseShift} except for additional
tiny blips in the  $\omg$ and $\phi$ energy regions as can also be seen in Figure 5
of \cite{ExtMod5}.

In the case of BHLS$_2$ (red curve in the top panel of Figure \ref{Fig:phaseShift}), the 
phase shift superimposes almost exactly on the data points \cite{Ochs1} and onto the JS11 
curve up to the $\phi$ mass region. For this comparison
with the external theoretical expectations \cite{RoyEq,Fred11}, one should cancel out
the $\omg$ and $\phi$ terms from  the pion form factor $F_\pi^e(s)$ given in 
Equation (\ref{eq1-47}) as, indeed, these expectations do not carry such signals; one
should remind that,  within the BHLS/BHLS$_2$ frameworks, these signals are generated 
via the mixing angles $\alpha(s)$ and $\beta(s)$ (see Equations (\ref{eq1-48})) which
are proportional to the charged and neutral kaon loop difference and this vanishes in the
isospin symmetry limit of equal kaon masses.

\begin{figure}[!t]
\centering
\includegraphics[width=0.99\textwidth]{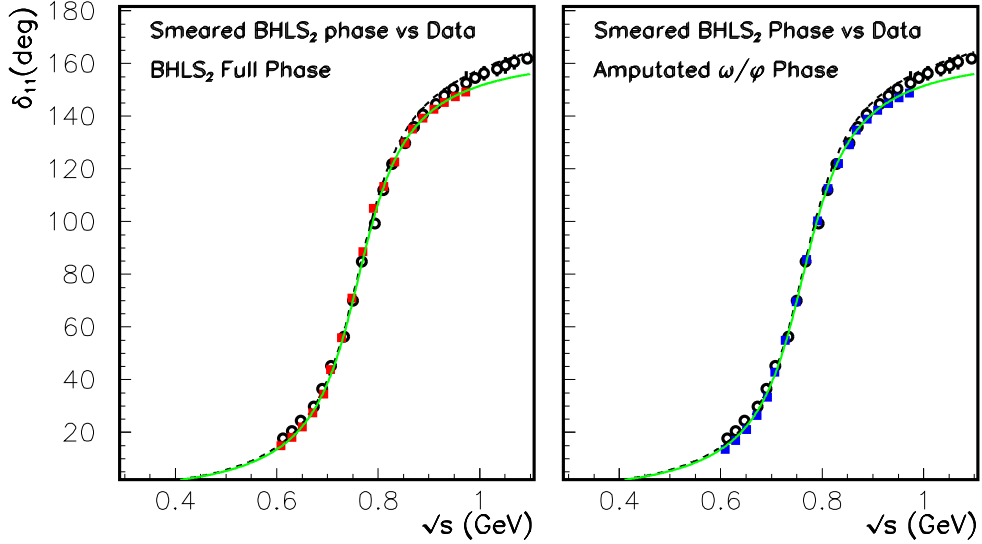}
\caption{\label{Fig:phaseShift2} BHLS$_2$  $\delta_{11}$ $\pi \pi$ phase shift versus data 
\cite{Ochs1}~:
The empty black circles display the experimental data points, the red (left panel)
and blue (right panel) square symbols  show the smeared BHLS$_2$ phase as indicated in the respective
panels. The green full line represents the JS11 phase shift   \cite{Fred11} while the
black dashed line is the classical phase shift from \cite{RoyEq}.
}
\end{figure}

This argument might not fully apply when comparing the BHLS$_2$ phase to the 
experimental data \cite{Ochs1}. However, the  CERN--Munich 
 Collaboration reported \cite{Ochs2} a tiny $\rho^0-\omg$ mixing effect -- actually 
 unsignificant (see Table 1 in ths Reference) -- in the \cite{Ochs1} phase shift data.
 Nevertheless, let us quickly consider this aspect of the 
 experimental phase shift for completeness.

For a well-grounded comparison between the BHLS$_2$ phase and the  CERN--Munich data, 
one should rather use the BHLS$_2$ phase smeared out over the experimental (20 MeV) 
 bin width. The purpose of Figure \ref{Fig:phaseShift2} is to display 
the smeared BHLS$_2$ phase together with the CERN-Munich data \cite{Ochs1}.
In its leftmost panel the smeared BHLS$_2$ phase plotted is those of the full pion form factor
$F_\pi^e(s)$ just as given in Equation (\ref{eq1-47}); in the rightmost panel, the smeared
BHLS$_2$ phase is those of the $(\omg/\phi)$-amputated $F_\pi^e(s)$. In both panels, the
CERN-Munich data points are also plotted but their symbols generally vanish underneath those for
the smeared BHLS$_2$ data. One clearly observes a good agreement between BHLS$_2$ and the
CERN-Munich data up to the $\phi$ mass region. Comparing the phase shift of the full and 
amputated BHLS$_2$ pion form factors as displayed in Figure \ref{Fig:phaseShift2} clearly
shows that the effects of the $\omg$ and $\phi$ signals on  $\delta_{11}$ is quite 
marginal as expected from the  \cite{Ochs2} analysis.

\vspace{0.5cm}

In a large energy interval
(about $\simeq 200$ MeV wide) centered on the $\omg$ mass, the 3 curves for the BHLS/BHLS$_2$, Roy and JS11 phase
shifts overlap within the thickness of the curves and, so, it is impossible 
to distinguish any one from each other. Then, 
both JS11 and BHLS$_2$ start to depart from the data increasingly with increasing energies. 
Put together with the remarks already stated
about the behavior of the dikaon cross sections slightly above the $\phi$ mass, this gives a hint
for an  onset of higher mass vector mesons presently not accounted for in the HLS framework.

Thus, the agreement between the  BHLS/BHLS$_2$ predictions on the one hand and
 the theoretical expectations \cite{RoyEq,Fred11} and the CERN-Munich data on the 
 other hand, looks quite satisfactory\footnote{As illustrated in Figure 2 of ~\cite{Colangelo:2003yw} 
 the solution of the Roy equation requires a normalization of the energy scale at
about 1 GeV.  In the BHLS approach, one could adjust $g_{\rho\pi\pi}$
to get a better agreement for the high energy tail of the Roy solution.}. 

One should also note that the bottom panel in Figure \ref{Fig:phaseShift} exhibits
a fair agreement with the datat much above the validity range of the BHLS model (the
$\phi$ mass region); this can be traced back to having allowed the pion loop in the
$V-\gamma$ transition amplitude $F_\rho^e$ and in the $\rho$ self-mass amplitude  to carry
different subtraction polynomials (see Subsection \ref{form_factors} above). This clearly accounts for
the onset of the high mass vector mesons as the agreement remains valid up to 1.3 GeV and somewhat above.
\begin{table}[!ptbh!]
\begin{minipage}{0.8\textwidth}
\begin{center}
\begin{tabular}{|| c  || c  || c | c || c ||}
\hline
\hline
\hhhd ~~~~ Channels 	& \hhhv BHLS \cite{ExtMod3}   &  \hhhv BHLS$_2$ (BS)  	 &  \hhhv BHLS$_2$ (RS)&  \hhhv BHLS$_2$ (RS)\\
\hhhd ~~~~  	        & \hhhv (model B)    &  \hhhv (no $\tau$) 	 &  \hhhv  ($z_V$ free) & \hhhv  ($z_V \equiv 1$)\\
\hline
\hline
 \hhhv $a_{\rm HLS}$    &   $2.620\pm 0.001$   &  $1.970\pm 0.002$     & $1.590 \pm 0.001$       & $2.658 \pm 0.002$ \\
\hline
 \hhhv $g$          &   $5.570\pm 0.001$       &  $6.855\pm 0.003$     & $6.707 \pm 0.002$	 & $6.702 \pm 0.003$\\
\hline
 \hhhv $(c_3+c_4)/2$&   $0.886 \pm 0.003$      & $0.762 \pm 0.003$     & $0.756\pm  0.003$	 & $0.755\pm 0.003$\\
\hline
 \hhhv $c_1-c_2$    &   $1.130\pm0.022$        &  $0.771\pm 0.020$     & $0.833 \pm 0.021$       & $ 0.789 \pm 0.018 $ \\
\hline
 \hhhv $10^2\times  z_3 $  & $\bf{\times}$     &  $-0.373 \pm 0.004$   & $-0.424 \pm 0.003$	 & $-0.428 \pm 0.003$\\
\hline
\hline
 \hhhv $z_A$        &    $1.57\pm0.01$      &  $1.55 \pm 0.01$         & $1.55 \pm .01$          & $1.48 \pm .01$\\
\hline
 \hhhv $z_V$        &    $1.204\pm0.001$      &  $1.403 \pm 0.002$     & $1.674 \pm 0.002$       & $ \bf{1.00 }$\\
\hline
 \hhhv $10^2 \times \Delta_A$ &    $1.955\pm0.517$    & $0.11\pm 0.51$   & $0.83 \pm 0.51$       & $0.82 \pm 0.50$\\
\hline
 \hhhv $10^2 \times \Sigma_V$ &    $-15.63 \pm0.06$   &  $-7.69\pm 14.8$   & $10.42 \pm 0.15$    & $-34.53 \pm 0.09$ \\
\hline
\hline
 \hhhv $10^2 \times  \xi_0 $  &    $\bf{\times}$      &  $-6.40  \pm 0.04$  &$-3.60 \pm 0.03$    & $-1.85 \pm 0.04$\\
\hline
 \hhhv $10^2 \times  \xi_3$ &   $\bf{\times}$         &  $2.34 \pm 0.16$  & $2.92 \pm 0.15$     & $ 2.02\pm 0.080$\\
\hline
\hline
 \hhhv $10^2 \times  \psi_0 $  &    $\bf{\times}$     &  $\bf{\times}$	  &$-3.91 \pm 0.27$	 & $-4.24 \pm 0.24$\\
\hline
 \hhhv $10^2 \times  \psi_\phi$ &   $\bf{\times}$     &  $\bf{\times}$	  &$-0.80 \pm 0.09$	 & $-0.63 \pm 0.09$\\
\hline
 \hhhv $10^2 \times  \psi_\omg$ &   $\bf{\times}$     & $\bf{\times}$	  & $-3.20 \pm 0.11$   & $-3.39 \pm 0.12$\\
\hline
\hline
 \hhhv $\chi^2/N_{\rm pts}$      &    $949.1/1056$     &  $1054/1152$      & 1133/1237   & 1137/1237 \\
 \hhhw Probability               &         96.7\%      &  93.9 \%          & 93.3\%      & 92.4\%\\
\hline
\hline
\end{tabular}
\end{center}
\end{minipage}
\caption {
\label{Table:T6}
Model parameter values in
 BHLS and in the BS and RS variants of BHLS$_2$~; second line in the Table title
 indicates the running conditions for what concerns the data samples submitted to fit.
The last lines display  the  $\chi^2$, $N_{\rm pts}$ and the probability of the fits.}
\end{table}
\section{Lagrangian Parameter Values in BHLS and BHLS$_2$}
\label{LagVal}
\indent \indent Table \ref{Table:T6} reports on the Lagrangian parameter values\footnote{The  full list of
parameter values and uncertainties, especially   the subtraction polynomial coefficients, are not given;
they can be provided upon request.}
derived when running BHLS  with updated kaon data (first data column). Running the basic variant BS of BHLS$_2$
with $\tau$ data discarded gives parameters values displayed in the second data column, while the third data column
reports on the  numerical results from the BHLS$_2$ Reference Solution (RS) using the largest set of consistent
data samples.

Clearly, the  Lagrangian parameters common to the various modelings cover a wide range of
 values; this does not prevent  the fit properties (reminded in the last two data lines) to
 always exhibit a fairly good description of the data.

\subsection{The ($z_V \div a$) and ($\Sigma_V \div a$) Anticorrelations within BHLS$_2$}
\indent \indent
The value obtained for $z_V$ from fitting  with the
RS  variant of BHLS$_2$ may raise  questions -- see Section \ref{RefSol} -- as $z_V-1$
is numerically large\footnote{
\label{degen} Even if only a $ 2.4 \epsilon$ effect, referring to the numerical
estimate for generic $\epsilon$'s given in Footnote \ref{epsilon}.},
even larger than $z_A-1$. For
this purpose, it has been worthwhile to explore the RS variant behavior in fits performed
discarding the $3 \pi$ final state data; this has revealed a numerical anticorrelation between the  HLS
parameter $a$ and the BKY breaking parameter $z_V$, undetectable in the parameter error
correlation matrix  as $<\!\delta a ~ \delta z_V\!> \simeq -0.24$. Likewise, but of less concern,
one observes a similar
numerical anticorrelation between $a$ and $\Sigma_V$ as $<\!\delta a ~ \delta \Sigma_V\!>\simeq -0.18$.
Several solutions with comparable total $\chi^2$ were found and the  most striking difference among them
just concerns the values for $a$, $z_V$ and $\Sigma_V$. This is well illustrated by the three following
parameter sets~:
\be
\hspace{-0.9cm}
\left \{
\begin{array}{llll}
 a = 2.340 \pm 0.003 , &  z_V=1.122 \pm 0.003, & \Sigma_V = -0.272 \pm 0.001,  & \chi^2/N_{\rm pts}= 981.2/1079\\[0.5cm]
 a = 2.083 \pm 0.003 , &  z_V=1.266 \pm 0.005, & \Sigma_V = -0.182 \pm 0.001,  & \chi^2/N_{\rm pts}= 981.5/1079\\[0.5cm]
 a = 1.606 \pm 0.001 , &  z_V=1.653 \pm 0.002, & \Sigma_V = +0.068 \pm 0.002,  & \chi^2/N_{\rm pts}= 983.5/1079
\end{array}
\right .
\label{degeneracy}
\ee
 which exhibit  clear anticorrelations between the numerical values for $a$ and $z_V$ on the one hand and
 for  $a$ and $\Sigma_V$ on the other hand.  Despite the large variations of each of $a$, $z_V$ and
 $\Sigma_V$,  the ranges covered by the central values for $a z_V$ and $a (1+\Sigma_V)$ 
 are much narrower than those for  $a$, $z_V$ and $\Sigma_V$; indeed, one obtains 
 $a z_V = (2.63 \div 2.66)$ and $a (1+\Sigma_V) = (1.70 \div 1.72)$,  {\it i.e.}
 both products exhibit spreads at not more than the percent level. 

 In order to track back the source of these anticorrelations, it is worth considering the
 $\rho^0 \pi^+ \pi^-$  and $\phi K  \overline{K}$ couplings. These are given by (see Appendix \ref{AA})~:
  \be
\begin{array}{ll}
\displaystyle g^\phi_{K  \overline{K}} = \mp \left[a z_V\right]~~\frac{g\sqrt{2}}{4 z_A} 
\left(1 \pm \frac{\Delta_A}{4}\right) (1+\xi_0)~~{\rm and} &
\displaystyle g^{\rho^0}_{\pi^+ \pi^-} =\left[a (1+\Sigma_V) \right] \frac{g}{2} (1+\xi_3).
\end{array}
\label{phikk}
\ee
\noindent which exhibit the dependences upon $\left[a z_V\right]$ and $\left[a (1+\Sigma_V) \right]$.

On the other hand, given the measured cross sections for $e^+ e^- \ra \pi^+ \pi^-$ and 
$e^+ e^- \ra K  \overline{K} $, the observables  
 $\Gamma(\rho^0 \ra  e^+ e^-) \times \Gamma(\rho^0 \ra  \pi^+ \pi^-)$ and 
 $\Gamma(\phi \ra  e^+ e^-)\times \Gamma(\phi \ra   K  \overline{K})$ should be weakly model/parameter
 dependent within {\it successful} (global) fits. Substantially, these products are 
 the squares of 
  $H_\rho=| F^e_{\rho \gam}(m_{\rho^0}^2) ~g^{\rho^0}_{\pi^+ \pi^-}|$ and
 $H_\phi=|F^e_{\phi \gam}(m_{\phi}^2) ~ g^\phi_{K  \overline{K}}|$,
  the $F^e_{V \gam}(s)$'s  being given by Equation (\ref{eq43q}); these quantities  stretch
  ranges of the order $\pm 3\%$
  only~:$H_\rho=0.753 \div 0.803$ and $H_\phi=0.124 \div 0.136$.
  So, as the fits reported by Equations (\ref{degeneracy}) are successful, 
  they all reproduce successfully the resonance peak values  
for the $e^+ e^- \ra \pi^+ \pi^-$ and  $e^+ e^- \ra K  \overline{K} $ cross sections, 
which implies  that the products  $a z_V$ and $a (1+\Sigma_V)$ weakly vary within different
fits, even if each of these parameters scans a wider interval\footnote{ The other parameters
involved in the expressions for $g^\phi_{K  \overline{K}}$ and $g^{\rho^0}_{\pi^+ \pi^-}$
-- {\it i.e.} $g$,$\Delta_A$, $z_A$ and the $\xi$'s  --  undergo variations at the percent level or
better.}.
 \vspace{0.5cm}
 
One has also run the BHLS$_2$ RS variant by fixing $z_V=1$ and found the results shown in
the last data column of Table \ref{Table:T6}. The 6 units difference between
the minimum  $\chi^2$ obtained fixing $z_V=1$ and the best  fit
leaving $z_V$ free is almost entirely concentrated in $\chi^2(e^+e^- \ra 3 \pi)$
which becomes 146.3 (compared
to 140.4 for 158 data points). Running also  the BHLS$_2$ RS variant  imposing via {\sc minuit}, a lower bound $a\ge 2$ returns~:
\be
\begin{array}{llll}
 a = 2.0 \pm 0.001 ~, &  z_V=1.330 \pm 0.001  ~, &  \Sigma_V=-0.128 \pm 0.001~,& \chi^2/N_{\rm pts}= 1133.2/1237~~. \\[0.5cm]
\end{array}
\label{degeneracy2}
\ee

 This fit also   returns  $a z_V \simeq 2.66$ and  $a (1+\Sigma_V) \simeq 1.75 $
 and  shows fairly good quality (93.6\% probability).  Moreover, all solutions
return consistent  information about the physics observables as will be exemplified below.
Therefore, the relatively large
  value for $z_V-1$ obtained in the best fit with the (unconstrained) RS variant of BHLS$_2$ is not
  a real issue. One should also note that the multiplicity of solutions
reflected by   Equations (\ref{degeneracy}) reduces to a single one once the 3-pion data are accounted for.
\subsection{A Remark on the HLS Parameter $a$}
\indent \indent
Correlated with the remark on $z_V$, one should note the value for the standard HLS model
parameter $a$ coming out of the best RS fit~: $1.590 \pm 0.001$. This is much smaller
than any of its previous determinations. One should also note the unexpected value for 
$a$ naturally returned by 
the BS variant fit~: $a=1.970 \pm  0.002$, very close to 2   (see Table \ref{Table:T6}).
Therefore, one may conclude that, within the BS variant of BHLS$_2$,  breaking parameters
supply  the needed $\gamma \pi^+\pi^-$, $\gamma K^+K^-$ and $\gamma K_LK_S$ couplings 
which, otherwise, would vanish when the $a=2$ constraint is imposed.

\subsection{The ${\cal O}(p^4)$ HLS Lagrangian~: Effects of the Parameter $z_3$}
\indent \indent The present work is the first one where a manifest and
thorough use of the 
$z_3$ term from the ${\cal O}(p^4)$ HLS Lagrangian is performed; Appendix \ref{AA-3}
displays its full expression in the BHLS$_2$ context in terms of the ideal combinations
of the vector fields. Actually, the existence of such a term
 was implicitly accounted for in the former BHLS \cite{ExtMod3} by disconnecting
the subtraction polynomial for the $\gam \ra V$ transitions from the subtraction
polynomials accompanying the pion and kaon loops occuring elsewhere  -- notably
inside the $\rho$ propagators.

Within the BHLS$_2$ framework, the loop corrections to the $\gam \ra V$ transition amplitudes
and to the $\rho^0$ self-mass are tightly related\footnote{Compare Subsection 
\ref{pionFF} and Appendix \ref{AA-5}.} and additional terms
 of the form $c_{V\gam} z_3 s$ come into play -- see Equations (\ref{eq1-51b}).
 
 Table \ref{Table:T6} shows that $z_3$ is small but significant.  As the $\gam \ra V$
 transition amplitudes, where solely $z_3$ occurs, affect all the cross sections
 considered,  the most appropriate way to examine its effects is to compare fits 
 where $z_3$ is let  free and fits performed by fixing $z_3=0$.
 
 The   overall picture coming from these fits is  reported just below. To be concise,
 one denotes here by BHLS$_2(z_3)$ and BHLS$_2(z_3=0)$ the RS based  fitting framework
 running with resp. $z_3$ free and $z_3=0$.
  \begin{itemize}
 \item When all data and channels considered in this paper  are used (the 3--pion channel 
 is discarded, for simplicity),
 one gets $\chi^2(\mathrm{BHLS}_2(z_3)/N_{data} =982/1079$ (93.1\% probability),
 whereas $\chi^2(\mathrm{BHLS}_2(z_3=0)/N_{data}=1049/1079$ (50.6\%  probability),
 a significant degradation.
 
 A careful comparison  of the fit results obtained in both cases
  shows that the partial $\chi^2$ for the $e^+e^-$ annihilation channels 
  ($\pi^+ \pi^-$, $K^+ K^-$, $K_L K_S$, $\pi^0 \gam$, $\eta \gam$
 and even the $\tau$ dipion spectra) are comparable\footnote{ BHLS$_2(z_3)$ 
 remains, nevertheless, favored against BHLS$_2(z_3=0)$ in these channels.}. 
 
 \item The bulk of the degradation when going from BHLS$_2(z_3)$ to BHLS$_2(z_3=0)$
 is located in the account by the latter of the spacelike sector where the NA7 data 
 impose accurate constraints, hardly absorbed by BHLS$_2(z_3=0)$.
 
 Indeed, for the spacelike experimental data, the fit performed with BHLS$_2(z_3=0)$
 returns  $\chi^2(\mathrm{spacelike}~\pi)/N_{data} =78/59$
 and $\chi^2(\mathrm{spacelike}~K)/N_{data} =29/25$, whereas the corresponding numbers
 in the  (standard) BHLS$_2(z_3)$ framework
 are resp. $56/59$ and $18/25$, in perfect agreement
 with the information already reported in Table \ref{Table:T4}.
 
 To fully compare with the information reported in Table  \ref{Table:T4}, a  
 BHLS$_2(z_3=0)$ fit, based on only the timelike data ({\it i.e. }
discarding the spacelike data), is useful as it provides  
the  corresponding BHLS$_2(z_3=0)$ {\it predictions} for the spacelike sector.
 
This BHLS$_2(z_3=0)$  global fit has been performed and displays an improved 
fit quality, as $\chi^2(\mathrm{BHLS}_2(z_3=0)/N_{data} =939/995$ 
(72.6\% probability), nevertheless lower than those for BHLS$_2(z_3)$. The $\chi^2$ distances 
of the $\pi$ and $K$ spacelike data to the corresponding BHLS$_2(z_3=0)$ {\it predictions}  
are $\chi^2(\rm{spacelike}~\pi)/N_{data} =85/59$ and
  $\chi^2(\rm{spacelike}~K)/N_{data} =30/25.$  
  
  These two numbers  should
   be compared with the first  data column in Table \ref{Table:T4}. 
   So, the nice feature of BHLS$_2(z_3)$ which predicts accurately the spacelike data
   is lost with BHLS$_2(z_3=0)$.
\end{itemize}
One should note that, when $z_3=0$ is imposed, the effects accounted for by $z_3 \ne 0$
are absorbed by the vector mesons propagators\footnote{For instance $g$ jumps from $\simeq 7$
to $\simeq 10$ and the pion and kaon loop subtraction polynomials are severely modified.  };
this is what degrades the continuation
of the form factors to the spacelike region and, fitting the spacelike data, does not solve 
the issue.

Stated otherwise, a non-zero $z_3$ is the best way to treat simultaneously the spacelike
and timelike data with high accuracy and comfortable fit probabilities. This may be considered
as an evidence pointing toward the relevance of the ${\cal O}(p^4)$ terms; a range for $z_3$
can be proposed  $z_3 = -([0.37\div 0.42] \pm 0.003_{fit})\times 10^{-2}$ 
for its  first experimental determination.

\subsection{Other Physics Results}
\indent \indent
As for other model parameters, focusing on BHLS and the RS variant of BHLS$_2$~:
\begin{itemize}
\item One may wonder that the value for the vector coupling $g$ increases by more than 20\%  when going from BHLS to
BHLS$_2$, keeping good fit qualities.
However, the relevant piece of information, closer to observables, being the coupling
$g^{\rho^0}_{\pi \pi} =ag(1+\Sigma_V)(1+\xi_3)/2$, it is interesting to compare BHLS to the RS
variant of BHLS$_2$~: BHLS returns $g^{\rho^0}_{\pi \pi} = 6.156 \pm 0.005$
while  the full BHLS$_2$ fits gets $g^{\rho^0}_{\pi \pi} = 6.06 \pm 0.013$ in reasonable
correspondence  with each other.
\item The parameters associated with the anomalous sectors show that\footnote{In the present work,
 one
assumes $c_3=c_4$.} $(c_3+c_4)/2$ yields comparable values  with BHL and with BHLS$_2$. However, for
$c_1-c_2$, the difference is large; in this case, however, one should remind that the
so-called "model B" variant of BHLS  excludes from the fit the $3 \pi$ data in the $\phi$ mass
region while the whole $3 \pi$ spectra are considered within BHLS$_2$ and are well
fitted.

\item The values for $z_A$ are also interesting as they are related with $f_K$. Indeed, one knows
that \cite{ExtMod3}~:
$$ r=\frac{f_{K^\pm}}{f_{\pi^\pm}} = \sqrt{z_A} \left [1+ \frac{\Delta_A}{4} \right]~~.$$
is still valid within the BHLS$_2$ framework. For this ratio, BHLS returns $r=1.257 \pm 0.003_{fit}$, 
while the RS variant of BHLS$_2$  finds $r=1.247 \pm 0.003_{fit}$. 
The systematic uncertainty has been estimated to $\pm 0.020$ by varying the running
conditions of the BHLS$_2$ fitting 
code\footnote{We made additional runs  fixing $z_V=1$ (see Table \ref{Table:T6}), 
including  the CMD-3 dikaon data and, finally, varying also the running of $\alpha_{em}$.
}.

 This result should be compared  with the
Review of Particle Properties \cite{RPP2010} which recommends $r=1.197 \pm 0.002 \pm 0.006\pm 0.001$ and
with the LQCD determination for $N_f=2+1+1$ flavors \cite{FLAG_2016} $r=1.195 \pm 0.005$.

\item The additional parameters introduced by the CD breaking and the Primordial Mixing are all found
at the few percent level. The difference for $\xi_0$ and $\xi_3$ between the (full) RS variant 
and the BS variant should be noted; it may have to be revisited later on, when a
complementary mechanism may allow one to absorb satisfactorily the $\tau$ data within the 
BS global fit procedure. This may also influence the $3 \pi$ sector.
\end{itemize}

\section{The muon LO-HVP Evaluation}
\label{LO-HVP}
\indent \indent  BHLS$_2$ (or BHLS) fits provide the contributions of
the various hadronic channels\footnote{All along  this Section, one may use
interchangeably $a_\mu^{\rm HVP-LO}$ or, preferably,  $a_\mu$ to lighten writing; the context will be
unambiguous.} it covers
 to  $a_\mu(\sqrt{s}\le 1.05 ~{\rm GeV})$. An overview of the main
results, obtained in different running conditions is given in Table \ref{Table:T7}.
These have been derived by submitting to fit most of the
data samples commented in Section \ref{DataSamples} which define a set of
consistent samples covering the six annihilations considered;
altogether, these 6 channels and the 10 radiative meson decay partial widths represent 1237 data points.

\subsection{Comparison of BHLS$_2$ Estimates for  $a_\mu^{\rm HVP-LO}$ with Others }
\indent \indent  Without going lengthily into detailed comparisons, it looks worthwhile to
provide some elements. For instance, one can extract from Figure 3 in \cite{ExtMod5},  values
for $G=a_\mu^{\pi\pi}(0.630 \!<\!\sqrt{s}\!<\!0.958~{\rm GeV} )$;  our updated
estimate for  $G$ is  $G({\rm BHLS})= (355.59 \pm 0.58)\times 10^{-10}$, very close to its former
evaluation (see Fig. 3 in \cite{ExtMod5}) while the numerical integration of the NSK+KLOE
data was estimated $G({\rm exp})= (356.67 \pm 1.69)\times 10^{-10}$; BHLS$_2$
gives\footnote{As the $(0.630 <\sqrt{s}<0.958)$ GeV  interval is relatively far
from both the threshold and the $\phi$ mass regions, one
could indeed expect similar results from BHLS and BHLS$_2$. }
$G({\rm BHLS}_2)= (356.50 \pm 0.75)\times 10^{-10}$.

In a quite different theoretical context, the recent analysis \cite{Colangelo:rpi}, provides the estimate
$G=a_\mu^{\pi\pi}(\sqrt{s} <0.630~{\rm GeV} )=(132.5  \pm 0.4_{\rm stat}
\pm 1.0_{\rm syst})\times 10^{-10}$ while the
BHLS gives $G({\rm BHLS})=a_\mu^{\pi\pi}(\sqrt{s}<0.630~{\rm
GeV})=(130.80 \pm 0.22_{\rm fit})\times 10^{-10}$.
In the same region
one also gets $G({\rm BHLS}_2)=a_\mu^{\pi\pi}(\sqrt{s}<0.630~{\rm
GeV})=(130.03 \pm 0.18_{\rm fit}
+[~^{+0.25}_{-0.15}]_{\rm syst})\times 10^{-10}$.

Reference  \cite{Colangelo:rpi} also provides
$H=a_\mu^{\pi\pi}(\sqrt{s}<1.0~{\rm GeV})= (494.7 \pm 1.5_{\rm stat} \pm
2.0_{\rm syst})\times 10^{-10}$
while the updated BHLS gives~ $H({\rm BHLS})=(490.65\pm 0.84_{\rm fit})\times 10^{-10}$ 
and BHLS$_2$ leads to
$H({\rm BHLS}_2)=(490.75\pm 0.90_{\rm fit}+[~^{+0.31}_{-0.26}]_{\rm syst}))\times 10^{-10}$. The method in
 \cite{Colangelo:rpi} and ours are different, but more importantly for this energy range, the
  dipion BaBar data included in the study \cite{Colangelo:rpi} and discarded
 in ours, certainly contribute to the difference.

One may also compare the BHLS$_2$ results with others for the contribution to $a_\mu^{\rm HVP-LO}(\pi^+ \pi^-)$ from
the energy region $(0.6 \le \sqrt{s} \le 0.9) $  GeV.  KNT18 \cite{Teubner3} gets $(369.41 \pm 1.32)\times 10^{-10}$ 
and  CHS \cite{Colangelo:rpi} $(369.8 \pm  1.3_{fit}  \pm 1.3_{syst})\times 10^{-10}$ using 
BaBar data at low energies. BHLS2 running
with KLOE10/12, CMD-2, SND, BESIII and Cleo-c data in the $\pi^+ \pi^-$ channel yields
$(367.81 \pm 0.77)\times 10^{-10}$; replacing the KLOE10/12 data by the KLOE85 combination
\cite{KLOEcomb}, one gets $(367.61 \pm 0.64)\times 10^{-10}$. So the
observed effect from BaBar in this energy region amounts to $\simeq 2\times 10^{-10}$. One also
observes the gain on accuracy by using BHLS$_2$~: A factor of 2 compared to  \cite{Teubner3},
a factor of 3  compared to  \cite{Colangelo:rpi}.

\subsection{Estimates for  $a_\mu^{\rm HVP-LO}(\sqrt{s} \le 1.05~{\rm GeV})$}
\label{a_mu_le_105}
\indent \indent  In order to get the full contribution to $a_\mu^{\rm HVP-LO}$ of the energy
region up to 1.05 GeV,
one should complement the  contribution from the channels covered by BHLS/BHLS$_2$ by those
of the channels presently outside this framework ($4 \pi$, $2 \pi \eta$, \ldots). The most recent
evaluation of this quantity, estimated by direct numerical integration of data  is~:
$$10^{10} \times a_\mu({\rm non~HLS}, \sqrt{s} \le 1.05~{\rm GeV}) = 1.28 \pm 0.17\epo$$

In view of model effect studies, it has been found motivated to update and correct
the data samples submitted to the BHLS fit  in the way stated in Section \ref{DataSamples}. One
has thus derived  the updated BHLS evaluations given in the first data column of
Table \ref{Table:T7}; this supersedes the BHLS results formerly given in Table 4 of
\cite{ExtMod5}. Differences between BHLS and the RS BHLS$_2$ variant for some channel contributions 
are observed and, {\it in fine},
$a_\mu({\rm HLS}, \sqrt{s} \le 1.05~ {\rm GeV})$ only increases by  $\simeq 1 \times 10^{-10}$ while
its uncertainty slightly decreases ($1.03 \ra 0.92$, in units of $10^{-10}$).
An evaluation of  possible additional systematics, specific of the BHLS modeling, has
been performed and summarizes by \cite{ExtMod5}~:
\be
10^{10} \times \delta_{\rm syst} a_\mu({\rm BHLS}) = \left [~^{+1.3}_{-0.6} \right]_\phi + \left [~^{+1.4}_{-0.0} \right]_{VNSB}
+\left [~^{+0.0}_{-0.9} \right]_{\tau}~~~.
\label{systBHLS}
\ee

Each piece shown here was found to  act by shifting the central value for $a_\mu$  rather than enlarging its
uncertainty. The first term refers to the uncertainty in the treatment of
the $\phi$ mass region in the 3-pion spectra, the second term takes into account
the uncertainty on the $\pi \pi$ threshold  behavior within BHLS (see Subsection \ref{pi0g_loop}
above and, especially, Figure
\ref{Fig:alpha_angle}); finally, the third piece reflects differences observed by running
the BHLS fit procedure \cite{ExtMod3,ExtMod4} with the $\tau$ data included
and excluded.

As already stated, the BHLS$_2$ modeling has been motivated by the aim to cancel out the
first two sources of uncertainties reported in Equation (\ref{systBHLS}) by a better
understanding of the $s=0$ region and a full account of the 3-pion data
up to the $\phi$ region.

For this purpose, the three data columns in Table \ref{Table:T7} referring to BHLS$_2$ show
that~:
\begin{itemize}
\item The evaluation for the various contributions to $a_\mu(\sqrt{s} \le 1.05~ {\rm GeV})$
derived using the BS and both RS variants of   BHLS$_2$ carries differences at a
few  $10^{-12}$ level.
\item Using  the $\tau$ data has some  effect. Indeed, Table \ref{Table:T7}
shows that running the RS variant of BHLS$_2$ including the $\tau$ data improves
the uncertainty in the dipion channel by $\simeq 10 \%$ while leaving the central 
value almost unchanged. This illustrates that the  $e^+e^-$ and $\tau$ 
data are not conflicting within the RS variant of   BHLS$_2$.
\end{itemize}
Another property exhibited in Table \ref{Table:T7} should be noted~:
\begin{itemize}
\item Whereas the sum of the various contributions to $a_\mu$ derived using
several variants of the BHLS$_2$ fit almost coincides with the estimate derived
by direct integration of the data, the $\pi^+ \pi^-$ contributions differ by 3.72
in units of $10^{-10}$, a significant difference.  

In the direct integration method, the normalization of each sample is the nominal one 
and {\it all}  uncertainties participate to the definition of its weight in the combination
with the other samples involved. Instead, in the global fit method, the absolute
normalization uncertainty is treated specifically and in 
consistency for all the data samples involved in all the annihilation channels considered
($\pi \pi$, $K \overline{K}$, $\pi \pi \pi$ and $(\pi^0/\eta) \gamma$). It has beeen
shown \cite{ExtMod5} that the iterative method which underlies the HLS global fit procedures
gives unbiased results \footnote{This is well reflected, in the present work, by the nice
matching observed between the predicted pion form factor in the spacelike region and the NA7 and ETMC
data.}. 

Therefore, that values of definite contributions to $a_\mu$ 
may significantly depend on the method used to derive them is not totally
unexpected. That the
sum of them may almost coincide  is, however, accidental\footnote{ 
In this respect, it is worth noting  that a quite 
similar behavior is observed in the comparison of the KNT18 and DHMZ17 data reported in 
\cite{Teubner3}. Indeed, Table 5 herein shows that the estimated dipion contributions (based on
the same data) differ by $3.40 \times 10^{-10}$, whereas the total sums giving the  
HVP-LO's differ by only
$0.2\times 10^{-10}$.}.
\end{itemize}

The data update having been performed, BHLS and BHLS$_2$ have been run on the same
data and, thus, the differences between their respective evaluations reflect modeling effects.
One can also compare these differences to
the  numbers listed in Equation (\ref{systBHLS}).

As shown by the
various topics examined in Section \ref{spacelike}, one can legitimately consider
the  $\pi^+ \pi^-$ threshold region accurately treated.
Therefore, the difference
$494.08-493.73=0.35$ between the BHLS$_2$ and BHLS estimates for
$a_\mu(\pi^+ \pi^-, \sqrt{s} \le 1.05~ {\rm GeV})$  is a fairly well 
motivated evaluation
of the unaccounted for nonet symmetry breaking effects within the BHLS framework;
it is therefore justified  to replace the  former $+1.4$ estimate
in Equation (\ref{systBHLS}) by $+0.35     $.

Table \ref{Table:T3} shows that all variants of BHLS$_2$ take
well into account the $3 \pi$ annihilation data over the whole energy domain up to the
$\phi$ mass region, the fit quality of the RS variant being noticeable.
The difference
between BHLS$_2$ and BHLS for this channel $\simeq 0.5 \times 10^{-10}$ is in the range
expected from the estimate \cite{ExtMod5} reported by the first term in Equation (\ref{systBHLS}).
Nevertheless, the $0.06 \times 10^{-10}$  difference for the $3 \pi$ entry
between the RS variant of BHLS$_2$ runnings
including and excluding the $\tau$ data  might  be considered as
additional  systematics\footnote{There might be a numerical effect
reflecting the  physics correlation between the $\pi^+ \pi^- \pi^0 $ annihilation
channel and the $\tau$ spectrum due to their common $\rho^\pm$ exchanges
taking place in their intermediate states. }.

It is interesting to compare the outcomes from fits to the values derived by the direct integration
of the same data sample set treated in the BHLS$_2$ framework. One may note that the sum of all HLS channel
contributions differs from the fit expectations by only
$\simeq 0.6 \times 10^{-10}$; however, this hides the fact that the dipion
channel is fitted
$3.7 \times 10^{-10}$ smaller and the 3-pion channel is fitted $1.70 \times 10^{-10}$ larger.

In summary~:
\be
10^{10} \times a_\mu({\rm HLS}, \sqrt{s} \le 1.05~ {\rm GeV}) = 571.98 \pm 1.20_{fit}
\label{amu_hls}
\ee
is the most conservative  final  BHLS$_2$ answer, where the uncertainty collects statistical and systematic
errors as reported by the various experiments. The remarks just stated concerning the
3-pion channel
allows to keep an additional uncertainty  of $-0.06 \times 10^{-10}$,
playing as a bias.

The issue is to examine whether more significant additional sources of systematics
are not at work. For this purpose, the consequence on $a_\mu({\rm HLS})$ of the
various fits referred to in the previous Section carry a relevant piece of information,
as also for the fits underlying the results given in Equations (\ref{degeneracy})
and (\ref{degeneracy2})  or those
displayed in the last data column in Table \ref{Table:T6}. These fits, covering a wide
range of parameter values, are relevant for systematics estimates.

 \begin{table}[!ptbh!]
{\small
\begin{tabular}{|| c  || c  | c || c | c|| c ||}
\hline
\hline
\hhhd Channel  & \hhhv BHLS   & \hhhv  BHLS$_2$ (BS) &  \hhhv  BHLS$_2$ (RS)  &  \hhhv  BHLS$_2$ (RS) &  \hhhv Data Direct  \\ 
\hhhd~~~~  &  \hhhw ~~~~    &  \hhhw (excl. $\tau$)  &  \hhhw (incl. $\tau$)  &  \hhhw (excl. $\tau$) &  \hhhw Integration \\ 
\hline							    							 
\hline
 \hhhv $\pi^+ \pi^-$    &     $493.73 \pm 0.70 $     &  $494.04 \pm 0.76$        & $494.08\pm 0.90$      & $494.10\pm  1.0$         & $497.82\pm 2.80$     \\
\hline
 \hhhv $\pi^0 \gam$     &     ~~$4.42 \pm 0.03$      &  ~~$4.46 \pm 0.03$	 & ~~$4.45 \pm 0.03$     & ~~$4.45 \pm 0.03$      &  ~~$3.47 \pm 0.11$   \\
\hline
 \hhhv $\eta \gam$      &     ~~$0.63 \pm 0.01$      & ~~ $0.63 \pm 0.01$	 & ~~$0.64 \pm 0.01$     & ~~$0.64 \pm 0.01$      & ~~$0.55 \pm 0.02$         \\
\hline
 \hhhv $\pi^+ \pi^- \pi^0 $ & ~$42.56\pm 0.54$        &  ~$42.97 \pm 0.53$	 & ~$43.02 \pm 0.54$     & ~$43.08 \pm 0.48$    & ~$41.38 \pm 1.28$
       \\
\hline
 \hhhv $K^+ K^- $      &      ~$18.10 \pm 0.14$       &  ~$18.04 \pm 0.14$	 & ~$18.01\pm 0.14$       & ~$17.99\pm 0.14$    & ~$17.37 \pm 0.55$       \\
\hline
 \hhhv $K_L K_S$       &      ~$11.53 \pm 0.08$       &  ~$11.70 \pm 0.08$ 	 & ~$11.71 \pm 0.08 $	   & ~$11.72 \pm 0.09 $   & ~$11.98 \pm 0.36$    \\
\hline
\hline
 \hhhv HLS Sum         &      $570.97 \pm 0.92$       &  $571.84 \pm 1.06$	 & $571.90\pm 1.10$    & $571.98\pm 1.20$       & $572.57 \pm 3.15$     \\
\hline
\hline
 \hhhv $\chi^2/N_{\rm pts}$ &     $949/1056$           &  $1054/1152$    & 1133/1237  & 1052/1152 & $\times$   \\
 \hhhw Probability      &           96.7\%             & 93.9 \%        & 93.3 \%         & 93.6 \%    &  $\times$     \\
\hline
\hline
\end{tabular}
}
\caption {HLS contributions to $10^{10} \times a^{\rm HVP-LO}$
integrated up to 1.05 GeV, including FSR. The first data column displays the results
using the former BHLS \cite{ExtMod3,ExtMod5} and, the second one, those derived from the Basic
Solution for BHLS$_2$,  the $\tau$ decay data being discarded.
The next two data columns refer to the results obtained using the Reference Solution for
BHLS$_2$ using the largest set of data samples, keeping or discarding   the $\tau$ data.
The last data column refers to the  numerical integration for each channel of the same set of
data which are used in the BHLS/BHLS$_2$ fits.
\label{Table:T7}
}
\end{table}

The runs leaving aside the 3-pion data briefly reported in Equations (\ref{degeneracy}), show
that the  dominant $\pi \pi$ contribution can be as low as
$(493.80\pm 0.89)\times 10^{-10}$, {\it i.e.}
$0.30$ units of $10^{-10}$
smaller than the corresponding reference entry in Table \ref{Table:T7}. Nevertheless, the middle
entry line in Equations (\ref{degeneracy}) is very close to Equations (\ref{degeneracy2}), where the
3-pion data have been included and exhibits negligible differences with respect to our
reference in Table \ref{Table:T7}. Finally, the fit corresponding to the last data column
in Table \ref{Table:T6} has also been analyzed and returns  a value for
$a_\mu(\pi^+ \pi^-, \sqrt{s} \le 1.05~ {\rm GeV}) = (494.32 \pm 0.90)\times 10^{-10}$.
This may indicate another additional systematic
able to shift the central value by, at most,  $0.22 \times 10^{-10}$
affecting our estimate    for   $a_\mu^{\rm HVP-LO}(\sqrt{s}\le 1.05~{\rm GeV})$ given in Equation (\ref{amu_hls}).

The possible sources of additional systematics just emphasized can be considered 
as a model uncertainty bounded by $(+0.22,-0.36) \times 10^{-10}$. On the other hand, running the 
fitting code using the available parametrizations for $\alpha_{em}(s)$, returns differences for
the estimates for $a_\mu({\rm HLS})$ of $\simeq 0.4 \times 10^{-10}$, mostly located in
in the $\phi$ mass region.

Finally, the question of using the CMD-3 data can be raised. We have run our code
using the CMD-3 data \cite{CMD3_KpKm,CMD3_K0K0b} for both $K \overline{K}$ decay
channels, discarding the kaon data from BaBar, CMD-2 and SND and using only the diagonal part
of their error covariance matrices (see Section \ref{kkb_data} above).
In total, the change for
$a_\mu(\sqrt{s} \le 1.05~ {\rm GeV})$ is noticeable $(572.52 \pm 1.12)\times 10^{-10}$,
$0.54\times 10^{-10}$ larger than our reference in Equation (\ref{amu_hls}), essentially coming from
 the $K^+ K^-$ and $K^0 \overline{K}^0$ channels. On the other hand, combining CMD-3,
 BaBar, (corrected) CMD-2 \cite{Mainz2018} and SND, leads to
$a_\mu(\sqrt{s} \le 1.05 ~ {\rm GeV})=(572.52 \pm 1.04)\times 10^{-10}$. This leads us to complete
our estimate Equation (\ref{amu_hls})~:
\be
\displaystyle
10^{10} \times a_\mu({\rm BHLS_2}, \sqrt{s} \le 1.05~{\rm GeV}) =
571.98 \pm 1.20_{\rm fit} +
\left[^{+1.16}_{-0.75} \right]_{\rm syst}\;\;,
\label{amu_hls2_f}
\ee
where the additional systematics come from model variations, $\alpha_{em}$
parametrizations, and data sample consistency issues. The model uncertainties include the
marginal effect attributable to  the $\tau$ data, as this could reflect some
(marginal) isospin breaking shortcoming.
We have chosen conservatively as reference  the RS variant fit excluding the $\tau$ data.
One should also note that the tension between dikaon data samples introduces
non-negligible systematics which  contribute a bias. This may have to be revisited
when new data will arise.

Taking into account the data upgrade, the corresponding quantity for BHLS is~:
\be
\displaystyle
10^{10} \times a_\mu({\rm BHLS}, \sqrt{s} \le 1.05 ~\rm{GeV}) = 570.97
\pm 0.92_{\rm fit} +
\left[^{+1.16}_{-0.75} \right]_{\rm syst}+\left[^{+0.0}_{-0.9} \right]_{\tau}\;\;,
\label{amu_hls_f}
\ee
using  the present findings to go beyond the former systematics evaluation
reminded in Equation (\ref{systBHLS}). The various additional uncertainties just 
estimated  represent conservative upper bounds.
 
\subsection{Dispersive vs LQCD Methods~: Additional Systematics}  
\indent \indent
Experimental $e^+e^-\to {\rm hadrons}$ data are dressed by photon
radiative corrections and disentangling strong interaction from the
electromagnetic effects is a non-trivial task. What is ideally needed
as input to the dispersion integrals is the one-particle
irreducible (1PI) hadronic part. It is given in QCD by the correlator
between two hadronic currents, and this is what is primarily
calculated as the LO part of order $O(\alpha^2)$ in lattice QCD. We
remind that the bare cross-section undressed from photon radiation
effects is not an observable by itself. It requires some theory input,
primarily the photon radiative corrections, for its extraction from
the physical dressed data. Mandatorily, one has to  drop out the photon
vacuum polarization effects to get the undressed bare
cross-section. By convention, one also includes final state radiation (FSR)
in the dipion channel by adopting scalar QED (sQED) for its
calculation. Similarly, the radiative decay channels $\pi^0\gamma$ and $\eta\gamma$ 
are included in our evaluation. Like FSR, they are related to 
light-by-light scattering insertions.

However, there are other important QED effects like the
$\rho^0-\gamma$ interference which is substantial in the dipion
channel because this QED effect is magnified by the resonance
enhancement in the $\rho$ region. In the standard dispersive approach,
$\rho^0-\gamma$ mixing effects -- inherent in the $e^+e^-$ data -- are
treated as part of the LO-HVP contribution and are included as well in
the calculations of the higher-order contributions. An updated
$e^+e^-$ data-based evaluation of the LO and the NLO parts yields the
results shown in Table~\ref{tab:LO+NLOold}\footnote{For the present
exercise, the BaBar dipion data have been taken into account}.
\begin{table}[h]
\centering
\begin{tabular}{lrrrrr}
\hline
        &   & $a_\mu\times10^{-10}$& stat & syst & tot  \\
\noalign{\smallskip}\hline\noalign{\smallskip}
LO  &           &   690.93 & 0.71 & 3.83 & [3.90]\\
NLO & diagram a &   -20.71 & 0.03 & 0.13 & [0.13]\\
NLO &         b &    10.39 & 0.01 & 0.06 & [0.06]\\
NLO &         c &     0.34 & 0.00 & 0.01 & [0.01]\\
NLO & sum       &    -9.98 & 0.03 & 0.07 & [0.08]\\
LO+NLO &        &   680.95 & - ~~   & - ~~   & [3.82]\\
\noalign{\smallskip}\hline
\end{tabular}
\caption{LO- and NLO-HVP results in the standard dispersive approach
based on $e^+e^-$ data. The NLO diagrams are shown in
Fig.~\ref{Fig:amu_hvp_ho}.}
\label{tab:LO+NLOold}
\end{table}

\noindent
Here we have to remind that, in contrast to the usual dispersive
estimations, LQCD calculations are based on the hadronic current
correlator such that $\rho^0-\gamma$ mixing is not included in the
leading order LQCD results (unless it is added by hand from
phenomenological analyses, if not calculated separately).  

So the
question of how to disentangle the QCD effects from QED ones, has to be
reconsidered under this aspect. One possibility is to apply effective
field theory methods like the resonance Lagrangian approach, the other
is lattice QCD, the only method which can answer the question from
first principles.  In fact, to disentangle better QCD from QED effects
in the $e^+e^-$ data-based dispersive approach, one has to subtract
the $\rho^0-\gamma$ interference from the $e^+e^-\to \pi^+\pi^-$
data\footnote{The similar $\omega-\gamma$ and $\phi-\gamma$ effects
are much smaller and can be neglected.}, expecting to get the effect
back as a part of the higher-order contributions.

\begin{figure}[!hptb]
\hspace{0.5cm}
\begin{minipage}{0.9\textwidth}
\begin{center}
\resizebox{1.0\textwidth}{!}
{\includegraphics*{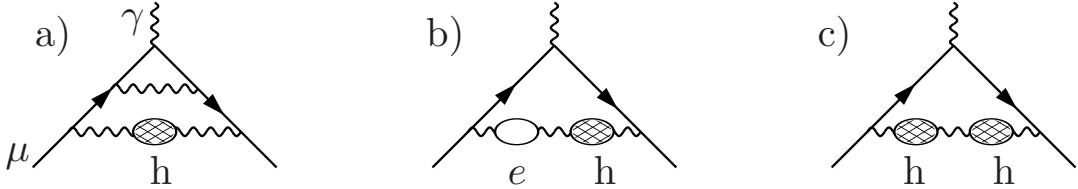}}
\end{center}
\end{minipage}
\begin{center}
\caption{\label{Fig:amu_hvp_ho} Higher order contributions
to the hadronic VP; the shadded blob depicts the LO hadronic VP,
the empty blob the LO leptonic VP.  }
\end{center}
\end{figure}
Here  is the place to remind that the $\rho^0-\gamma$ mixing is
responsible for the discrepancy between the isospin rotated isospin
breaking corrected $\tau$ spectral-function and the $e^+e^-$
spectral-function~\cite{DavierPrevious1,DavierPrevious2,DavierHoecker};
this has been shown in a straightforward ``VMD-II+sQED'' simple
effective resonance Lagrangian model \cite{Fred11} as well as in the
BHLS approach \cite{taupaper,ExtMod3} as utilized in the present
analysis. By convention, to bring $\tau$ data closer to $e^+e^-$-data,
it looked natural to correct the $\tau$-data to conform with the
commonly adopted dispersive approach based on the $e^+e^-$ data. 

In
fact, the measured $\tau$-decay dipion spectra are devoid of photon
interference effects. According to what we just have been arguing,
 one should actually not apply this mixing correction to the
$\tau$-decay data, as it has been done within our BHLS global fit
framework or in the much simpler model applied in~\cite{Fred11}. One
 should better subtract the $\rho^0-\gamma$ mixing from the $e^+e^-$ data,
to get an object closer to a pure hadronic ``blob'' and then
proceed as usual. As a possible subtlety, this requires to work
in the ``VMD-I'' basis where the $\rho^0-\gamma$ system is
parametrized in terms of fields where a direct lepton-$\rho$ coupling
is absent. This is the basis  of the BHLS Lagrangian
formulation. Being proportional to the hadronic production
cross-section, the hadronic 1PI amplitude and the corresponding $R(s)$
(see Eq.~(\ref{eq1}))
do not depend on rotations in the $(\gamma,\rho^0)$ field space. 

The
field rotation invariance has been directly confirmed by comparing, on
the one hand, the ``VMD-II + sQED'' modeling adopted
in~\cite{Fred11}, which uses the mass eigenstate basis where a
direct $\rho$-lepton coupling is induced by the necessary field
rotation~\cite{Kroll:1967it} and,  on the other hand, our BHLS modeling
where a direct $\rho$-lepton coupling is absent. One thus can apply the
standard diagrammatic expansion of Fig.~\ref{Fig:amu_hvp_ho}
with the modified effective $R(s)$-function and
obtains the results presented in
Table~\ref{tab:LO+NLOnew}~:
\begin{table}[h]
\centering
\begin{tabular}{lrrrrrcr}
\hline
         &  & $a_\mu\times10^{-10}$& stat & syst & tot & shift \\
\noalign{\smallskip}\hline\noalign{\smallskip}
LO  &          &  694.11 & 0.71 & 3.86 & [3.93]  & +3.18 \\
NLO &diagram a &  -20.85 & 0.03 & 0.13 & [0.13]  & -0.14 \\
NLO &        b &   10.45 & 0.01 & 0.06 & [0.06]  & +0.05 \\
NLO &        c &    0.34 & 0.00 & 0.01 & [0.01]  & +0.00 \\
NLO & sum      &  -10.06 & 0.03 & 0.07 & [0.08]  & -0.09\\
LO+NLO &       &   684.05 & - ~~   & - ~~   & [3.85] & 3.10\\
\noalign{\smallskip}\hline
\end{tabular}
\caption{LO- and NLO-HVP results in the dispersive approach
based on $e^+e^-$ data, after having removed the $\rho^0-\gamma$ mixing 
 from the hadronic blob.}
\label{tab:LO+NLOnew}
\end{table}

\noindent
which corresponds to what, in principle, lattice QCD is doing. 

The LQCD
results displayed in Fig.~\ref{Fig:amu_hvp_lo} are full HVP results, 
which include $ud$
light quark, strange and charm contributions including the connected
and disconnected parts as well as isospin breaking contributions
either calculated on the lattice or adopted from phenomenological
estimates in terms of data. They do not include $\gamma-\rho$
mixing to our knowledge, however.  In lattice QCD this effect is counted 
as a NLO effect, which has to be evaluated separately\footnote{
Contributions from FSR and radiative decays should have been added 
in a consistent comparison with data-based dispersive results.  
}.

According to our Tables, the commonly adopted dispersive set up gets a
shift\footnote{The systematic uncertainties in Tables \ref{tab:LO+NLOold} and 
\ref{tab:LO+NLOnew} are 100\% correlated, except for the sub-dominant
diagram c.} of $\delta a_\mu [\pi\pi]_{\rho\gamma} \simeq +(3.10\pm 0.03)
\times10^{-10}$ at NLO, {\it i.e.}  a 1 $\sigma$ increase.
See also the results presented in Table 3
of~\cite{Jegerlehner:2017zsb} where a $\rho^0-\gamma$ mixing shift
of $2.74\times10^{-10}$ on the I=1 component at LO has been reported
as a difference between LQCD data and the 
phenomenological dispersive approach.The upgraded result 
for  $\delta a_\mu [\pi\pi]_{\rho\gamma}$
is based on an actualized data compilation, with mixing correction 
applied to the full vector form-factor and including NLO corrections. 

More or less surprisingly, 
the sum of LO+NLO differs in the two settings, {\it i.e.} the mixing effect
subtracted from the LO result does not show up as a higher-order
effect of comparable magnitude. It is actually not surprising because the
$\rho^0-\gamma$ mixing effect, if included in the LO dispersion
integral,  is larger than if taken into account in higher-order diagrams
where the effect is weighted by the higher-order kernel 
functions. 
Our result based on correcting the $e^+e^-$ di\-pion channel in the range up
to 1 GeV, where sQED estimates can be trusted, indirectly compares with
the deviation $(9.1 \pm 5.0)\times 10^{-10} (1.8
\sigma)$ between the $\tau$-based hadronic contribution to $a_\mu$ and
the $e^+e^-$-based one reported in~\cite{Davier_amu}. Our result does
not include $\tau$ data while the $\tau$ based result of~\cite{Davier_amu}
includes $\tau$ data up to 1.8 GeV which are combined with $e^+e^-$
data which still include $\rho^0-\gamma$ mixing. That the results are
increasing in both cases is as expected, but there is no reason why the
numerical results should agree.

Our suggestion to base the standard bookkeeping such as Figure \ref{Fig:amu_hvp_ho} 
on a purer QCD $\gamma-\rho$ undressed hadronic blob is motivated by the
fact that lattice QCD does not include any QED effects if not taken
into account as an extra contribution and also by the fact that
$\tau$-data are free of similar mixing effects, and it looks somewhat
artificial to add effects to $\tau$-decay spectral data that are
absent in the corresponding measurements. Nevertheless, as already
said, surprisingly we subtract a non-negligible contribution, which we
do not get back in higher orders. It is important to note that the
shift $\delta a_\mu [\pi\pi]_{\rho\gamma}$ is real as it derives from
the manifest discrepancy observed between $\tau$ and $e^+e^-$
spectral-function data in the corresponding time-like $\pi\pi$
production processes. As mentioned before, the $\gamma-\rho$
interference is substantial because of the manifest resonance enhancement 
in the time-like regime. Lattice QCD is taking advantage
of the fact that, for the calculation of $a_\mu^{\rm had}$, only the
space-like VP-function is needed. However, there is no $\rho$-resonance 
peak  in the Euclidean region which could enhance a
photon-exchange interference effect; the $\rho$-resonance is
completely smeared out and a resonance-enhancement cannot be localized
there in any obvious way. Because of a model-independent clean separation of
QCD is only possible in lattice QCD,  one has to wait for the relevant
LQCD results to get a better understanding of QED-QCD interference effects. 

One may conclude that {\it all } standard dispersive evaluations should 
either be upgraded by adding $+(3.10\pm 0.31) \times 10^{-10}$ to the 
LO+NLO results\footnote{Concerning the model dependence of
the prediction for the $\rho^0-\gamma$ interference, we remind that this
effect depends on only one extra parameter, the well known  
leptonic width of the $\rho^0$ or, equivalently, on the coupling
$g_{\rho ee}$, also well known experimentally. Since the effect is
small, a few units in $10^{-10}$, we can generously assign a very
conservative 10\% model uncertainty, without substantially increasing
the overall uncertainties coming with the $e^+e^- \to \pi^+ \pi^-$
data.}, or accounted for in the systematic errors by adding there
the amount of this shift. For the time being, we have adopted this second solution.

\subsection{BHLS$_2$ Evaluation of  $a_\mu^{\rm HVP-LO}$}
\indent \indent With the BHLS$_2$ and BHLS evaluations of $a_\mu^{\rm
HVP-LO}({\rm HLS}, \sqrt{s}\le 1.05$ GeV
displayed in Equations (\ref{amu_hls2_f}) and (\ref{amu_hls_f}) and the 
information given in the left
part of Table \ref{Table:T8}, one can derive the corresponding values
for  the full  $a_\mu^{\rm HVP-LO}$.
One thus gets~:
\be
\left \{
\begin{array}{lll}
\displaystyle
10^{10} \times a_\mu({\rm BHLS_2})& = 686.65 \pm 3.01_{\rm fit} +
\left[^{+1.16}_{-0.75} \right]_{\rm syst} + \left[^{+3.10}_{-0.0} \right]_{\rho \gamma}  \,,\\[0.5cm]
\displaystyle
10^{10} \times a_\mu({\rm BHLS })& = 685.64 \pm 2.91_{\rm fit} +
\left[^{+1.16}_{-1.65} \right]_{\rm syst} + \left[^{+3.10}_{-0.0} \right]_{\rho \gamma} \epo
\end{array}
\right .
\label{amu_hvplo}
\ee
In the case of BHLS, due to the inclusion of $\tau$ channel within the global fit
procedure, the corresponding systematics have been added linearly
to the others. Indeed, as stated several times for BHLS and BHLS$_2$, the observed systematics rather
come as biases and so, should be treated separately from the fit error which fully absorbs
the reported statistical and systematics  from the various experiments encompassed inside
each of our HLS frameworks. 

The additional systematics absorb uncertainties which can be attributed
to the model workings {\it and} to the (marginal) tension observed between some of the selected
experimental data samples. For clarity, the error coming from the $\rho^0-\gamma$
interference effect discussed just above is given separately.

\begin{figure}[!hptb]
\hspace{0.5cm}
\begin{minipage}{0.9\textwidth}
\begin{center}
\vspace{-1.5cm}
\resizebox{1.0\textwidth}{!}
{\includegraphics*{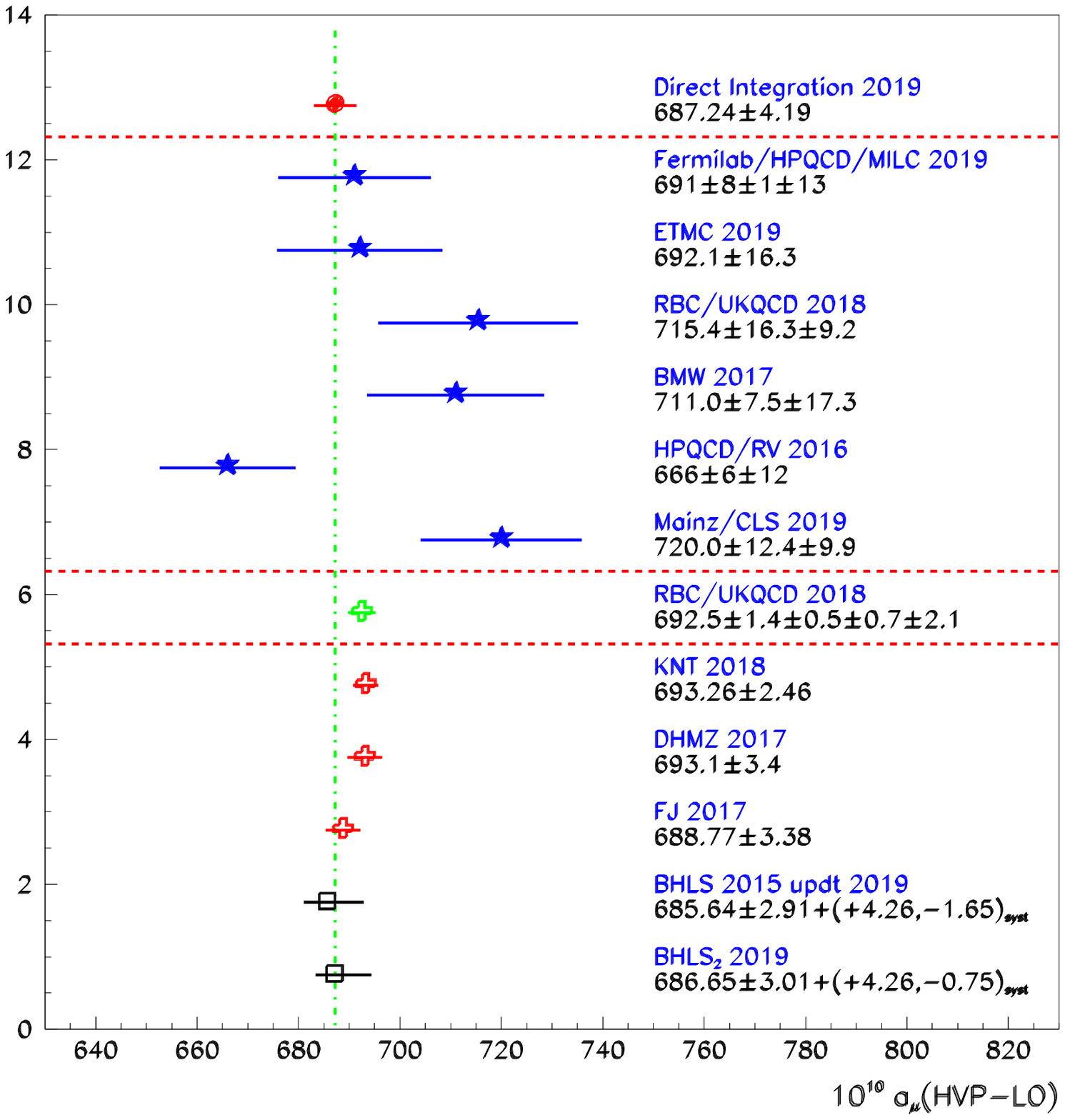}}
\end{center}
\end{minipage}
\begin{center}
\vspace{-1.cm}
\caption{\label{Fig:amu_hvp_lo}  Recent evaluations of $10^{10} \times
a_\mu^{\rm HVP-LO}$:
On top, the result  derived by direct integration of the data combined with perturbative
QCD; the next six points display some recent evaluations derived by LQCD methods
and reported in resp. \cite{FmHPQCD_MILC_amu}, \cite{ETMC_amu_up}, \cite{RBC_UKQCD_amu},
\cite{BMW_amu}, \cite{Chakraborty2016} and  \cite{Gerardin:2019rua}
with $N_f=2+1+1$. The second point from 
\cite{RBC_UKQCD_amu} displayed has been derived by supplementing lattice data with some
phenomenological information. These are followed by the evaluations
from \cite{Teubner3},\cite{Davier_amu} and \cite{Jegerlehner_2017}. The value derived
using BHLS \cite{ExtMod5} -- updated with the presently available data -- and the evaluation
from BHLS$_2$ are given with their full systematic uncertainties, including the $\rho^0 - \gamma$
mixing effect (see text). }
\end{center}
\end{figure}

Our evaluations for $a_\mu^{\rm HVP-LO}$, given by Equations (\ref{amu_hvplo}), are shown in
Figure \ref{Fig:amu_hvp_lo} together with some other recent results derived in different
phenomenological contexts \cite{Jegerlehner_2017,Davier_amu,Teubner3} or via LQCD data samples  with $N_f=2+1+1$. 
As for their specific
HLS modeling parts, the corresponding fit probabilities shown by the down-most entries in
Table \ref{Table:T7} are always fairly good.

\begin{table}[!htpb!]
\begin{center}
{\scriptsize
\begin{tabular}{|| c  | c || c  ||||c  | c || }
\hline
\hhhv Contribution from	&  Energy Range           &  $10^{10} \times
a_\mu^{\rm HVP-LO} $ & Contribution from & $10^{10} \times a_\mu $\\
\hline
\hline
\hhhq missing channels 	&  $\sqrt{s} \le 1.05$   &  $1.28 \pm 0.17 $
& LO-HVP  &    $686.65 \pm 3.01 +\left[^{+1.16}_{-0.75} \right]_{\rm syst} + \left[^{+3.10}_{-0.0} \right]_{\rho \gamma}$ \\
\hline
\hhhv $J/\psi$ 		& ~~~  			  &  $8.94\pm 0.59$   & NLO  HVP\cite{Fred11} & $-9.927  \pm 0.072 $\\
\hline
\hhhv $\Upsilon$        & ~~~  			  &  $0.11 \pm 0.01$   & NNLO HVP \cite{NNLO} &$~1.24  \pm 0.01 $\\
\hline
\hhhv hadronic		& (1.05, 2.00)		  &  $62.21 \pm 2.53$  & LBL \cite{LBL} &  $10.34 \pm 2.88$\\
\hline
\hhhv hadronic		& (2.00, 3.20)		  &  $21.63 \pm 0.93 $  & NLO-LBL \cite{LbLNLO} & $0.3 \pm 0.2$ \\
\hline
\hhhv hadronic		& (3.20, 3.60)		  &  $3.81 \pm 0.07 $   &  QED \cite{Passera06,FJ2013} &$11~658~471.8851 \pm 0.0036$\\
\hline
\hhhv hadronic		& (3.60, 5.20)		  &  $7.59 \pm 0.07$  &  EW \cite{Fred09} &$15.36\pm 0.11 $\\
\hline
\hhhq  pQCD	        & (5.20, 9.46)		  &  $6.27 \pm 0.01$  &  Total Theor. &$11~659~175.96\pm 4.17 
 +\left[^{+1.16}_{-0.75} \right]_{\rm syst}+\left[^{+3.10}_{-0.0} \right]_{\rho \gamma}$ \\
\hline
\hhhv hadronic     	& (9.46, 11.50)		  &  $0.87 \pm 0.05$  & Exper. Aver. &$11~659~209.1 \pm 6.3$ \\
\hline
\hhhq  pQCD		& (11.50,$\infty$)	  &  $1.96 \pm 0.00$ &
$10^{10} \times \Delta a_\mu$ &$33.14 \pm 7.56 -\left[^{+1.16}_{-0.75} \right]_{\rm syst}  - 
							\left[^{+3.10}_{-0.0} \right]_{\rho \gamma}$ \\
\hline
\hhhv Total		& 1.05 $\to \infty$	  &  {$114.67 \pm 2.76$}   & Significance ($n \sigma$) &$4.38 \sigma$\\
\hhhe ~~~		& + missing chann.	  &  {~~~} &  {~~~}		&~~~~~	\\
\hline
\hline
\end{tabular}
}
\end{center}
\caption{
\label{Table:T8} The left-side Table displays the contributions to
$a_\mu^{\rm HVP-LO} $  from the various energy regions,
including the contribution of the non-HLS channels from the $\sqrt{s} < 1.05$ GeV region; only total errors are shown.
The right-side Table provides the various contributions to $a_\mu $ and information for
$\Delta a_\mu=a_\mu({\rm exp})-a_\mu({\rm th})$ for the case of the RS variant of BHLS$_2$. The statistical significance
of $\Delta a_\mu$ is given at central value (see text).
 }
\end{table}

 As for modeling effects, the central values for BHLS and BHLS$_2$ -- running on the same set of data samples --
 are shifted apart by $\simeq 1.5 \times 10^{-10} $; however, if one takes into account the magnitude
 of the additional systematics, Figure \ref{Fig:amu_hvp_lo} clearly indicates that this shift is not significant
 and so no noticeable modeling effect shows up. Moreover, Figure \ref{Fig:amu_hvp_lo} clearly shows that
 both HLS based estimates for $a_\mu^{\rm HVP-LO} $ are consistent with the evaluation derived by numerical
 integration of the data\footnote{See the top point in Figure \ref{Fig:amu_hvp_lo}, derived by
 excluding the BaBar dipion sample for a consistent comparison with the BHLS entries. }.
 
 Actually, the estimates for the muon HVP-LO derived by the various dispersive methods (the five data points 
 displayed in the bottom  part of Figure \ref{Fig:amu_hvp_lo}) are quite comparable with each other 
 as they use as input to evaluate  $a_\mu^{HVP-LO}$
 the $e^+e^-$ annihilation cross sections as derived from the data or as reconstructed from fit. Therefore, 
 the results thus provided do not split out the electromagnetic corrections; 
 in particular the $\gamma - \rho^0$  mixing effects is merged in the plotted systematics 
only for the BHLS/BHLS$_2$ entries, not for the other dispersive results plotted.  

\subsection{BHLS$_2$ Evaluation of  Muon Anomalous Magnetic Moment}
\indent \indent
Having at hand our estimate   Equation (\ref{amu_hls2_f}), one can
derive the BHLS$_2$ evaluation for $a_\mu=(g-2)/2$, the muon anomalous magnetic
moment. The left part of Table \ref{Table:T8} collects the pieces to add up
with $a_\mu({\rm BHLS_2})$ to get the full leading  order HVP, its value is shown
as the top entry in the right-hand part of  Table \ref{Table:T8} where
the other contributions are included.  The value for
$\Delta a_\mu=a_\mu({\rm exp})-a_\mu({\rm th})$ and its statistical significance are 
the bottom entries there.
Taking into account all additional systematics, one thus gets~:
\be
\displaystyle
10^{10} \times a_\mu= 11\,659\,175.96 \pm 4.17_{\rm th} 
+\left[^{+1.16}_{-0.75} \right]_{\rm syst} +\left[^{+3.10}_{-0.0} \right]_{\rho \gamma} 
\label{amu_th_2}
\ee
and, taking into account the shifts produced by all systematics, one gets~:
\be
\displaystyle
28.88\pm 4.17 \le
10^{10} \times \Delta a_\mu (= a_\mu^{exp.}-a_\mu^{theor.})
\le 33.89  \pm 4.17\;\;,
\label{amu_th_2b}
\ee
which means that the statistical significance of $\Delta a_\mu$ is
predicted greater than $3.82 \sigma$. The shift just reported is partly due
to the estimates of the various modeling effects involved, partly due to somewhat conflicting
aspects of some data samples and parly due to the identified $\gamma-\rho^0$ mixing 
effects which add up linearly.

Our estimate for $a_\mu$ given in Equation (\ref{amu_th_2}) relies on a variant of the
 traditional estimate  \cite{LBL} for the Light-by-Light (LBL) contribution
($(10.34 \pm 2.88)\times 10^{-10}$). Danilkin, Redmer and Vanderhaegen (DRV)
have quite recently published a comprehensive analysis of all ingredients participating
to the LBL term and proposed for it the much more precise value \cite{Danilkin:LbL}
$(8.7 \pm 1.3)\times 10^{-10}$ which leads to~:
$$ 10^{10} \times a_\mu^{\rm DRV}= 11\,659\,174.78 \pm 3.28_{\rm th}
+\left[^{+1.16}_{-0.75} \right]_{\rm syst}  +\left[^{+3.10}_{-0.0} \right]_{\rho \gamma} $$
and to $\Delta a_\mu^{\rm DRV}=(34.32 \pm 7.10)\times 10^{-10}$. This makes
the significance for $\Delta a_\mu$ jumping from $(3.8\div 4.5)\, \sigma$ to $ (4.2\div 4.7)\,\sigma$.

\section{Summary and Conclusions}
\label{summary}
\indent \indent The present paper has addressed several topics covering modeling,
phenomenology and the evaluation of the muon anomalous magnetic moment which puts
forward, as well known, a hint for a physics effect beyond the Standard Model expectations.
For the sake of clarity,  each of these topics is considered separately.
\subsection*{From BHLS to BHLS$_2$}
\indent \indent The BHLS framework  \cite{ExtMod3}, although quite successful
\cite{ExtMod4,ExtMod5,ExtMod6}, has been shown to exhibit a non-optimum behavior
in limited and well identified kinematical regions~: The threshold region
and some aspects of the $\phi$ mass region; the former issue renders delicate the
continuation of form factors across the chiral point, the latter has led
to discard the $\phi$ mass region for 3-pion final state -- the so-called Model B   \cite{ExtMod3}.

In order to cure these diseases, one  has  initiated a new breaking
procedure of the general HLS framework \cite{HLSRef}.
For this purpose, besides the BKY mechanism \cite{BKY,Heath,Hashimoto}, one
has introduced a breaking mechanism taking place at the level of the covariant
derivative (CD) itself, {\it i.e.} inside the basic ingredient  of the HLS Model.
This (seemingly novel) CD breaking mechanism allows naturally the $\rho^0$, $\omg$ and
$\rho^\pm$ fields to carry different Higgs-Kibble masses. This takes its importance when
going to loop corrections to the vector meson mass matrix \cite{taupaper,ExtMod3} which
 generate  a dynamical mixing of the vector mesons  calling for an $s$-dependent
redefinition of the vector meson fields to recover mass eigenstates.

Thanks to its nonet symmetry breaking piece, the CD breaking scheme generates
a non-zero  $m_{\rho^0}- m_\omg$ difference which  naturally makes
the three $s$-dependent mixing angles to vanish at the chiral point. In this way,
the timelike to spacelike analytic continuations become smooth without any {\it ad hoc} trick.

Relying on BKY and the CD breaking, the Basic Solution (BS) is defined and represents
the first variant of BHLS$_2$.
Another mechanism can also be invoked which turns out to state that the neutral vector
fields involved in physical processes are not their  ideal combinations but mixtures of
these, constructed via a rotation. This Primordial Mixing (PM)
mechanism\footnote{The Primordial Mixing
of neutral vector mesons resembles the mixing scheme for the
neutral pseudoscalar mesons developed in \cite{leutw96}.} of vector mesons
assumes a ${\bf 1}+{\cal O}(\epsilon)$ rotation. Working at first order in
breaking parameters, the merging of the BKY and CD breakings  together
with the PM mechanism define our Reference Solution (RS), the second
variant of BHLS$_2$  covering the full HLS scope  more easily than the
BS variant (as it stands presently).

In both variants of BHLS$_2$, the ${\cal L}_V$ piece of the non-anomalous Lagrangian
is modified  while its ${\cal L}_A$ piece remains
unchanged\footnote{This might have to
be revisited in order to improve the working of BS in the $\tau$ and $\pi^+ \pi^- \pi^0$
sectors of the broken HLS Lagrangian.}  compared to the former BHLS \cite{ExtMod3}.

\subsection*{The $e^+e^-$ Annihilation Phenomenology}
\indent \indent The BKY and CD breaking schemes, together with the Primordial Mixing, allow
to construct the non-anomalous and anomalous pieces of BHLS$_2$, a new (broken) HLS Lagrangian.
This allows for  a simultaneous  study of the $e^+ e^-$ annihilation into  6 final states ($\pi^+\pi^-$,
$\pi^0\gamma$,  $\eta \gamma$, $\pi^+\pi^-\pi^0$, $K^+K^-$, $K_L K_S$) within a unified
framework where global fits can be worked out.

The fits have been performed by taking into account the
fact that most of the data samples carry normalization uncertainties, frequently dominant;
these
are appropriately considered in order to avoid biased results
for physics quantities. The iterative method developed in \cite{ExtMod5}
is thus intensively used in the global fits reported here; the convergence
criterion chosen to stop the iteration at order $n+1$ was
$ \delta \chi^2= |\chi^2_{\rm tot}(n+1)-\chi^2_{\rm tot}(n)| < (0.3 \div 0.5)$. As
 $\chi^2_{\rm tot} \simeq 1000 \div 1200$ for $\simeq 1150 \div 1240$ data points,
this convergence criterion is clearly constraining.

The fitting procedure has allowed to detect (and discard) a very few data samples which
hardly accommodate the global framework fed with almost all the existing
data samples. Beside the already identified \cite{ExtMod5} two $\pi^+\pi^-$
samples, one also found that, out of the 7 available $\pi^+ \pi^- \pi^0$ data samples,
one  \cite{CMD2-2006} hardly fits the global context, as also reported
by others  \cite{Kubis_3pi}. Figures \ref{Fig:PIPI}, \ref{Fig:PIPIPI} and
\ref{Fig:PSG} together with Table \ref{Table:T3} prove the fairly good quality of the
global description for the $\pi^+\pi^-$, $(\pi^0/\eta)\gamma$ and $\pi^+\pi^-\pi^0$
channels. As for the 3-pion channel, one should note that BHLS$_2$ encompasses
easily  the samples collected in the $\phi$ mass region and so, solves the difficulty
encountered there  by the former BHLS.

A thorough study of the available data samples has been performed concerning
both the  $K^+K^-$ and $K_L K_S$ channels for which some tension is expected,
 as reported
in \cite{CMD3_KpKm}. The analysis in Section \ref{kkb_data} has shown that
the SND, BaBar \cite{BaBarKK} and the updated \cite{Mainz2018}  CMD-2 data
are in fairly good agreement with each other and accommodate easily the HLS context
where all other kinds of data are encompassed. Tension has indeed been observed
between the CMD-3 data \cite{CMD3_KpKm,CMD3_K0K0b} and the other samples.
This tension has been traced back to the off-diagonal part of the CMD-3 error covariance
matrices for both  the $K^+K^-$ and $K_L K_S$ spectra;
 this analysis is summarized by
 Table \ref{Table:T2}. In order to include the BaBar (and/or CMD-3 data) within the global framework,
energy shifts relative to the CMD-2/SND energy scale  (chosen as energy scale
reference) are
mandatory and have been fitted at values consistent with the reported
expectations \cite{BaBarKK,CMD3_KpKm}.
To perform this exercise, when using the CMD-3 data, the error covariance
matrices of these (only) have been amputated from their off-diagonal part.

One can finally remark that, once the energy scales
of the CMD-3 and BaBar \cite{BaBarKK} data  relative to CMD-2/SND are
applied, one does not observe a visible tension among them any longer as illustrated
by Figure  \ref{Fig:KLKS}, where all data samples are nicely superimposing
onto the same
fit function.

\subsection*{Meson Form Factors in the Spacelike Region}
\indent \indent The BHLS$_2$ form factors supplied with the parameter values derived
from global fits to the {\it timelike region data only } have been
shown to nicely predict the
behavior of the $\pi^\pm$ and $K^\pm$ form factors  {\it in the close spacelike region}, as they
can be expected from   NA7 and Fermilab data.
Moreover, including these (model-independent) spacelike data within the BHLS$_2$ fit procedure
does not exhibit  any gain in the description of the spacelike region as obvious from
Table \ref{Table:T4} and from Figure \ref{Fig:spacelike_1}.

Our derived   $\pi^\pm$ and $K^\pm$ form factors,  which perform fairly well
simultaneously in the spacelike and timelike regions,  strongly supports the validity of
the BHLS$_2$ predictions in the $s$-region located in between. If the fairly good continuation
of $F_{\pi^\pm}(s)$ is actually performed over a limited gap in $s$ to reach the spacelike region, one may
wonder  about the so successful continuation of $F_{K^\pm}(s)$ over an $s$-gap greater than
1 GeV$^2$. This indicates that our modified Breit-Wigner formulas are performing (unexpectedly) well
far beyond the $\phi$ energy region where their parameters are determined.

The LQCD Collaborations HPQCD and ETM have provided $F_{\pi^\pm}(s)$,
the former as a para\-metri\-zation, the latter
by corresponding spectra, covering the same space-like $s$-region as the NA7 experiment.
Figure \ref{Fig:LQCD} shows the superposition of the BHLS$_2$ prediction (without any free parameter)
together with the
 parametrization provided by  HPQCD  and the two ETM spectra; the accord is very good for  HPQCD
 and simply perfect for both the ETM  spectra. There is, unfortunately, no reported LQCD data
 for the kaon form factors to which one could compare.

 The success just emphasized for the BHLS$_2$ $\pi^\pm$ and $K^\pm$ form factors compared
 to spacelike experimental or LQCD data gives confidence in other physics information involving
 the chiral point region. Let us limit oneself, in this Subsection,
 to briefly comment on the neutral kaon
 electromagnetic form factor. As reminded in the main text, the slope for the $F_{K^0}(s<0)$
 form factor at $s\ra 0$ is reported negative. This is qualitatively consistent with the BHLS$_2$ finding for
 $F_{K^0}(s)$ shown in Figure \ref{Fig:spacelike_k0}, as it is found to carry a negative slope at negative $s$ and
a  positive slope at  positive $s$.
 However,  the effects of the neglected ${\bf O}(\epsilon^2)$ corrections actually prevent to provide
 confidently accurate  slope predictions in the neighborhood of $s=0$ which is a
 delicate region for our ${\bf O}(\epsilon)$ approximations.

The pion form factor measured by the NA7 and Fermilab Collaborations are model independent and,
actually, it cannot be measured in a model-independent way above $|s| \simeq 0.25$ GeV$^2$.
To go beyond, one should accept to consider data where the extraction of  $F_{\pi^\pm}(s)$
is model-dependent. Figure \ref{Fig:spacelike_1} already indicates that the two {\it unfitted}
{\sc desy} measurements fall (almost) exactly onto the fitting curve together with
 the model-independent  NA7 and Fermilab measurements. The
JLAB data points are more or less closer to the BHLS$_2$ expectation depending on the method
used to extract $F_{\pi^\pm}(s)$. Nevertheless, our analysis  indicates that
the BHLS$_2$ prediction could well be valid down to $\simeq -1$ GeV$^2$, possibly slightly beyond.

In summary, our analysis clearly indicates
that, {\it inherently}, BHLS$_2$ only supplied with consistent $e^+e^-$ annihilation data,
leads precisely to the spacelike expectations supported by the precise NA7 data or the available LQCD input
as well. Nevertheless, LQCD information on $F_{K^\pm}(s)$ (and $F_{K^0}(s)$) would
be welcome to  better check the open strangeness sector.

\subsection*{Physical Quantities and Model Dependence Effects}
\indent \indent BHLS$_2$ has derived evaluations of several physics quantities (see Sections
\ref{spacelike}, \ref{phaseshift} and \ref{LagVal}) and some   deserve special emphasis.
As shown in Table \ref{Table:T5}, one reaches precise values for the charged pion and kaon
charge radii (fm$^2$)~:
$$<r_{\pi^\pm}^2>=0.430 \pm 0.002_{\rm mod} \pm 0.001_{\rm fit}~~~,~~
<r_{K^\pm}^2>=0.268\pm 0.004_{\rm mod} \pm 0.001_{\rm fit}~~,$$
our estimate for $<r_{\pi^\pm}^2>$ is in fairly good agreement with the most
recent evaluations \cite{Colangelo:rpi,Caprini}. Our evaluation for $<r_{K^\pm}^2>$
is in accord with the reported values shown in Table \ref{Table:T5} and is more precise.
The Reference Solution of BHLS$_2$ also leads to~:
$$\displaystyle \frac{f_{K^\pm} }{f_{\pi^\pm}}=1.247 \pm 0.020_{\rm
syst} \pm 0.003_{\rm fit}\,,$$
where the quoted systematic error reflects different variants of the modeling.

\vspace{0.5cm}

Table \ref{Table:T6} collects the values of the BHLS$_2$ model parameters obtained
under several fit conditions and discussed in Section \ref{LagVal}. An interesting
numerical correlation is observed between the fundamental HLS parameter $a$  and the
BKY breaking parameter $z_V$ within the BHLS$_2$ framework. This can be approximated
by  $a z_V \simeq 2.6$; good fit qualities are thus observed with  $a$  varying over
a large interval, namely from 1.6 to 2.6. The rightmost two data columns in
Table \ref{Table:T6}, which collect the fit values of a large sample of breaking parameters
in  two extreme cases ($a\simeq 1.6$ and $a\simeq 2.6$), also show correlatedly important
changes in some of the other breaking parameter values.  Table \ref{Table:T6} also shows that
the fit qualities reached in all cases are almost identical.

One can herefrom conclude  that
the various model parameter sets efficiently conspire to provide almost identical
cross-sections and form factors which are, actually, the real observables feeding BHLS$_2$.
This has been substantiated by comparing the running of BHLS$_2$ fed with our consistent set
of data samples under the various conditions
underlying the fits reported in Table \ref{Table:T6}.

A motivated piece of information to detect parametrization effects are
the integrals $a_\mu(\sqrt{s} \le 1.05$~GeV$)$ given in Table \ref{Table:T7}.
Running the BS variant (excluding the $\tau$ data) provides
$a_\mu(BS, \sqrt{s} \le 1.05$~GeV$)= (571.84 \pm 1.06) \times 10^{-10}$, while
running the RS variant running on the same data (with the Primordial Mixing angles free) returns
$a_\mu(RS, \sqrt{s}\le 1.05$~GeV$)= (571.98 \pm 1.20) \times 10^{-10}$. So the difference
between the various parametrizations happens indeed to be marginal and quite comparable
in magnitude to the uncertainties generated by using somewhat conflicting data samples.
Nevertheless, external information
constraining or relating  some of the model parameters could be useful in order to reduce
the parameter freedom in fits and so the parameter correlations.

\vspace{0.5cm}

As for the important muon HVP topic~: Complementing the BHLS$_2$ contribution by 
the whole hadronic contribution not covered by the (BHLS$_2$) model and reported in the left-hand side
part of Table \ref{Table:T8}, one gets~:
 $$ 10^{10} \times a_\mu^{\rm HVP-LO} = 686.65 \pm 3.01
 +\left[^{+1.16}_{-0.75} \right]_{\rm syst} +\left[^{+3.10}_{-0.0} \right]_{\rho \gamma}~~.$$
 Its (statistical + model) uncertainty  (3.01) is  sharply dominated by those of 
the whole non-HLS contribution~: $2.76 \times 10^{-10}$ -- reported in Table \ref{Table:T8} --
versus the BHLS$_2$ uncertainty $(1.06 \div 1.20)\times 10^{-10}$ (see Table \ref{Table:T7}). 

So, a significant
 improvement of the uncertainty for $a_\mu^{\rm HVP-LO}$
 can only follow from an improved accuracy of the hadronic contributions of higher energies, especially
 those from the $(1.05-2.00)$ GeV region.
  Therefore, presently, as global fit methods like BHLS$_2$
   constrain the central value for $a_\mu^{\rm HVP-LO}$(HLS),
 they do alike for $a_\mu^{\rm HVP-LO}$(HLS+non-HLS). The effect of the $\rho^0-\gamma$ mixing vs LQCD
 is shown separately for clarity.

Our estimate for the full muon anomalous moment is~:
 $$ 10^{10} \times a_\mu= 11\,659\,175.96 \pm 4.17_{\rm th}
+\left[^{+1.16}_{-0.75} \right]_{\rm syst}+\left[^{+3.10}_{-0.0} \right]_{\rho \gamma}\:,$$
which is, at least, at  $ 3.8 \sigma$ from the BNL measurement. The systematics are actually
upper bounds to a possible bias affecting $a_\mu$.

\clearpage
\section*{\Large{Appendices}}
\appendix
\section{The Full HLS Non-Anomalous Lagrangian in the New Scheme}
\label{AA}
\indent \indent For clarity,  the full non-anomalous HLS Lagrangian is written~:
\be
 \displaystyle  {\cal L}_{\rm HLS} ={\cal L}_{A}+a {\cal L}_{V}+{\cal L}_{\tau}+{\cal L}_{p^4}
\label{eqAA-1}
\ee
and  the various pieces will be given just below.

\subsection{The ${\cal L}_{A}+a {\cal L}_{V}$ Lagrangian Piece}
\label{AA-1}
\indent \indent
First, one displays the part of ${\cal L}_{\rm VMD}={\cal L}_{A}+ a {\cal L}_{V}$ relevant for the physics
we address,  especially $e^+ e^-$ annihilations. This is~:
\be
\begin{array}{ll}
{\cal L}_{\rm VMD} &=  \displaystyle  \partial \pi^+ \cdot \partial \pi^-
+ \frac{1}{2}  \partial \pi^0 \cdot  \partial \pi^0
+ \partial K^+ \cdot  \partial K^- +  \partial K^0 \cdot  \partial \overline{K}^0
\\[0.5cm]
\hspace{-2.5cm} ~& \displaystyle
+ ie \left[ 1 -\frac{a}{2} (1+ \Sigma_V +\frac{\Delta_V}{3}) \right] A \cdot \pi^- \parsym \pi^+
\\[0.5cm]
\hspace{-2.5cm} ~& \displaystyle
+ i e \frac{a}{6z_A} (1+\frac{\Delta_A}{2})(z_V-1 - \Sigma_V +\Delta_V) A \cdot K^0 \parsym \overline{K}^0
\\[0.5cm]
\hspace{-2.5cm} ~& \displaystyle
+ ie \left[ 1 -\frac{a}{6 z_A} \left \{ 2+z_V + 2  \Sigma_V +2  \Delta_V
\right \}
(1-\frac{\Delta_A}{2})\right]
 A \cdot K^- \parsym K^+
\\[0.5cm]

\hspace{-2.5cm} ~& \displaystyle +\frac{1}{2} \left [
m_{\rho^0}^2  (\rho^0_I)^2 + m_{\omg}^2 \omg^2_I + m_{\phi}^2 \phi^2_I \right]
-e
\left [f_{\rho\gamma}^I \rho^0_I+ f_{\omg \gamma}^I \omg_I + f_{\phi \gamma}^I\phi_I\right] \cdot A
\\[0.5cm]
\hspace{-2.5cm} ~& \displaystyle
 + \frac{i a g}{2} (1+ \Sigma_V)\left\{ \left[ 1+\xi_3 \right]\rho^0_I  + \Delta_V \omg_I
\right\} \cdot\pi^- \parsym \pi^+

\\[0.5cm]
\hspace{-2.5cm} ~& \displaystyle
+\frac{i a g}{4 z_A} [1-\frac{\Delta_A}{2}]
\left[ 1+ \Sigma_V +\Delta_V +\xi_3 \right] \rho^0_I \cdot  K^- \parsym K^+
\\[0.5cm]
\hspace{-2.5cm} ~& \displaystyle
+\frac{i a g}{4 z_A} [1-\frac{\Delta_A}{2}]
\left[
1+ \Sigma_V +\Delta_V +\frac{2}{3} (1-z_V)\xi_0 +\frac{1}{3} (1+2z_V)\xi_8
\right]\omg_I \cdot  K^- \parsym K^+
\\[0.5cm]

\hspace{-2.5cm} ~& \displaystyle
+\frac{i a g}{4z_A}[1+\frac{\Delta_A}{2}] [1+ \Sigma_V -\Delta_V+\xi_3]
\rho^0_I \cdot K^0 \parsym \overline{K}^0
\\[0.5cm]

\hspace{-2.5cm} ~& \displaystyle
-\frac{i a g}{4z_A} [1+\frac{\Delta_A}{2}]
\left[   1+ \Sigma_V -\Delta_V+
\frac{2}{3} (1-z_V)  \xi_0+\frac{1}{3} (1+2z_V)\xi_8
\right]  \omg_I \cdot  K^0 \parsym \overline{K}^0
\\[0.5cm]

\hspace{-2.5cm} ~& \displaystyle
-\frac{i a g\sqrt{2}}{4z_A}  (1-\frac{\Delta_A}{2})
\left[z_V  -\frac{1}{3} (1-z_V) \xi_0 +\frac{1}{3} (1+2z_V) \xi_8
\right]  \phi_I \cdot  K^- \parsym K^+
\\[0.5cm]

\hspace{-2.5cm} ~& \displaystyle
+\frac{i a g\sqrt{2}}{4z_A}  (1 + \frac{\Delta_A}{2})\left[
z_V  -\frac{1}{3} (1-z_V) \xi_0 +\frac{1}{3} (1+2z_V)\xi_8
\right] \phi_I \cdot
K^0 \parsym \overline{K}^0\epo
\end{array}
\label{eqAA-2}
\ee
The pseudoscalar fields occurring there are the renormalized ones defined exactly as
in \cite{ExtMod3}.  For simplicity, we use  ideal vector fields.
The  $V-\gamma$ couplings are~:
 \be
\left\{
\begin{array}{ll}
f^{I}_{\gamma \rho}&= \displaystyle \frac{m^2}{g}
[1+ \Sigma_V  +\frac{\Delta_V}{3}+  \xi_3]
\,,\\[0.5cm]
f^{I}_{\gamma \omg}&= \displaystyle \frac{m^2}{3g}
\left[1 + \Sigma_V +3\Delta_V  +
\frac{2}{3}(1-z_V) ~\xi_0 +\frac{ (1+2z_V)}{3} ~\xi_8 \right]\,,\\[0.5cm]
f^{I}_{\gamma \phi}&= \displaystyle -\frac{\sqrt{2} m^2}{3g}
\left[z_V-\frac{\xi_0}{3}(1-z_V) + \frac{(1+2z_V)}{3}\xi_8
\right]\epo
\end{array}
\right .
\label{eqAA-3}
\ee
One has discarded the (irrelevant) photon mass term. The vector meson masses
are given in Eq.
(\ref{eq1-26}). $\Delta_V$ and $\Sigma_V$ have been defined
in  Section \ref{BKYbrk}.

In practical use, one should perform the change of field given in Eq. (\ref{eq1-23})
and collect the terms corresponding to each (neutral) vector meson coupling.

\subsection{The ${\cal L}_{\tau}$ Lagrangian Piece}
\label{AA-2}
\indent \indent
The new  ${\cal L}_{\tau}$ is given below (after performing an integration by parts
to remove the terms with  $\partial W^\pm(=0)$~:
\be
\begin{array}{ll}
{\cal L}_{\tau}& = \displaystyle a g^2  f_\pi^2 (1+ \Sigma_V)\rho^+ \cdot \rho^-
- \frac{i V_{ud}~g_2}{2\sqrt{2}} W^+ \cdot
\left [1 - \frac{a}{2 z_A}   (1+ \Sigma_V )\right ] K^0 \parsym K^-
\\[0.5cm]
 ~& \hspace{-0.5cm} \displaystyle
-\frac{1}{2} a g g_2 f_\pi^2(1+ \Sigma_V)  \left [V_{ud} W^+ \cdot \rho^-
+\overline{V}_{ud}W^- \cdot \rho^+ \right] -\frac{i a g }{2\sqrt{2}z_A} (1+ \Sigma_V)
\rho^+ \cdot  K^0 \parsym K^-
 \\[0.5cm]
 ~& \hspace{-0.5cm} \displaystyle
+ \frac{i V_{ud}g_2}{2} W^+ \cdot
\left[ 1 -\frac{a}{2} (1+ \Sigma_V ) \right]
\left [
\pi^0  - \left( \epsilon_1 -\frac{\Delta_A}{2}\sin{\delta_P} \right )  \eta
-\left( \epsilon_2 -\frac{\Delta_A}{2}\cos{\delta_P} \right ) \etp
\right]\parsym\pi^-
\\[0.5cm]

 ~&\hspace{-0.9cm} \displaystyle - i \Delta_A \frac{V_{ud}g_2}{2} W^+ 
\left [  \sin{\delta_P} ~ \eta - \cos{\delta_P}~
 \etp \right]  \parsym \pi^-  - \frac{ f_\pi g_2}{4}~ W^+
\left [ V_{ud} \partial \pi^- +  V_{us}   \sqrt{z_A} (1+\frac{\Delta_A}{4}) \partial K^-\right]
\\[0.5cm]
~&\hspace{-0.5cm}  \displaystyle - \frac{i a g }{2}
\rho^+ \cdot \left [(1+ \Sigma_V) \pi^- \parsym \pi^0
- \left( \epsilon_1 -\frac{\Delta_A}{2}\sin{\delta_P} \right ) \pi^- \parsym \eta
-\left( \epsilon_2 -\frac{\Delta_A}{2}\cos{\delta_P} \right ) \pi^- \parsym \etp
\right]\,,
\end{array}
\label{eqAA-4}
\ee
{\it where all fields are renormalized}. $\delta_P$ is related with the PS mixing in 
the one angle mixing scheme \cite{WZWChPT}; its definition is reminded in Appendix 
\ref{BB} below together with those for the $\epsilon_i$. 

The appearance of 
$W^\pm (\pi^0/\eta/\eta^\prime)\pi^\pm$ couplings should be noted;
one should also note, about the $W^\pm (\eta/\eta^\prime) \pi^\pm$ couplings,  that
the contributions from ${\cal L}_A$ and ${\cal L}_V$  are not proportional
in contrast with the $W^\pm \pi^0 \pi^\pm$ case.  Finally, it is also worth noting
that the $\rho^\pm (\eta/\eta^\prime) \pi^\pm$ couplings in Equation (\ref{eqAA-4})
are needed in the transitions amplitudes for $\eta/\eta^\prime \ra \pi^+ \pi^- \gamma$ 
in order to yield the right behavior at the chiral point.

\subsection{The ${\cal L}_{p^4}$ Lagrangian Piece ${\cal L}_{z_3}$}
\label{AA-3}
\indent \indent
As for the ${\cal L}_{z_3}$ Lagrangian, it writes in terms of ideal fields~:
\be
\begin{array}{lll}

\displaystyle 
{\cal L}_{z_3}=& \displaystyle e g z_3 ~s~ A_\mu  \left [ (1+\xi_3)
\rho^{0~\mu}_I +\frac{1}{3}(1+\xi_8) \omg^\mu_I
 -\frac{\sqrt{2}}{3}(1+\xi_8)
  ~\phi^\mu_I \right ] \crn &\displaystyle 
+\frac{g_2 g z_3~s}{2} \left [\overline{V}_{ud} ~\rho^+_\mu ~W^-_\mu +
	V_{ud} ~\rho^-_\mu ~W^+ _\mu\right ]\epo
\end{array}
\label{eqAA-5}
\ee
It clearly allows for different $V\gamma$ and $WV$ couplings, especially for the $\rho$ meson.

\subsection{The Renormalized Couplings of Vector Mesons to $K \overline{K}$}
\label{AA-4}
\indent \indent
Let us first remind the coupling of the fully renormalized vector mesons to
$\pi^+\pi^-$; having defined these as $g^V_{\pi^\pm}=ag \widetilde{g}_{\pi\pm}/2$,
one has~:
\be
\begin{array}{lll}
\displaystyle \widetilde{g}_{\pi\pm}^\rho=[1+\Sigma_V+\xi_3]~~,&
\displaystyle \widetilde{g}_{\pi\pm}^\omg=-[\psi_\omg+\alpha(s)]~~,&
\displaystyle \widetilde{g}_{\pi\pm}^\phi=[\psi_\phi+\beta(s)]\epo
\end{array}
\label{eqAA-6}
\ee

Using the Lagrangian in Section \ref{AA-1} and Eq. (\ref{eq1-33}) one can derive the
renormalized coupling of vector mesons to $K \overline{K}$ pairs. Defining:
\be
\left \{
\begin{array}{lll}
\displaystyle
\widetilde{g}_{K\pm}^\rho =\left (1-\frac{\Delta_A}{2} \right )
\left \{
1 +\Sigma_V+\xi_3+ [\psi_\omg+\alpha(s)] +\sqrt{2} z_V [\psi_\phi+\beta(s)]
\right \}\,,\\[0.5cm]
\displaystyle
\widetilde{g}_{K\pm}^\omg =\left (1-\frac{\Delta_A}{2} \right )
\left \{
1 +\Sigma_V+\frac{2}{3} (1-z_V)\xi_0 + \frac{1+2z_V}{3}\xi_8
- [\psi_\omg+\alpha(s)] +\sqrt{2} z_V [\psi_0+\gamma(s)]
\right\}\,,\\[0.5cm]
\displaystyle
\widetilde{g}_{K\pm}^\phi=\left (1-\frac{\Delta_A}{2} \right )
\left \{-\sqrt{2}
\left[
z_V-\frac{1}{3} (1-z_V)\xi_0 +\frac{1+2z_V}{3}\xi_8
\right]
+ [\psi_\phi+\beta(s)] + [\psi_0+\gamma(s)]
\right\}\,,
\end{array}
\right .
\label{eqAA-7}
\ee
the couplings involved in form factors and cross-sections
are given by $G^V_{K^+ K^-}=ag /[4 z_A]~ \widetilde{g}_{K\pm}^V$. One also has~:
\be
\left \{
\begin{array}{lll}
\displaystyle
\widetilde{g}_{K0}^\rho =~\left (1+\frac{\Delta_A}{2} \right )
\left \{
1 +\Sigma_V+\xi_3- [\psi_\omg+\alpha(s)] -\sqrt{2} z_V [\psi_\phi+\beta(s)]
\right \}\,,\\[0.5cm]
\displaystyle
\widetilde{g}_{K0}^\omg =- \left (1+\frac{\Delta_A}{2} \right )
\left \{
1 +\Sigma_V+\frac{2}{3} (1-z_V)\xi_0 + \frac{1+2z_V}{3}\xi_8
+ [\psi_\omg+\alpha(s)] +\sqrt{2} z_V [\psi_0+\gamma(s)]
\right\}\,,\\[0.5cm]
\displaystyle
\widetilde{g}_{K0}^\phi=~\left (1+\frac{\Delta_A}{2} \right )
\left \{\sqrt{2}
\left[
z_V-\frac{1}{3} (1-z_V)\xi_0 +\frac{1+2z_V}{3}\xi_8
\right]
+ [\psi_\phi+\beta(s)] - [\psi_0+\gamma(s)]
\right\}\,,
\end{array}
\right .
\label{eqAA-8}
\ee
and the full couplings are given by
 $G^V_{K^0 \overline{K}^0}=ag /[4 z_A]~ \widetilde{g}_{K0}^V$. One  cancels out
 here and in the following the dependence upon $\Delta_V$.
One should also note that all these couplings depend on $s$, the square energy flowing through
the relevant vector line.
\subsection{The $V-\gamma$ Transition Loop Terms}
\label{AA-5}
As seen in the main text, the $V-\gamma$ transitions amplitudes write~:
$$F_{V\gamma}^e(s)= f_{V\gamma} - c_{V\gamma} z_3 s - \Pi_{V\gamma}(s) $$
for each of $ V=\rho^0_R,~\omg_R,~\phi_R$. In this Section, one displays
the expression for the loop terms $\Pi_{V\gamma}(s)$ in terms of the basic loop
functions denoted $\Pi_\pi^e(s)$, $\Pi_{K\pm}^e(s)$ and $\Pi_{K^0}^e(s)$
and defined in Subsection \ref{loopfc}. For the sake of conciseness, it is convenient
to define~:
\be
\begin{array}{lll}
\displaystyle \widetilde{g}_{\pi\pm}^\gam=
\left[1-\frac{a}{2}\left[1+\Sigma_V \right ]\right ]~~,\\[0.5cm]
\displaystyle \widetilde{g}_{K\pm}^\gam=
\left[1-\frac{a}{6 z_A}\left[2+z_V+2\Sigma_V -\frac{3}{2} \Delta_A\right ]\right ]~
~~,&
\displaystyle \widetilde{g}_{K0}^\gam=
-\frac{a}{6 z_A}(1-z_V+\Sigma_V)\,,
\end{array}
\label{eqAA-9}
\ee
neglecting the ${\cal O}(\epsilon^2)$ terms. These couplings do not depend
on the "angles" $\alpha(s)$, $\beta(s)$ and $\gamma(s)$.

Using these definitions and those displayed in the Subsection just above, on finds~:
\be
\left \{
\begin{array}{lll}
\displaystyle \Pi_{\rho \gam}=
\frac{\widetilde{g}_{\pi\pm}^\gam  \widetilde{g}_{\pi\pm}^\rho}{G_\pi} \Pi^e_\pi(s)
+\frac{\widetilde{g}_{K\pm}^\gam  \widetilde{g}_{K\pm}^\rho}{G_K} \Pi^e_{K\pm}(s)
+\frac{\widetilde{g}_{K0}^\gam  \widetilde{g}_{K0}^\rho}{G_K} \Pi^e_{K0}(s) \,,\\[0.5cm]
\displaystyle \Pi_{\omg \gam}=
\frac{\widetilde{g}_{\pi\pm}^\gam  \widetilde{g}_{\pi\pm}^\omg}{G_\pi} \Pi^e_\pi(s)
+\frac{\widetilde{g}_{K\pm}^\gam  \widetilde{g}_{K\pm}^\omg}{G_K} \Pi^e_{K\pm}(s)
+ \frac{\widetilde{g}_{K0}^\gam  \widetilde{g}_{K0}^\omg}{G_K} \Pi^e_{K0}(s)
\,,\\[0.5cm]
\displaystyle \Pi_{\phi \gam}=
\frac{\widetilde{g}_{\pi\pm}^\gam  \widetilde{g}_{\pi\pm}^\phi}{G_\pi} \Pi^e_\pi(s)
+\frac{\widetilde{g}_{K\pm}^\gam  \widetilde{g}_{K\pm}^\phi}{G_K} \Pi^e_{K\pm}(s)
+ \frac{\widetilde{g}_{K0}^\gam  \widetilde{g}_{K0}^\phi}{G_K} \Pi^e_{K0}(s)\epo
\end{array}
\right.
\label{eqAA-10}
\ee
Expanding Equations (\ref{eqAA-10}) in order to rather parametrize in terms of
$\Pi^e_{K~{\rm sum}}$ and $\Pi^e_{K~{\rm diff}}$
(see Subsection \ref{loopfc}) allows to exhibit the terms of order ${\cal O}(\epsilon^2)$
in breakings which can be dropped out.

\section{Pseudoscalar Field Renormalization~: A Brief Reminder}
\label{BB}
\indent \indent
As already stated, one goes on using unchanged the breaking procedure in the ${\cal L}_A$ sector
of the HLS model as defined in \cite{ExtMod3}. The  broken ${\cal L}_A$ Lagrangian
displayed in Eq. (\ref{eq1-6}) implies a first step field redefinition in order to make the pseudoscalar
kinetic energy term canonical. Additionally, as BHLS and BHLS$_2$ also include the 't~Hooft determinant
terms \cite{tHooft}, the  pseudoscalar kinetic energy receives an additional piece~:
\be
\displaystyle
{\cal L}_{\rm 'tHooft}=\frac{1}{2} \lambda \pa_\mu \eta_0\pa^\mu \eta_0 + \cdots\:,
\label{BB1}
\ee
where $\eta_0$ is the singlet PS field and
$\lambda$ is a parameter to be fixed. This term
imposes a second step renormalization \cite{ WZWChPT} of the ($\pi^0,~\eta,~\etp$) sector.

The first step renormalization turns out to define the (step one) renormalized pseudoscalar
field matrix $P^{R_1}$ in term of the bare one $P$ by \cite{ExtMod3}~:
 \be
\begin{array}{lll}
P^{R_1}=X_A^{1/2} P X_A^{1/2}~~,& \displaystyle  {\rm where} ~~~
X_A={\rm Diag}(1 +\frac{\Delta_A}{2},1 -\frac{\Delta_A}{2},z_A) \epo
\end{array}
\label{BB2}
\ee

Some pseudoscalar (PS) fields happen to be fully renormalized after this first step; they are
 related to their  bare  partners by~:
\be
\left \{
\begin{array}{lll}
\displaystyle
\pi^\pm_{\rm bare} &= \displaystyle \pi^\pm\,, \\[0.5cm]
K^{\pm}_{\rm bare}&=\displaystyle \frac{1}{\sqrt{z_A}}(1-\frac{\Delta_A}{4})
K^{\pm}\,, \\[0.5cm]
K^{0}_{\rm bare}&=\displaystyle \frac{1}{\sqrt{z_A}}(1+\frac{\Delta_A}{4}) K^{0}\epo
\end{array}
\right.
\label{BB3}
\ee
These renormalized fields have already been used to derive the Lagrangians
given in Appendix \ref{AA}. As for the other PS fields, the first step
renormalization leads to the $R1$ renormalized PS fields~:
\be
\left \{
\begin{array}{ll}
\pi_0^{bare} =  \displaystyle \pi^{R_1}_0
-\frac{\Delta_A}{2\sqrt{3}}\eta_8^{R_1}
-\frac{\Delta_A}{\sqrt{6}}\eta_0^{R_1}\,,\\[0.5cm]
\eta_0^{bare} =
\displaystyle -\frac{\Delta_A}{\sqrt{6}}\pi^{R_1}_0
+\frac{\sqrt{2}}{3}\frac{z_A -1}{z_A} \eta_8^{R_1}
+\frac{1}{3}\frac{2 z_A+1 }{z_A} \eta_0^{R_1}\,,\\[0.5cm]
\eta_8^{bare} =
\displaystyle -\frac{\Delta_A}{2\sqrt{3}}\pi^{R_1}_0
+\frac{1}{3}\frac{z_A  +2}{z_A} \eta_8^{R_1}
+\frac{\sqrt{2}}{3}\frac{z_A -1 }{z_A} \eta_0^{R_1}\epo
\end{array}
\right.
\label{BB4}
\ee
The second step renormalization required by the 't~Hooft term
gives the $R$ renormalized fields \cite{WZWChPT}~:
\be
\left \{
\begin{array}{lll}
\displaystyle
\pi^{R_1}_0=\pi^{R}_0\,,\\[0.5cm]
\displaystyle
\eta^{R_1}_8=\frac{1+v\cos^2{\beta}}{1+v}\eta^R_8
-\frac{v\sin{\beta}\cos{\beta}}{1+v}\eta^R_0\,,\\[0.5cm]
\displaystyle
\eta^{R_1}_0=-\frac{v\sin{\beta}\cos{\beta}}{1+v}\eta^R_8
+ \frac{1+v\sin^2{\beta}}{1+v}\eta^R_0\,,
\end{array}
\right.
\label{eqBB5}
\ee
where~:
\be
\left \{
\begin{array}{ccc}
\displaystyle
\cos{\beta}=\frac{2z_A+1}{\sqrt{3(2 z_A^2+1)}}~~,~~
\displaystyle
\sin{\beta}=\frac{\sqrt{2}(z_A-1)}{\sqrt{3(2 z_A^2+1)}}\\[0.5cm]
\displaystyle
v=\sqrt{1+\lambda \frac{(2 z_A^2+1)}{3z_A^2}} -1
\simeq \frac{\lambda}{2}\frac{(2 z_A^2+1)}{3z_A^2}~~ .
\end{array}
\right.
\label{BB6}
\ee
The parameters affecting the PS fields submitted to fit are thus $\Delta_A$, $z_A$
and $\lambda$ (or alternatively $v$).

The $\pi^0-\eta-\eta^\prime$ mixing is supplemented by defining
the physically observable PS fields in terms of their $R$ renormalized
partners by \cite{leutw96,ExtMod3}~:
 \be
\left \{
\begin{array}{lll}
\displaystyle
\pi_R^0= \pi^0-\epsilon_1 ~\eta-\epsilon_2 ~\eta^\prime\,,\\[0.5cm]
\displaystyle
\eta_R^8= \cos{\theta_P} (\eta+\epsilon_1 ~\pi^0)+
\sin{\theta_P} (\eta^\prime+\epsilon_2 ~\pi^0)\,, \\[0.5cm]
\displaystyle
\eta_R^0 = -\sin{\theta_P} (\eta+\epsilon_1 ~\pi^0)+
\cos{\theta_P} (\eta^\prime+\epsilon_2 ~\pi^0)\,,
\end{array}
\right.
\label{BB7}
\ee
which exhibits three more parameters to be determined by fit.
The singlet-octet  mixing angle $\theta_P$ occurring here is
 expected in the range $(-10^\circ \div -15^\circ)$ and is
also algebraically related to the usual $\theta_8$ \cite{WZWChPT}.
In order to express the forthcoming $VVP$ and $AVP$ coupling constants, the mixing angle $\delta_P$
defined by  \cite{ExtMod3}~:
 \be
\left \{
\begin{array}{ll}
\displaystyle \sin{\theta_P}=\frac{1}{\sqrt{3}}
(\cos{\delta_P}+\sqrt{2}\sin{\delta_P})\,,\\[0.5cm]
\displaystyle \cos{\theta_P}=\frac{1}{\sqrt{3}}
(\sqrt{2} \cos{\delta_P}-\sin{\delta_P})\,,
\end{array}
\right.
\label{BB8}
\ee
happens to be more appropriate than $\theta_P$. On the other hand, it is also
useful to define the combinations \cite{ExtMod3}~:
\be
\left \{
\begin{array}{ll}
\displaystyle
x=1-\frac{3 z_A^2}{2 z^2_A+1} v\,,\\[0.5cm]
\displaystyle
x^\prime=1-\frac{3 z_A}{2 z^2_A+1} v\,,\\[0.5cm]
\displaystyle
x^\second=1-\frac{3}{2 z^2_A+1} v\,,
\end{array}
\right.
\label{BB9}
\ee
where $v$ is the nonet symmetry breaking parameter  defined in Equations(\ref{BB6}).
$x$, $x^\prime$ and $x^\second$  reflect the various ways by which
 nonet symmetry breaking occurs in the PS sector of broken HLS Lagrangians.

\section{The $VVP$ Lagrangian in Terms of Ideal Vector Fields}
\label{CC}
\indent \indent The $\pi^0$, $\eta$ and $\etp$ pieces of the
${\cal L}_{VVP}$  Lagrangian are given separately just below.
If the PS fields occurring in this Appendix are renormalized by the procedure
reminded in Appendix \ref{BB}, it is simpler to express the various
$VVP$ Lagrangian pieces in terms of the ideal vector fields.
The derivation of the couplings involving the renormalized fields is given in the main text.

We split up the $VVP$ Lagrangian into the pieces involving the renormalized
$\pi^0$, $\pi^\pm$, $\eta$ and $\etp$ pieces, below.  As for the $\pi^0$, we have~:
\ba
\displaystyle
{\cal L}_{VVP}(\pi^0)=&
\displaystyle
\frac{C}{2} \epsilon^{\mu \nu \alpha \beta} \Biggl\{
 \left [
 \left ( 1+\frac{2 \xi_0}{3} + \frac{\xi_8}{3} \right )
\partial_\mu \omg_\nu^I
+\frac{\sqrt{2}}{3} (\xi_0-\xi_8)
\partial_\mu \phi^I_\nu                            
 \right] \times \non \\[0.5cm]
~~& \displaystyle \hspace*{2.5cm} \left[ \partial_\alpha \rho^+_\beta \pi^- +\partial_\alpha \rho^-_\beta \pi^+
 +\left ( 1+ \xi_3 \right)\partial_\alpha \rho^{I}_\beta \pi^0 \right] \\[0.2cm]
~~& 
\displaystyle 
\hspace{-1.5cm}
\displaystyle 
+  \left[\widetilde{g}_{\omg \pi^0}~
\partial_\mu \omg^I_\nu \partial_\alpha \omg^I_\beta +
\widetilde{g}_{\rho \pi^0}~
 \partial_\mu \rho^I_\nu\partial_\alpha \rho^I_\beta+
 \widetilde{g}_{\phi\pi^0 }~
 \partial_\mu \phi^I_\nu  \partial_\alpha \phi^I_\beta +
 \widetilde{g}_{\rho^\pm \pi^0 }~
 \partial_\mu \rho^{+}_\nu  \partial_\alpha \rho^{-}_\beta
\right] \pi^0  \Biggr\}\:, \non
\label{CC1}
\ea
where~:
\be
\hspace{-1.cm}
 \left \{
\begin{array}{ll}
\displaystyle C=-\frac{N_c g^2 c_3}{4 \pi^2 f_\pi}\,,
\\[0.5cm]
\displaystyle
\widetilde{g}^0_{\rho^\pm \pi^0} =
2  \widetilde{g}^0_{\rho \pi^0 }=
\epsilon_2 ~\cos{\delta_P}
-\epsilon_1 ~\sin{\delta_P}
-  \frac{\Delta_A}{2}\,,
\\[0.5cm]
\displaystyle
\widetilde{g}^0_{\omg \pi^0 }=\frac{1}{2}
\left [ \epsilon_2 ~\cos{\delta_P}
-\epsilon_1 ~\sin{\delta_P}
-  \frac{\Delta_A}{2}  \right]\,,
\\[0.5cm]
\displaystyle
\widetilde{g}^0_{\phi \pi^0}=-\frac{1}{z_A\sqrt{2}}
\left [ \epsilon_1 ~\cos{\delta_P}
+\epsilon_2 ~\sin{\delta_P} \right ]\epo
\end{array}
\right.
\label{CC2}
\ee

The $VV\eta$ Lagrangian is given by~:
\be
\begin{array}{ll}
\displaystyle
{\cal L}_{VVP}(\eta)=&
\displaystyle
\frac{C}{2} \epsilon^{\mu \nu \alpha \beta} \left \{
\widetilde{g}^0_{\omg \phi \eta} ~ \partial_\mu \omg_\nu^I
\partial_\alpha \phi^I_\beta
 +
 \widetilde{g}^0_{\rho  \omg \eta} ~
 \partial_\mu \omg_\nu^I\partial_\alpha \rho^{I}_\beta
+ \widetilde{g}_{\rho^\pm \eta }~
 \partial_\mu \rho^{+}_\nu  \partial_\alpha \rho^{-}_\beta
\right .
\\[0.5cm]
~~& \displaystyle 
\left.
+\widetilde{g}_{\omg \eta}~
\partial_\mu \omg^I_\nu \partial_\alpha \omg^I_\beta +
\widetilde{g}_{\rho \eta}~
 \partial_\mu \rho^I_\nu\partial_\alpha \rho^I_\beta+
 \widetilde{g}_{\phi\eta }~
 \partial_\mu \phi^I_\nu  \partial_\alpha \phi^I_\beta
  \right  \} \eta \,,
\end{array}
\label{CC3}
\ee

with~:

\be
\hspace{-1.5cm}
 \left \{
\begin{array}{ll}
\displaystyle \widetilde{g}^0_{\omg \phi \eta}=\frac{1}{3z_A}
\left [2 \cos{\delta_P}+ z_A \sqrt{2}  \sin{\delta_P}\right ] (\xi_8-\xi_0)
\,,\\[0.5cm]
\displaystyle   \widetilde{g}^0_{\rho  \omg \eta} =
\left [
\frac{\Delta_A}{2} \sin{\delta_P} -\epsilon_1 \right ]
\,,\\[0.5cm]
\displaystyle
\widetilde{g}^0_{\rho^\pm \eta} = \frac{1}{3}
\left [\sqrt{2} (1-x^\prime)  \cos{\delta_P}- (1+2 x)\sin{\delta_P}
\right ]
\,,\\[0.5cm]
\displaystyle
\widetilde{g}^0_{\rho \eta}=~\frac{1}{6}
\left [ \sqrt{2}  (1-x^\prime)   ~\cos{\delta_P}
-(1+2 x)\sin{\delta_P}  -6 \xi_3 \sin{\delta_P} \right ]
\,,\\[0.5cm]
\displaystyle
\widetilde{g}^0_{\omg \eta}=\frac{1}{6}
\left [ \sqrt{2}  (1-x^\prime)   ~\cos{\delta_P}
-(1+2 x+ 4\xi_0+2\xi_8)\sin{\delta_P} \right ]
\,,\\[0.5cm]
\displaystyle
\widetilde{g}^0_{\phi \eta}=-\frac{1}{6z_A}
\left [ \sqrt{2}  (2+x^\second +2 \xi_0+4\xi_8) ~\cos{\delta_P}
- 2 (1-x^\prime) \sin{\delta_P}
\right ]
\epo\\[0.5cm]
\end{array}
\right.
\label{CC4}
\ee

The $VV\etp$ Lagrangian is given by~:
\be
\begin{array}{ll}
\displaystyle
{\cal L}_{VVP}(\etp)=&
\displaystyle
\frac{C}{2} \epsilon^{\mu \nu \alpha \beta} \left \{
\widetilde{g}^0_{\omg \phi \etp} ~ \partial_\mu \omg_\nu^I
\partial_\alpha \phi^I_\beta
 +
 \widetilde{g}^0_{\rho  \omg \etp} ~
 \partial_\mu \omg_\nu^I\partial_\alpha \rho^{I}_\beta
+ \widetilde{g}_{\rho^\pm \etp }~
 \partial_\mu \rho^{+}_\nu  \partial_\alpha \rho^{-}_\beta

\right .
\\[0.5cm]
~~& \displaystyle 
\left.
+  \widetilde{g}_{\omg \etp}~
\partial_\mu \omg^I_\nu \partial_\alpha \omg^I_\beta +
\widetilde{g}_{\rho \etp}~
 \partial_\mu \rho^I_\nu\partial_\alpha \rho^I_\beta+
 \widetilde{g}_{\phi\etp }~
 \partial_\mu \phi^I_\nu  \partial_\alpha \phi^I_\beta
  \right  \} \etp \,,
\end{array}
\label{CC5}
\ee
with~:
\be
\hspace{-1.5cm}
 \left \{
\begin{array}{ll}
\displaystyle \widetilde{g}^0_{\omg \phi \etp}= \frac{1}{3z_A}
\left [2 \sin{\delta_P}- z_A \sqrt{2}  \cos{\delta_P}\right ] (\xi_8-\xi_0)
\,,\\[0.5cm]
\displaystyle   \widetilde{g}^0_{\rho  \omg \etp} =-
\left [
\frac{\Delta_A}{2} \cos{\delta_P} +\epsilon_2 \right ]
\,,\\[0.5cm]
\displaystyle
\widetilde{g}^0_{\rho^\pm \etp} =  \frac{1}{3}
\left [\sqrt{2} (1-x^\prime)  \sin{\delta_P} + (1+2 x)\cos{\delta_P}
\right ]
\,,\\[0.5cm]
\displaystyle
\widetilde{g}^0_{\rho \etp}=~\frac{1}{6}
\left [ \sqrt{2}  (1-x^\prime)   ~\sin{\delta_P}
+ (1+2 x)\cos{\delta_P} + 6 \xi_3 \cos{\delta_P} \right ]
\,,\\[0.5cm]
\displaystyle
\widetilde{g}^0_{\omg \etp}=~\frac{1}{6}
\left [ \sqrt{2}  (1-x^\prime)   ~\sin{\delta_P}
+ (1+2 x +4 \xi_0+ 2 \xi_8) ~\cos{\delta_P} \right ]
\,,\\[0.5cm]
\displaystyle
\widetilde{g}^0_{\phi \etp}=-\frac{1}{6z_A}
\left [ \sqrt{2}  (2+x^\second +2 \xi_0+4 \xi_8) ~\sin{\delta_P}
+ 2 (1-x^\prime) \cos{\delta_P}
\right ]\epo
\end{array}
\right.
\label{CC6}
\ee
Finally, it is worth extracting out from Eq. (\ref{CC1}) the part
involving a charged $\rho$ meson for illustrative purposes. This write~:
\be
 \begin{array}{ll}
\displaystyle
{\cal L}_{VVP}(\rho^\pm,\pi^0)=&
\displaystyle
\frac{C}{2} \epsilon^{\mu \nu \alpha \beta}
 \left [
\widetilde{g}^\pm_{\omg \rho^\pm}
\partial_\mu \omg_\nu^I
+\widetilde{g}^\pm_{\phi \rho^\pm}
\partial_\mu \phi^I_\nu
 \right ]
\left[ \partial_\alpha \rho^+_\beta \pi^- +
\partial_\alpha \rho^-_\beta \pi^+\right]\,,
\end{array}
\label{CC7}
\ee
with~:
\be
\begin{array}{lll}
\displaystyle
\widetilde{g}^\pm_{\omg \rho^\pm}=\left ( 1+\frac{2 \xi_0}{3} +\frac{\xi_8}{3}\right )
~~,&
\displaystyle
\widetilde{g}^\pm_{\phi \rho^\pm}= \frac{\sqrt{2}}{3} (\xi_0-\xi_8)\epo
\end{array}
\label{CC8}
\ee
\section{The $AAP$ and $APPP$ Lagrangians in the HLS Framework}
\label{DD}
\indent \indent For the reader's convenience, we have found appropriate
to remind the expressions for the $AAP$ Lagrangian and for the part
of the $APPP$ Lagrangian relevant for the annihilation process
$e^+ e^- \ra \pi^0\pi^+\pi^-$. These have been derived in \cite{ExtMod3}.
One first have~:
\be
\displaystyle
{\cal L}_{AAP}=-\frac{3 \alpha_{\rm em}}{ \pi f_\pi} (1-c_4)\epsilon^{\alpha \beta \mu \nu}
\partial_\alpha A_\beta \partial_\mu A_\nu \left[
g_{\pi^0 \gamma \gamma} \frac{\pi^0}{6} + g_{\eta \gamma \gamma} \frac{\eta}{2\sqrt{3}}
 + g_{\eta^\prime \gamma \gamma} \frac{\eta^\prime}{2\sqrt{3}}
\right]\,,
\label{DD1}
\ee
with~:
\be
\left \{
\begin{array}{ll}
g_{\pi^0 \gamma \gamma}=&  \displaystyle 1-\frac{5 \Delta_A}{6} +
 \frac{\epsilon_1}{\sqrt{3}}
\left \{\frac{5z_A-2}{3z_A}\cos{\theta_P}-\sqrt{2}\frac{5z_A+1}{3z_A}\sin{\theta_P}
\right\} \crn & \displaystyle + \frac{\epsilon_2}{\sqrt{3}}\left \{\frac{5z_A-2}{3z_A}\sin{\theta_P}+
\sqrt{2}\frac{5z_A+1}{3z_A}\cos{\theta_P}\right\} \,,\\[0.5cm]
g_{\eta \gamma \gamma} =&  \displaystyle \frac{\cos{\theta_P}}{3}
\left \{\frac{5z_A-2}{3z_A(1+v)}+v \frac{1+2z_A}{1+2z_A^2}-\frac{\Delta_A}{2}
\right\}\crn&\displaystyle -\sqrt{2} \frac{\sin{\theta_P}}{3} \left \{
\frac{5z_A+1}{3z_A(1+v)}+v \frac{1-z_A}{1+2z_A^2} -\frac{\Delta_A}{2}\right\}
-\frac{\epsilon_1}{\sqrt{3}}\,,\\[0.5cm]
g_{\eta^\prime \gamma \gamma} =&  \displaystyle \frac{\sin{\theta_P}}{3}
\left \{\frac{5z_A-2}{3z_A(1+v)}+v \frac{1+2z_A}{1+2z_A^2}
-\frac{\Delta_A}{2}
\right\} \crn&\displaystyle+\sqrt{2} \frac{\cos{\theta_P}}{3} \left \{
\frac{5z_A+1}{3z_A(1+v)}+v \frac{1-z_A}{1+2z_A^2}-\frac{\Delta_A}{2}\right\}-
\frac{\epsilon_2}{\sqrt{3}}\,,
\end{array}
\right.
\label{DD2}
\ee
using the renormalized PS fields defined in Appendix \ref{BB} and
their specific parameters.

The pion sector of the $APPP$ Lagrangian is \cite{ExtMod3}~:
\be
\begin{array}{ll}
\displaystyle
{\cal L}_{APPP}=  -iE  g_{\pi^0 \pi^+\pi^-\gam}
\epsilon^{\mu \nu \alpha \beta} A_\mu
 \partial_\nu \pi^0
 \partial_\alpha  \pi^- \partial_\beta  \pi^+~~,&
\end{array}
\label{DD3}
\ee
with
\be
\displaystyle
E=-\frac{e }{ \pi^2 f_\pi^3} \left[ 1- \frac{3}{4}(c_1-c_2+c_4)
\right]
\ee
and~:
\be
\displaystyle g_{\pi^0 \pi^+\pi^-\gam} =\frac{1}{4}
\left[ 1-\frac{\Delta_A}{2} + \frac{\cos{\theta_P}}{\sqrt{3}}
\left \{\epsilon_1+\sqrt{2}\epsilon_2\right \}
 -\frac{\sin{\theta_P}}{\sqrt{3}}
\left \{\sqrt{2}\epsilon_1-\epsilon_2\right \} \right]\epo
\label{DD4}
\ee
One may prefer  reexpressing  this formula by~:
\be
\displaystyle
g_{\pi^0 \pi^+\pi^-\gam} =\frac{1}{4}\left [
1-\frac{\Delta_A}{2} +\epsilon_2 \cos{\delta_P}-\epsilon_1 \sin{\delta_P}
\right ]\,,
\label{DD5}
\ee
using the angle $\delta_P$ defined by Equations (\ref{BB8}).

\section{The $VPPP$ Lagrangian and its Renormalization}
\label{EE}
\indent \indent The $VPPP$ anomalous Lagrangian in terms of {\it ideal}
 vector fields can be written~:
 \be
\left \{
\begin{array}{ll}
\displaystyle
{\cal L}_{VPPP}=iD\epsilon^{\mu \nu \alpha \beta}  \left\{
\left [ g_{\rho \pi}^0 \partial_\nu \pi^0+g_{\rho \eta}^0 \partial_\nu \eta+
+g_{\rho \eta^\prime}^0 \partial_\nu \eta^\prime \right ] ~\rho^{I}_\mu\right.\,,\\[0.5cm]
\displaystyle
~~~+\left [ g_{\omg \pi}^0 \partial_\nu \pi^0+g_{\omg \eta}^0 \partial_\nu \eta+
+g_{\omg \eta^\prime}^0 \partial_\nu \eta^\prime \right ]~\omega_\mu^{I}
+ g_{\phi \pi}^0\partial_\nu \pi^0 ~\phi_\mu^{I}
 \left\}\, \partial_\alpha\pi^-  \partial_\beta \pi^+ \right.\,,\\[0.5cm]
\displaystyle
{\rm ~~with~~} D=-\frac{ 3 g (c_1-c_2-c_3)}{4 \pi^2 f_\pi^3}\,,
\end{array}
\right.
\label{EE1}
\ee
where one has limited oneself to display the $P_0\pi^+\pi^- $ sector. The leading terms
of these couplings are~:
\be
 \left \{
\begin{array}{ll}
\displaystyle
g^0_{\rho \pi^0 }=\frac{1}{4} \left [
 \frac{\Delta_A}{2} +\epsilon_1 ~\sin{\delta_P}
-\epsilon_2 ~\cos{\delta_P} \right ]
\,,\\[0.5cm]
\displaystyle
g^0_{\omg \pi^0 }= -\frac{3}{4} \left [ 1+ \frac{2\xi_0 + \xi_8}{3}    \right ]
\,,\\[0.5cm]
\displaystyle
g^0_{\phi \pi^0}=-\frac{\sqrt{2}}{4} \left [\xi_0 -\xi_8 \right ]\,,
\end{array}
\right.
\label{EE2}
\ee

\be
\hspace{-1.5cm}
 \left \{
\begin{array}{ll}
\displaystyle
g^0_{\rho \eta }=-\frac{1}{12} \left [ (1-x^\prime)~\sqrt{2} \cos{\delta_P}-
\left \{ (1+ 2 x)+ 3 \xi_3 \right \}~\sin{\delta_P}
\right ]  \,,\\[0.5cm]
\displaystyle
g^0_{\omg \eta }= -\frac{3}{4} \left [\frac{ \Delta_A }{2}~ \sin{\delta_P}-\epsilon_1
\right ]
\,,\\[0.5cm]
\displaystyle
g^0_{\phi \eta}= 0 \,,
\end{array}
\right.
\label{EE3}
\ee

\be
\hspace{-1.5cm}
 \left \{
\begin{array}{ll}
\displaystyle
g^0_{\rho \etp }=-\frac{1}{12} \left [  (1-x^\prime)~\sqrt{2} \sin{\delta_P}
+\left \{ (1+ 2 x)+ 3 \xi_3 \right \}~\cos{\delta_P}
\right ]
\,,\\[0.5cm]
\displaystyle
g^0_{\omg \etp }= \frac{3}{4} \left [\frac{ \Delta_A }{2}~\cos{\delta_P}+\epsilon_2
\right ]\,,\\[0.5cm]
\displaystyle
g^0_{\phi \etp}= 0 \epo
\end{array}
\right.
\label{EE4}
\ee
One may  note, once again  the clarity reached by using $\delta_P$
instead of $\theta_P$.

The couplings to the renormalized fields can be derived  from  those to the ideal fields
by means of the ${\cal R}(s)$ matrix defined in the body of the text~:
 \be
 g^R_P = \widetilde{{\cal R}}(s) g^0_{P_0}
\label{EE5}
\ee
the elements of the $g^0_{P_0}$ vectors being defined in  Equations (\ref{EE2}),
Equations (\ref{EE3}) or Equations (\ref{EE4}) for resp. the $P_0=~\pi^0$ or $~\eta,~\etp$
mesons.

For the couplings of the $\phi$ meson, at leading  order in breaking, one gets~:
 \be
 \displaystyle g^R_{\phi \pi^0}(s) = -\frac{3}{4} \left \{  [\psi_0 +\gamma(s)]
-\frac{\sqrt{2}}{3} [\xi_0-\xi_8] \right \}\epo
\label{EE6}
\ee
One clearly sees that, at the $\phi$ mass,  one obtains
a non-vanishing direct coupling of the $\phi$ to 3 pions, additionally
$s$-dependent.

                  \bibliographystyle{h-physrev}
                     \bibliography{vmd2_z3}

\end{document}